\documentclass[12pt]{iopart}

\usepackage{iopams}
\usepackage{graphicx}
\usepackage{mathrsfs}
\usepackage{color}
\usepackage[table]{xcolor}
\usepackage{hyperref}
\hypersetup{colorlinks, citecolor=black, filecolor=black, linkcolor=black, urlcolor=black}

\expandafter\let\csname equation*\endcsname\relax
\expandafter\let\csname endequation*\endcsname\relax
\usepackage{amsmath}

\definecolor{grayish}{RGB}{230,230,230}

\newcommand{\refEq}[1] {(\ref{#1})}
\newcommand{\superscript}[1]{\ensuremath{^{\textrm{#1}}}}

\newcommand{\Sin}[1]{\ensuremath{\sin \left( #1 \right)}}
\newcommand{\Cos}[1]{\ensuremath{\cos \left( #1 \right)}}

\newcommand{\ArcTan}[1]{\ensuremath{\text{arctan} \left( #1 \right)}}
\newcommand{\Exp}[1]{\ensuremath{\exp \left( #1 \right)}}
\newcommand{\Ln}[1]{\ensuremath{\ln \left( #1 \right)}}
\newcommand{\Order}[1]{\ensuremath{O \left( #1 \right)}}
\newcommand{\Nabla}{\ensuremath{\vec{\nabla}}}

\newcommand{\dlpdtheta}{\ensuremath{\frac{\partial l_{p}}{\partial \theta}}}
\newcommand{\dlpdthetaPrime}{\ensuremath{\frac{\partial l_{p}}{\partial \theta'}}}

\begin{document}

\title[Effect of the Shafranov shift and $\beta'$ on intrinsic momentum transport]{Effect of the Shafranov shift and the gradient of $\beta$ on intrinsic momentum transport in up-down asymmetric tokamaks}

\author{Justin Ball\superscript{1,2}, Felix I. Parra\superscript{1,2}, Jungpyo Lee\superscript{3}, and Antoine J. Cerfon\superscript{4}}

\address{\superscript{1} Rudolf Peierls Centre for Theoretical Physics, University of Oxford, Oxford OX1 3NP, United Kingdom}
\address{\superscript{2} Culham Centre for Fusion Energy, Culham Science Centre, Abingdon OX14 3DB, United Kingdom}
\address{\superscript{3} Plasma Science and Fusion Center, Massachusetts Institute of Technology, Cambridge, MA 02139, USA}
\address{\superscript{4} Courant Institute of Mathematical Sciences, New York University, New York, NY 10012, USA}
\ead{Justin.Ball@physics.ox.ac.uk}

\begin{abstract}

Tokamaks with up-down asymmetric poloidal cross-sections spontaneously rotate due to turbulent transport of momentum. In this work, we investigate the effect of the Shafranov shift on this intrinsic rotation, primarily by analyzing tokamaks with tilted elliptical flux surfaces. By expanding the Grad-Shafranov equation in the large aspect ratio limit we calculate the magnitude and direction of the Shafranov shift in tilted elliptical tokamaks. The results show that, while the Shafranov shift becomes up-down asymmetric and depends strongly on the tilt angle of the flux surfaces, it is insensitive to the shape of the current and pressure profiles (when the geometry, total plasma current, and average pressure gradient are kept fixed). Next, local nonlinear gyrokinetic simulations of these MHD equilibria are performed with GS2, which reveal that the Shafranov shift can significantly enhance the momentum transport. However, to be consistent, the effect of $\beta'$ (i.e. the radial gradient of $\beta$) on the magnetic equilibrium was also included, which was found to significantly reduce momentum transport. Including these two competing effects broadens the rotation profile, but leaves the on-axis value of the rotation roughly unchanged. Consequently, the shape of the $\beta$ profile has a significant effect on the rotation profile of an up-down asymmetric tokamak.

\end{abstract}



\section{Introduction}
\label{sec:introduction}

Current experiments generally rely on neutral beams, directed toroidally, in order to induce the plasma to rotate. This toroidal rotation has been experimentally proven to stabilize resistive wall modes (a class of MHD instabilities that can cause disruptions) thereby enabling sustained discharges with a plasma $\beta$ that violates the Troyon limit \cite{StraitExpRWMstabilizationD3D1995, SabbaghExpRWMstabilizationNSTX2002, deVriesRotMHDStabilization1996, ReimerdesRWMmachineComp2006}. Since most designs of reactor-scale devices violate the Troyon limit \cite{PolevoiITERscenario2002} and are intolerant to disruptions \cite{SugiharaITERdisruptionVDE2007}, driving fast toroidal rotation in large devices is critical. Throughout this paper we will use the ITER design \cite{AymarITERSummary2001} as an example of a large device in order to provide realistic numbers and a frame of reference. Here we note that numerical analysis indicates that the slowest rotation able to stabilize resistive wall modes in ITER has an on-axis Alfv\'{e}n Mach number around $0.5\% - 5\%$ \cite{LiuITERrwmStabilization2004}. The precise value depends on the exact numerical model used, but is significantly lower for broader rotation profiles.

Driving rotation in large plasmas is difficult because they have more inertia and require more energetic neutral beams to penetrate to the magnetic axis. Because of the velocity scalings of momentum versus energy, more energetic neutral beams inject less momentum per unit power \cite{ParraMomentumTransitions2011}. This explains why the neutral beams in ITER are not expected to drive substantial toroidal rotation \cite{LiuITERrwmStabilization2004, DoyleITERconfinement2007}. Therefore, unless the momentum pinch effect can be used to dramatically amplify the driven rotation \cite{WeisenMomPinch2012} or bring in rotation from the edge \cite{OmotaniEdgeNeutralRot2016, TalaMomPinch2011}, we must turn to ``intrinsic'' rotation (i.e. spontaneous rotation that is observed in the absence of external momentum injection \cite{RiceExpIntrinsicRotMeas2007}). This rotation is generated by the plasma through turbulent transport of momentum. Because it is generated by the plasma itself, intrinsic rotation would be expected to scale well to large devices. However, the gyrokinetic equation, which is thought to govern turbulence in the core of tokamaks, possesses a particular symmetry \cite{ParraUpDownSym2011, SugamaUpDownSym2011, CamenenPRLSim2009} that implies this intrinsic momentum flux must be small in $\rho_{\ast} \equiv \rho_{i} / a \ll 1$, the ratio of the ion gyroradius to the tokamak minor radius. Fortunately, there is one mechanism that breaks this symmetry and is capable of spontaneously generating rotation in the core of a stationary plasma: up-down asymmetry in the magnetic geometry.

If the flux surfaces in a tokamak are up-down asymmetric (i.e. do not have mirror symmetry about the midplane), then the momentum flux is no longer constrained to be small in $\rho_{\ast} \ll 1$. In principle, up-down asymmetric flux surfaces are no more difficult to create than up-down symmetric surfaces, but all existing devices have been designed with nearly up-down symmetric flux surface shapes in mind. Hence, the ability of a device to create a particular up-down asymmetric surface depends strongly on the specifics of the shaping coils and the vacuum vessel. The TCV tokamak \cite{HofmannTCVOverview1994}, which was designed to accommodate strong shaping, has been used to experimentally investigate flux surfaces with a single up-down asymmetric shaping mode \cite{CamenenPRLExp2010}. As expected, a large change in the rotation profile was observed when the up-down asymmetry of the plasma shape was varied. Subsequent gyrokinetic simulations \cite{BallMomUpDownAsym2014}, which give results consistent with the TCV experiments, indicate that up-down asymmetry is a feasible method to generate the current experimentally-measured rotation levels in reactor-sized devices.

Configurations with only a single up-down asymmetric shaping mode drive rotation through the direct interaction of toroidicity (which defines up versus down) and the shaping mode. Recent analytic work \cite{BallMirrorSymArg2016, BallDoctoralThesis2016} demonstrates that adding a second shaping effect introduces two new physical mechanisms that have the potential to enhance the rotation. First, the tilting symmetry presented in reference \cite{BallMirrorSymArg2016} shows that flux surfaces with only a single shaping mode $m$ must have momentum flux that is exponentially small in $m \gg 1$. Including two shaping effects allows them to beat together to produce an up-down asymmetric envelope on the connection length-scale that can interact with toroidicity to drive rotation. This breaks the tilting symmetry and permits the momentum flux to have a stronger scaling (i.e. polynomially small in $m \gg 1$). These scalings indicate that using low order shaping effects and combining different shaping effects to make asymmetric envelopes can effectively drive fast intrinsic rotation. Physically, high order shaping effects do not effectively drive rotation because the turbulent eddies, which are extended along the magnetic field line, average over small-scale variation in the magnetic equilibrium. Second, looking in the screw pinch limit (i.e. large aspect ratio limit) of a tokamak we learn that flux surfaces with mirror symmetry about any line in the poloidal plane do not drive any intrinsic rotation \cite{BallDoctoralThesis2016}. Including a second shaping effect can break mirror symmetry, allowing rotation to be driven through the direct interaction between the two shaping effects (completely independently of toroidicity). These two mechanisms dominate in certain regimes (i.e. the $m \gg 1$ and large aspect ratio limits) and bring in fundamentally new physics, but their importance in more realistic geometries is still unclear.

Together all of these results indicate that low order shaping effects are optimal for maximizing intrinsic rotation and it is important to explore non-mirror symmetric configurations with an up-down asymmetric envelope. In this context, there are two options. The first is to introduce up-down asymmetric elongation using external poloidal field coils and then rely on the Shafranov shift (i.e. the shift in the magnetic axis due to toroidicity) to break the mirror and tilting symmetries. This appears optimal because it makes use of the lowest possible shaping modes (i.e. $m = 1$ and $m = 2$). However, this strategy has the drawbacks that the effect of the Shafranov shift is formally small in aspect ratio and the direction and magnitude of the shift is a consequence of the plasma $\beta$ profile and the global MHD equilibrium. Hence it is not independently controlled by external coils. The second option is to use external coils to introduce both elongation and triangularity (i.e. $m = 2$ and $m = 3$ shaping) into the flux surface shape in order to directly break mirror symmetry and create an envelope that breaks the tilting symmetry. Both modes are lowest order in aspect ratio and can be directly controlled by external shaping magnets, but this relies on higher order shaping modes than the first option. Practically speaking these two strategies are intertwined as the divertor geometry nearly always introduces some triangularity into the flux surfaces and the Shafranov shift exists regardless of the shape of flux surfaces. Nevertheless, for simplicity it is useful to distinguish them and examine each option independently. In this work we will explore the former: the influence of the Shafranov shift and the effect of the $\beta$ profile on the turbulent momentum flux in the core of tokamaks.

In section \ref{sec:MHDequil} we use the Grad-Shafranov equation to estimate the magnitude and direction of the Shafranov shift in a tokamak with a tilted elliptical boundary. To do so we start in section \ref{subsec:analyticCalc} by expanding the Grad-Shafranov equation in the large aspect ratio limit to write the lowest and next order analytic solutions for a linear toroidal current profile as a Fourier series in poloidal angle. In section \ref{subsec:numCoeffCalc}, we calculate the Fourier coefficients needed to match the tilted elliptical boundary condition. In section \ref{subsec:ECOMcomparison}, we find the dependence of the Shafranov shift on the boundary tilt angle and show that the shift is insensitive to the shape of both the current and pressure profiles (when the geometry, total plasma current, and average pressure gradient are kept fixed). These analytic results are verified using equilibrium calculations performed with the numerical Grad-Shafranov solver ECOM \cite{LeeECOM2015}. Section \ref{sec:gyrokineticSims} contains the results from nonlinear gyrokinetic simulations of the equilibria calculated in section \ref{sec:MHDequil}. Section \ref{subsec:inputParameters} starts by using the results of the MHD analysis to generate local equilibria for the gyrokinetic simulations. Section \ref{subsec:results} details the results of several numerical scans aimed at illuminating the effect of the Shafranov shift and the $\beta$ profile on momentum transport. In section \ref{subsec:betaPrime} we discuss the sensitivity of the momentum transport to changes in the magnetic equilibrium caused by altering the local gradient of $\beta$. Furthermore, in section \ref{subsec:betaProfile} we consider the impact of changing the shape of the radial profile of $\beta$. Section \ref{sec:conclusions} contains a summary of the results and some concluding remarks.

\section{MHD equilibrium calculation of the Shafranov shift}
\label{sec:MHDequil}

In this section we will calculate a general analytic solution to the Grad-Shafranov equation for a linear (in poloidal flux) toroidal current profile to lowest and next order in an expansion in large aspect ratio. The zeroth and first order solutions are needed because the Shafranov shift does not appear to lowest order. The analytic solution will contain Fourier coefficients, which in general must be calculated numerically to achieve a tilted elliptical boundary flux surface. Making use of our numerically calculated Fourier coefficients, we will argue that varying the shape of the current profile and the shape of the pressure profile (while keeping the geometry, total plasma current, and average pressure gradient fixed) does not significantly affect the Shafranov shift. These theoretical results are verified against the equilibrium code ECOM.
Due to the insensitivity of the Shafranov shift to the exact current and pressure profiles, we are free to use the constant current case for input into the gyrokinetic simulations of section \ref{sec:gyrokineticSims}. This is helpful as the Fourier coefficients in the constant current equilibria can be calculated analytically.

\subsection{Analytic solution for a linear current profile}
\label{subsec:analyticCalc}

The geometry of a tokamak equilibrium is governed by the Grad-Shafranov equation \cite{GradGradShafranovEq1958},
\begin{align}
   R^{2} \Nabla \cdot \left( \frac{\Nabla \psi}{R^{2}} \right) = - \mu_{0} R^{2} \frac{d p}{d \psi} - I \frac{d I}{d \psi} , \label{eq:gradShaf}
\end{align}
where $R$ is the tokamak major radial coordinate, $\psi$ is the poloidal magnetic flux divided by $2 \pi$, $\mu_{0}$ is the permeability of free space, $p$ is the plasma pressure, $I \equiv R B_{\zeta}$ is the toroidal magnetic field flux function, $\vec{B}$ is the magnetic field, and $\zeta$ is the toroidal angle. Note that the effect of the $\beta \equiv 2 \mu_{0} p / B^{2}$ profile only enters through the gradient of the pressure. In order to investigate the behavior of the Shafranov shift in a tilted elliptical geometry we will expand in the large aspect ratio limit, i.e. $\epsilon \equiv a / R_{0} \ll 1$ where $a$ is the tokamak minor radius and $R_{0}$ is the major radial location of the center of the boundary flux surface. We will take the typical orderings for a low $\beta$, ohmically heated tokamak \cite{FreidbergIdealMHD1987pg126}:
\begin{align}
   \frac{B_{p}}{B_{0}} \sim \epsilon, \hspace{10pt}
   \frac{2 \mu_{0} p}{B_{0}^{2}} \sim \epsilon^{2} , \label{eq:gradShafOrderings}
\end{align}
where $B_{0}$ is the on-axis toroidal magnetic field and $B_{p} = | \Nabla \psi | / R$ is the poloidal magnetic field. Also, we must expand $\psi = \psi_{0} + \psi_{1} + \ldots$, $I = I_{0} + I_{1} + I_{2}$, and $p = p_{2}$, where the subscripts indicate the order of the quantity in $\epsilon$ relative to the lowest order contributions of $\psi_{0} \sim a R_{0} B_{p} \sim \epsilon^{2} R_{0}^{2} B_{0}$ and $I_{0} \sim R_{0} B_{0}$. To $O \left( \epsilon^{-1} B_{0} \right)$ we find that the Grad-Shafranov equation is
\begin{align}
   - I_{0} \frac{d I_{1}}{d \psi_{0}} - I_{1} \frac{d I_{0}}{d \psi_{0}} = 0 . \label{eq:gradShafNegOrder}
\end{align}
Since $I_{0} = R_{0} B_{0}$ is a constant, this requires that $I_{1}$ also be a constant. We are free to absorb $I_{1}$ into $I_{0}$ and set $I_{1} = 0$. Hence, using $r \sim a$ we find to $O \left( B_{0} \right)$ that
\begin{align}
   \frac{1}{r} \frac{\partial}{\partial r} \left( r \frac{\partial \psi_{0}}{\partial r} \right) + \frac{1}{r^{2}} \frac{\partial^{2} \psi_{0}}{\partial \theta^{2}} = - \mu_{0} R_{0}^{2} \frac{dp_{2}}{d \psi_{0}} - I_{0} \frac{dI_{2}}{d \psi_{0}}  \label{eq:gradShafLowestOrder}
\end{align}
and to $O \left( \epsilon B_{0} \right)$ that
\begin{align}
   \frac{1}{r} \frac{\partial}{\partial r} \left( r \frac{\partial \psi_{1}}{\partial r} \right) &+ \frac{1}{r^2} \frac{\partial^{2} \psi_{1}}{\partial \theta^{2}} = \psi_{1} \frac{d}{d \psi_{0}} \left( -\mu_{0} R_{0}^{2} \frac{d p_{2}}{d \psi_{0}} - I_{0} \frac{d I_{2}}{d \psi_{0}} \right)\label{eq:gradShafNextOrder} \\
   & - 2 \mu_{0} r R_{0} \frac{d p_{2}}{d \psi_{0}} \Cos{\theta} + \frac{\Cos{\theta}}{R_{0}} \frac{\partial \psi_{0}}{\partial r} - \frac{\Sin{\theta}}{r R_{0}} \frac{\partial \psi_{0}}{\partial \theta} , \nonumber
\end{align}
where $r \equiv \sqrt{\left( R - R_{0} \right)^{2} + Z^{2}}$ is the distance from the center of the boundary flux surface, $\theta \equiv \ArcTan{Z / \left( R - R_{0} \right)}$ is the usual cylindrical poloidal angle, and the axial location of the center of the boundary flux surface is assumed to be at $Z = 0$.

Like references \cite{BallMomUpDownAsym2014, BallMastersThesis2013}, we will develop our intuition by investigating how the Shafranov shift changes with three simple, but realistic toroidal current profiles: constant, linear peaked, and linear hollow (in poloidal flux). Using Ampere's law and $\vec{B} = I \Nabla \zeta + \Nabla \zeta \times \Nabla \psi$ one can show that the toroidal current is related to the right-hand side of the Grad-Shafranov equation through
\begin{align}
   - \mu_{0} R^{2} \frac{d p}{d \psi} - I \frac{d I}{d \psi} = \mu_{0} j_{\zeta} R , \label{eq:toroidalCur}
\end{align}
where $j_{\zeta}$ is the toroidal current density in the plasma. We will parameterize all three profiles (i.e. constant, peaked, and hollow) by
\begin{align}
   - \mu_{0} R_{0}^{2} \frac{dp_{2}}{d \psi_{0}} - I_{0} \frac{dI_{2}}{d \psi_{0}} = \mu_{0} j_{\zeta 0} R_{0} = j_{N} \left( 1 - f_{N} \psi_{0} \right) , \label{eq:currentProfiles}
\end{align}
where $j_{\zeta 0}$ is the lowest order current density in the aspect ratio expansion, $j_{N}$ is a positive constant, $f_{N} \in \left[ - \psi_{0 b}^{-1}, \psi_{0 b}^{-1} \right]$ determines the slope of the current profile, and $\psi_{0 b}$ is the lowest order value of the poloidal flux on the boundary flux surface. The constant current case is achieved by setting $f_{N} = 0$, while the hollow current case arises from allowing $f_{N}$ to be negative.

Additionally, from equation \refEq{eq:gradShafNextOrder} we see that it will be necessary to distinguish the contributions to the current from the pressure and magnetic field terms in equation \refEq{eq:toroidalCur}. Like the toroidal current, we will assume the pressure gradient has the form of
\begin{align}
   - \mu_{0} R_{0}^{2} \frac{dp_{2}}{d \psi_{0}} =& ~ j_{N p} \left( 1 - f_{N p} \psi_{0} \right) , \label{eq:pressureProfShape}
\end{align}
where $j_{N p}$ and $f_{N p} \in \left[ - \psi_{0 b}^{-1}, \psi_{0 b}^{-1} \right]$ are constants. By equation \refEq{eq:currentProfiles}, this implies that the toroidal magnetic field flux function term must be
\begin{align}
   - I_{0} \frac{dI_{2}}{d \psi_{0}} =& ~ j_{N I} \left( 1 - f_{N I} \psi_{0} \right) ,
\end{align}
where
\begin{align}
   j_{N I} \equiv& ~ j_{N} - j_{N p} \\
   f_{N I} \equiv& ~ \frac{1}{j_{N I}} \left( j_{N} f_{N} - j_{N p} f_{N p} \right)
\end{align}
are constants.

\subsubsection{Solutions to the $O \left( B_{0} \right)$ Grad-Shafranov equation.}
\label{subsubsec:GradShafSolLowestOrder}

Like references \cite{GreeneEquilibrium1971, LaoGradShafExpansion1981, RodriguesGradShafAspectRatio2004, RodriguesGradShafNoniterative2009}, we will solve the $O \left( B_{0} \right)$ Grad-Shafranov equation by Fourier analyzing the magnetic flux in poloidal angle as
\begin{align}
   \psi_{0} \left( r, \theta \right) =& ~ \psi_{0, 0}^{C} \left( r \right) + \sum_{m = 1}^{\infty} \left[ \psi_{0, m}^{C} \left( r \right) \Cos{m \theta} + \psi_{0, m}^{S} \left( r \right) \Sin{m \theta} \right] , \label{eq:gradShafLowestOrderSols}
\end{align}
where $m$ is an integer representing the poloidal flux surface shaping mode number. Using equation \refEq{eq:gradShafLowestOrderSols} we can rewrite equation \refEq{eq:gradShafLowestOrder} as
\begin{align}
   \frac{1}{r} \frac{d}{dr} \left( r \frac{d \psi_{0, m}^{T}}{dr} \right) &+ \left( f_{N} j_{N} - \frac{m^{2}}{r^{2}} \right) \psi_{0, m}^{T} \left( r \right) = j_{N} \delta_{m, 0} , \label{eq:gradShafFourierLowest}
\end{align}
where $m \geq 0$, $\delta_{i, j}$ is the Kronecker delta, and $T = C, S$ is a superscript that indicates the sine or cosine mode. The solutions to this equation with zero poloidal flux at the magnetic axis are
\begin{align}
   \psi_{0, 0}^{C} \left( r \right) =& - \frac{1}{f_{N}} \left( J_{0} \left( \sqrt{f_{N} j_{N}} r \right) - 1 \right) \label{eq:gradShafSol0modeFourierLowest} \\
   \psi_{0, m}^{C} \left( r \right) =& ~ C_{0, m} \frac{m! ~ 2^{m}}{\left( f_{N} j_{N} \right)^{m/2}} J_{m} \left( \sqrt{ f_{N} j_{N}} r \right) \\
   \psi_{0, m}^{S} \left( r \right) =& ~ S_{0, m} \frac{m! ~ 2^{m}}{\left( f_{N} j_{N} \right)^{m/2}} J_{m} \left( \sqrt{ f_{N} j_{N}} r \right) , \label{eq:gradShafSolMmodeFourierLowest}
\end{align}
where $m > 0$, $J_{m}$ is the $m^{\text{th}}$ order Bessel function of the first kind. The Fourier coefficients $C_{0, m}$ and $S_{0, m}$ are determined by the boundary conditions at the plasma edge, which is physically controlled by the locations and currents of external plasma shaping coils. Using trigonometric identities, equation \refEq{eq:gradShafLowestOrderSols} and equations \refEq{eq:gradShafSol0modeFourierLowest} through \refEq{eq:gradShafSolMmodeFourierLowest} can be rewritten as
\begin{align}
   \psi_{0} \left( r, \theta \right) =& - \frac{1}{f_{N}} \left( J_{0} \left( \sqrt{f_{N} j_{N}} r \right) - 1 \right)  \label{eq:gradShafLowestOrderSolsTilt} \\
   &+ \sum _{m=1}^{\infty} N_{0, m} \frac{m! ~ 2^{m}}{\left( f_{N} j_{N} \right)^{m/2}} J_{m} \left( \sqrt{f_{N} j_{N}} r \right) \Cos{m \left( \theta + \theta_{t 0, m} \right)} , \nonumber
\end{align}
where $N_{0, m} \equiv \sqrt{ C_{0, m}^{2} + S_{0, m}^{2}}$ is the magnitude of the Fourier mode and $\theta_{t 0, m} \equiv - \ArcTan{S_{0, m} / C_{0, m}} / m$ is the Fourier mode tilt angle.

Note that for the constant current case  (i.e. $f_{N} = 0$), equation \refEq{eq:gradShafLowestOrderSolsTilt} reduces to
\begin{align}
   \psi_{0} \left( r, \theta \right) =& ~ \frac{j_{N}}{4} r^{2} + \sum _{m=1}^{\infty} N_{0, m} r^{m} \Cos{m \left( \theta + \theta_{t 0, m} \right)} . \label{eq:gradShafLowestOrderSolsTiltConst}
\end{align}
To understand the hollow current case, it is useful to note the identity
\begin{align}
   J_{m} \left( i x \right) = i^{m} I_{m} \left( x \right) , \label{eq:modBesselFnFirstKind}
\end{align}
where $I_{m}$ is the $m^{\text{th}}$ order modified Bessel function of the first kind. From this we can demonstrate that equation \refEq{eq:gradShafLowestOrderSolsTilt} is equivalent to
\begin{align}
   \psi_{0} \left( r, \theta \right) =& \frac{1}{- f_{N}} \left( I_{0} \left( \sqrt{- f_{N} j_{N}} r \right) - 1 \right)  \label{eq:gradShafLowestOrderSolsTiltHollow} \\
   &+ \sum _{m=1}^{\infty} N_{0, m} \frac{m! ~ 2^{m}}{\left( - f_{N} j_{N} \right)^{m/2}} I_{m} \left( \sqrt{- f_{N} j_{N}} r \right) \Cos{m \left( \theta + \theta_{t 0, m} \right)} , \nonumber
\end{align}
which can be more easily applied to hollow toroidal current profiles (i.e. $f_{N} < 0$).

\subsubsection{Solutions to the $O \left( \epsilon B_{0} \right)$ Grad-Shafranov equation.}
\label{subsubsec:GradShafSolNextOrder}

In order to solve the $O \left( \epsilon B_{0} \right)$ equation we must first Fourier analyze the magnetic flux in poloidal angle. The lowest order Fourier-analyzed flux is given by equation \refEq{eq:gradShafLowestOrderSols} and equations \refEq{eq:gradShafSol0modeFourierLowest} through \refEq{eq:gradShafSolMmodeFourierLowest}. To next order, we can write
\begin{align}
   \psi_{1} \left( r, \theta \right) &= \psi_{1, 0}^{C} \left( r \right) + \sum_{m = 1}^{\infty} \left[ \psi_{1, m}^{C} \left( r \right) \Cos{m \theta} + \psi_{1, m}^{S} \left( r \right) \Sin{m \theta} \right] , \label{eq:gradShafNextOrderSols}
\end{align}
but we still must solve for $\psi_{1, m}^{C} \left( r \right)$ and $\psi_{1, m}^{S} \left( r \right)$ by substituting equations \refEq{eq:gradShafLowestOrderSols} and \refEq{eq:gradShafNextOrderSols} into equation \refEq{eq:gradShafNextOrder}. Since $\psi_{1, m}^{C} \left( r \right)$ and $\psi_{1, m}^{S} \left( r \right)$ do not depend on $\theta$, we can take each Fourier component of equation \refEq{eq:gradShafNextOrder} as a separate equation. This gives
\begin{align}
   \frac{1}{r} \frac{d}{dr} \left( r \frac{d \psi_{1, m}^{T}}{dr} \right) &+ \left( f_{N} j_{N} - \frac{m^{2}}{r^{2}} \right) \psi_{1, m}^{T} \left( r \right) = \Lambda_{m}^{T} \left( r \right) \label{eq:gradShafFourierNext}
\end{align}
for each Fourier mode $m$, where the inhomogeneous terms are given by $\Lambda_{m}^{T} \left( r \right)$. For $m = 0$ and $T = C$
\begin{align}
   \Lambda_{0}^{C} \left( r \right) &\equiv \frac{1}{R_{0}} \left[ \frac{1}{2} \frac{d \psi_{0, 1}^{C}}{dr} + \left( \frac{1}{2 r} - r f_{N p} j_{N p} \right) \psi_{0, 1}^{C} \left( r \right) \right] , \label{eq:inhomoTermsC0}
\end{align}
for $m = 1$ and $T =C$
\begin{align}
   \Lambda_{1}^{C} \left( r \right) &\equiv \frac{1}{R_{0}} \left[ \frac{1}{2} \frac{d \psi_{0, 2}^{C}}{dr} + \left( \frac{1}{r} - r f_{N p} j_{N p} \right) \psi_{0, 2}^{C} \left( r \right) \right. \label{eq:inhomoTermsC1} \\
      &+ \left. \frac{d \psi_{0, 0}^{C}}{dr} + 2 r j_{N p} \left( 1 - f_{N p} \psi_{0, 0}^{C} \left( r \right) \right) \right] , \nonumber
\end{align}
for $m = 1$ and $T = S$
\begin{align}
   \Lambda_{1}^{S} \left( r \right) &\equiv \frac{1}{R_{0}} \left[ \frac{1}{2} \frac{d \psi_{0, 2}^{S}}{dr} + \left( \frac{1}{r} - r f_{N p} j_{N p} \right) \psi_{0, 2}^{S} \left( r \right) \right] , \label{eq:inhomoTermsS1}
\end{align}
and for all other $m$ and $T = C, S$
\begin{align}
   \Lambda_{m}^{T} \left( r \right) &\equiv \frac{1}{R_{0}} \left[ \frac{1}{2} \frac{d \psi_{0, m + 1}^{T}}{dr} + \left( \frac{m+1}{2 r} - r f_{N p} j_{N p} \right) \psi_{0, m + 1}^{T} \left( r \right) \right. \label{eq:inhomoTermsTm} \\
      &+ \left. \frac{1}{2} \frac{d \psi_{0, m - 1}^{T}}{dr} - \left( \frac{m-1}{2 r} + r f_{N p} j_{N p} \right) \psi_{0, m - 1}^{T} \left( r \right)  \right] . \nonumber
\end{align}

Equation \refEq{eq:gradShafFourierNext} can be solved using the method of variation of parameters, yielding
\begin{align}
   \psi_{1, m}^{T} \left( r \right) &= - \frac{\pi}{2} J_{m} \left( \sqrt{f_{N} j_{N}} r \right) \int_{0}^{r} d r' ~ r' Y_{m} \left( \sqrt{f_{N} j_{N}} r' \right) \Lambda_{m}^{T} \left( r' \right) \nonumber \\
   &+ \frac{\pi}{2} Y_{m} \left( \sqrt{f_{N} j_{N}} r \right) \int_{0}^{r} d r' ~ r' J_{m} \left( \sqrt{f_{N} j_{N}} r' \right) \Lambda_{m}^{T} \left( r' \right) \label{eq:psiFourierNextOrderCoeffs} \\
   &+ T_{1, m} \frac{m! ~ 2^{m}}{\left( f_{N} j_{N} \right)^{m/2}} J_{m} \left( \sqrt{f_{N} j_{N}} r \right) , \nonumber
\end{align}
where we have imposed regularity at the origin, $Y_{m}$ is the $m^{\text{th}}$ order Bessel function of the second kind, and $T_{1, m} = C_{1, m}, S_{1, m}$ are Fourier coefficients determined by the boundary conditions at the plasma edge. Combining equations \refEq{eq:gradShafNextOrderSols}, \refEq{eq:inhomoTermsC0} through \refEq{eq:inhomoTermsTm}, and \refEq{eq:psiFourierNextOrderCoeffs} gives the complete solution to the $O \left( \epsilon B_{0} \right)$ Grad-Shafranov equation for an arbitrary boundary condition. 

To understand the hollow current case (i.e. $f_{N} < 0$), we will use equation \refEq{eq:modBesselFnFirstKind} and the identity
\begin{align}
   Y_{m} \left( i x \right) = i^{m+1} I_{m} \left( x \right) - \frac{2}{\pi} i^{-m} K_{m} \left( x \right) ,
\end{align}
where $K_{m}$ is the $m^{\text{th}}$ order modified Bessel function of the second kind. This enables equation \refEq{eq:psiFourierNextOrderCoeffs} to be reformulated as
\begin{align}
   \psi_{1, m}^{T} \left( r \right) =& ~ I_{m} \left( \sqrt{- f_{N} j_{N}} r \right) \int_{0}^{r} d r' ~ r' K_{m} \left( \sqrt{ - f_{N} j_{N}} r' \right) \Lambda_{m}^{T} \left( r' \right) \nonumber \\
   &- K_{m} \left( \sqrt{- f_{N} j_{N}} r \right) \int_{0}^{r} d r' ~ r' I_{m} \left( \sqrt{- f_{N} j_{N}} r' \right) \Lambda_{m}^{T} \left( r' \right) \label{eq:psiFourierNextOrderCoeffsHollow} \\
   &+ T_{1, m} \frac{m! ~ 2^{m}}{\left( - f_{N} j_{N} \right)^{m/2}} I_{m} \left( \sqrt{- f_{N} j_{N}} r \right) . \nonumber
\end{align}
For a constant current profile (i.e. $f_{N} = 0$), we can take the limit of equations \refEq{eq:gradShafNextOrderSols}, \refEq{eq:inhomoTermsC0} through \refEq{eq:inhomoTermsTm}, and \refEq{eq:psiFourierNextOrderCoeffs} as $f_{N} j_{N} \to 0$ to find
\begin{align}
   \psi_{1} \left( r, \theta \right) =& \frac{1}{4 R_{0}} \left[ \left( \frac{j_{N} + 4 j_{N p}}{4} r^{3} - \frac{j_{N} f_{N p} j_{N p}}{12} r^{5} \right) \Cos{\theta} \right. \nonumber \\
   &+ \sum_{m = 2}^{\infty} \left( r^{m+1} - \frac{f_{N p} j_{N p}}{2 \left( m + 1 \right)} r^{m+3} \right) N_{0, m} \Cos{ \left( m - 1 \right) \theta + m \theta_{t 0, m}} \nonumber \\
   &- \left. \sum_{m = 2}^{\infty} \frac{f_{N p} j_{N p}}{m + 2} r^{m+3} N_{0, m} \Cos{\left( m + 1 \right) \theta + m \theta_{t 0, m}} \right] \label{eq:psiNextOrderSolConst} \\
   &+ \sum_{m = 0}^{\infty} r^{m} N_{1, m} \Cos{m \left( \theta + \theta_{t 1, m} \right)} , \nonumber
\end{align}
where $N_{1, m} \equiv \sqrt{ C_{1, m}^{2} + S_{1, m}^{2}}$ is the magnitude of the next order Fourier mode, $\theta_{t 1, m} \equiv - \ArcTan{S_{1, m}/C_{1, m}} / m$ is the next order Fourier mode tilt angle, and we have used equation \refEq{eq:gradShafLowestOrderSolsTiltConst} along with
\begin{align}
   \lim_{f_{N} j_{N} \to 0} \frac{m! ~ 2^{m}}{\left( f_{N} j_{N} \right)^{m/2}} J_{m} \left( \sqrt{f_{N} j_{N}} r \right) =& ~ r^{m} \label{eq:BesselFnExpansionOn} \\
   \lim_{f_{N} j_{N} \to 0} Y_{m} \left( \sqrt{f_{N} j_{N}} r \right) =& - \frac{1}{m \pi} \frac{m! ~ 2^{m}}{\left( f_{N} j_{N} \right)^{m/2}} r^{-m} \label{eq:modifiedBesselFnExpansionOn}
\end{align}
for $m \neq 0$. The first line of equation \refEq{eq:psiNextOrderSolConst} contains the direct effect of toroidicity on the equilibrium, i.e. the Shafranov shift. The second and third lines show that a zeroth order shaping mode $m$ splits into two modes, $m-1$ and $m+1$, at first order. The last line contains the homogeneous solution, which enables an arbitrary boundary condition to be satisfied.

\subsection{Solution for a tilted elliptical boundary condition}
\label{subsec:numCoeffCalc}

\begin{figure}
 \centering
 \includegraphics[width=0.4\textwidth]{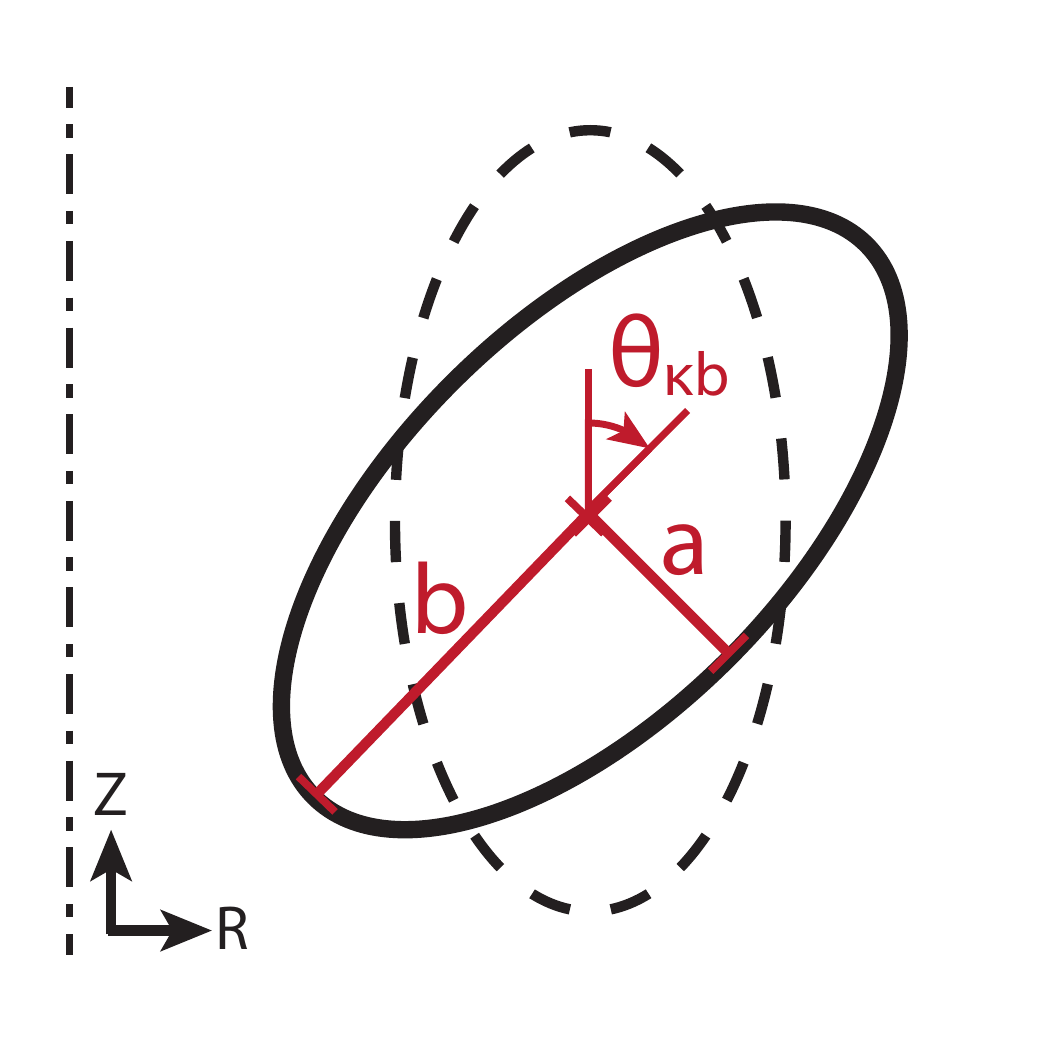}
 \caption{An illustration of the boundary flux surface (black, solid) with the untilted boundary surface (black, dashed) and the axis of axisymmetry (black, dash-dotted) shown for reference. Here $a$ is the tokamak minor radius (i.e. the minimum radial position on the flux surface of interest), $b$ is the maximum radial position, $\kappa_{b} \equiv b / a$ is the boundary elongation, and $\theta_{\kappa b}$ is the boundary tilt angle.}
 \label{fig:thetaKappaDef}
\end{figure}

In order to model realistic tilted elliptical tokamaks in our gyrokinetic simulations we must know how the Shafranov shift depends on the tilt angle of the elliptical boundary flux surface (parameterized by $\theta_{\kappa b}$ as shown in figure \ref{fig:thetaKappaDef}). We will argue that the Shafranov shift is insensitive to the shape of the current and pressure profiles (using linear profiles parameterized by equations \refEq{eq:currentProfiles} and \refEq{eq:pressureProfShape} respectively) when the geometry, plasma current, and average $d p / d \psi$ is kept fixed. Doing so makes the gyrokinetic simulations presented in section \ref{sec:gyrokineticSims} more widely applicable, as they use equilibria derived assuming constant current and pressure gradient profiles.

Together equations \refEq{eq:gradShafLowestOrderSolsTilt}, \refEq{eq:gradShafNextOrderSols}, \refEq{eq:inhomoTermsC0} through \refEq{eq:inhomoTermsTm}, and \refEq{eq:psiFourierNextOrderCoeffs} give the general solution of the Grad-Shafranov equation to $O \left( \epsilon B_{0} \right)$, which is sufficient to capture the behavior of the Shafranov shift. However, we still must determine the Fourier coefficients $N_{0, m}$, $\theta_{t 0, m}$, $C_{1, m}$, and $S_{1, m}$ in order to create a tilted elliptical boundary flux surface. To do so we require the poloidal flux to be constant on the boundary, parameterized in polar form by
\begin{align}
   r_{b} \left( \theta \right) = \frac{\sqrt{2} \kappa_{b} a}{\sqrt{\kappa_{b}^{2} + 1 + \left( \kappa_{b}^{2} - 1 \right) \Cos{2 \left( \theta + \theta_{\kappa b} \right)}}} , \label{eq:boundarySurf}
\end{align}
where figure \ref{fig:thetaKappaDef} shows the definitions of the various geometric parameters. Note that the tilt angle of the boundary $\theta_{\kappa b}$ is defined to increase in the clockwise direction, in contrast to the poloidal angle $\theta$.

To calculate $N_{0, m}$ and $\theta_{t 0, m}$ we substitute equation \refEq{eq:boundarySurf} into equation \refEq{eq:gradShafLowestOrderSolsTilt} to give
\begin{align}
   \psi_{0} \left( r_{b} \left( \theta \right), \theta \right) = \psi_{0 b} . \label{eq:gradShafBoundaryCondLowestOrder}
\end{align}
Since $\psi_{0 b}$ is a constant we know that $\psi_{0} \left( r_{b} \left( \theta \right), \theta \right)$ does not depend on $\theta$. In theory, ensuring that this is true for all values of $\theta$ determines all of the lowest order Fourier coefficients. However, the exact solution for these coefficients is not analytic, so we will resort to a numerical solution. Before we do so we will note that, because the lowest order Grad-Shafranov equation has cylindrical symmetry, the only angle intrinsic to the problem is $\theta_{\kappa b}$, which is introduced by the boundary condition. This implies that
\begin{align}
   \theta_{t 0, m} = \theta_{\kappa b} \label{eq:tiltAngleSol}
\end{align}
for all $m$, which suggests that it will be useful to define a new poloidal angle
\begin{align}
   \theta_{s} \equiv \theta + \theta_{\kappa b} . \label{eq:shiftThetaDef}
\end{align}
Furthermore, since an ellipse has mirror symmetry about exactly two axes, we know that $N_{0, m} = 0$ for odd $m$.

To determine $N_{0, m}$ for even $m$ we will take the Fourier series of $\psi_{0} \left( r_{b} \left( \theta_{s} \right), \theta_{s} \right) - \psi_{0 b}$. Truncating the series at a large mode number $m_{\text{max}}$ gives a long series of cosine terms. Requiring that the coefficient of each term must individually vanish gives a numerical approximation for all $N_{0, m}$ with $m \leq m_{\text{max}}$. In the limit that $m_{\text{max}} \to \infty$ this approximation approaches the exact solution, though in practice $m_{\text{max}} \approx 10$ was found to achieve sufficient precision for our purposes. This was determined by ensuring that the magnetic axis did not move significantly when $m_{\text{max}}$ was changed by $40\%$.

To next order we must determine $C_{1, m}$ and $S_{1, m}$ such that
\begin{align}
   \psi_{1} \left( r_{b} \left( \theta \right), \theta \right) = \psi_{1 b} \label{eq:gradShafBoundaryCondNextOrder}
\end{align}
is true, where $\psi_{1 b}$ is the next order value of the poloidal flux on the boundary flux surface. This is done in a similar manner to the lowest order equations, except the Grad-Shafranov equation no longer has cylindrical symmetry and we must evaluate the integrals in equation \refEq{eq:psiFourierNextOrderCoeffs}. The lack of symmetry means that we do not automatically know the tilt angle of the modes. However, since $\psi_{0}$ only has even Fourier mode numbers, it can be shown that equation \refEq{eq:gradShafNextOrder} only has odd Fourier modes. Hence, $C_{1, m} = S_{1, m} = 0$ for even $m$.

To calculate $C_{1, m}$ and $S_{1, m}$ for odd $m$ we construct $\psi_{1} \left( r, \theta \right)$ from equations \refEq{eq:gradShafNextOrderSols} and \refEq{eq:inhomoTermsC0} through \refEq{eq:psiFourierNextOrderCoeffs}. Taylor expanding this in $f_{N} j_{N} a^{2} \ll 1$ to $O \left( \left( f_{N} j_{N} a^{2} \right)^{f_{\text{max}}} \right)$ allows us to analytically calculate the integrals appearing in equation \refEq{eq:psiFourierNextOrderCoeffs} because the Bessel functions become summations of polynomials. We can now substitute equation \refEq{eq:boundarySurf} and find the Fourier series of $\psi_{1} \left( r_{b} \left( \theta \right), \theta \right) - \psi_{1 b}$ to mode number $m_{\text{max}}$. Again, we require that all of the Fourier coefficients must individually vanish, which produces a numerical approximation for each $C_{1, m}$ and $S_{1, m}$ with $m \leq m_{\text{max}}$. A value of $f_{\text{max}} \approx 10$ was found to give a sufficiently accurate solution. This was determined by ensuring that the magnetic axis did not move significantly when $f_{\text{max}}$ was changed by $40\%$.

For a hollow current profile, we repeat the entire above process except for using equation \refEq{eq:gradShafLowestOrderSolsTiltHollow} instead of equation \refEq{eq:gradShafLowestOrderSolsTilt} and equation \refEq{eq:psiFourierNextOrderCoeffsHollow} instead of equation \refEq{eq:psiFourierNextOrderCoeffs}. While the above process also works for the case of a constant toroidal current profile, this case actually has an analytic solution, which we derive in \ref{app:exactMagAxisLoc}.

In order to understand the effect of changing the current and pressure profiles in a single experimental device, we will choose to keep the major radial location of the center of the boundary flux surface ($R_{0}$), the minor radius ($a$), the edge elongation ($\kappa_{b}$), the total plasma current ($I_{p}$), and an estimate of the average pressure gradient ($p_{\text{axis}}/\psi_{0 b}$, i.e. the on-axis pressure divided by the edge poloidal flux) fixed. In order to keep these parameters fixed as we change the current and pressure profiles we must calculate how they enter into both $j_{N}$ and $j_{N p}$. Calculating $j_{N p}$ is straightforward, as we can directly integrate equation \refEq{eq:pressureProfShape} over poloidal flux to find
\begin{align}
   j_{N p} = \mu_{0} R_{0}^{2} \frac{p_{\text{axis}}}{\psi_{0 b}} \left( 1 - \frac{f_{N p} \psi_{0 b}}{2} \right)^{-1} . \label{eq:jNpEstimate}
\end{align}
To calculate $j_{N}$ we start with the definition of the plasma current,
\begin{align}
   I_{p} \equiv \int d S j_{\zeta} = \int_{0}^{2 \pi} d \theta_{s} \int_{0}^{r_{b} \left( \theta_{s} \right)} d r j_{\zeta} r , \label{eq:totalPlasmaCurrent}
\end{align}
where $S$ is the poloidal cross-sectional surface. Since we are only searching for a simple estimate, we will use equation \refEq{eq:currentProfiles} to rewrite equation \refEq{eq:totalPlasmaCurrent} as
\begin{align}
   I_{p} = \int_{0}^{2 \pi} d \theta_{s} \int_{0}^{r_{b} \left( \theta_{s} \right)} d r \frac{j_{N}}{\mu_{0} R_{0}} \left( 1 - f_{N} \psi_{0} \right) r , 
\end{align}
which is accurate to lowest order in aspect ratio. Substituting the boundary shape (i.e. equation \refEq{eq:boundarySurf}) and the constant current solution for $\psi_{0} \left( r, \theta_{s} \right)$ (i.e. equations \refEq{eq:gradShafLowestOrderSolsTiltConst}, \refEq{eq:tiltAngleSol}, \refEq{eq:boundaryFluxConst}, and \refEq{eq:fourierShapingConst}) allows us to directly take the integral to find
\begin{align}
   j_{N} = \mu_{0} \frac{I_{p}}{\pi a^{2} \kappa_{b}} R_{0} \left( 1 - \frac{f_{N} \psi_{0 b}}{2} \right)^{-1} + \Order{f_{N}^{2} j_{N}^{2} a^{4}} . \label{eq:jNestimate}
\end{align}
The $\Order{f_{N}^{2} j_{N}^{2} a^{4}}$ error arises from the fact that we used the constant current solution for $\psi_{0} \left( r, \theta_{s} \right)$, which is only accurate to lowest order in $f_{N} j_{N} a^{2} \ll 1$. This means that as we change $f_{N p}$ and $f_{N}$ we must change $j_{N p}$ and $j_{N}$ according to equations \refEq{eq:jNpEstimate} and \refEq{eq:jNestimate} respectively.

\begin{figure}
 \centering
 \includegraphics[width=0.6\textwidth]{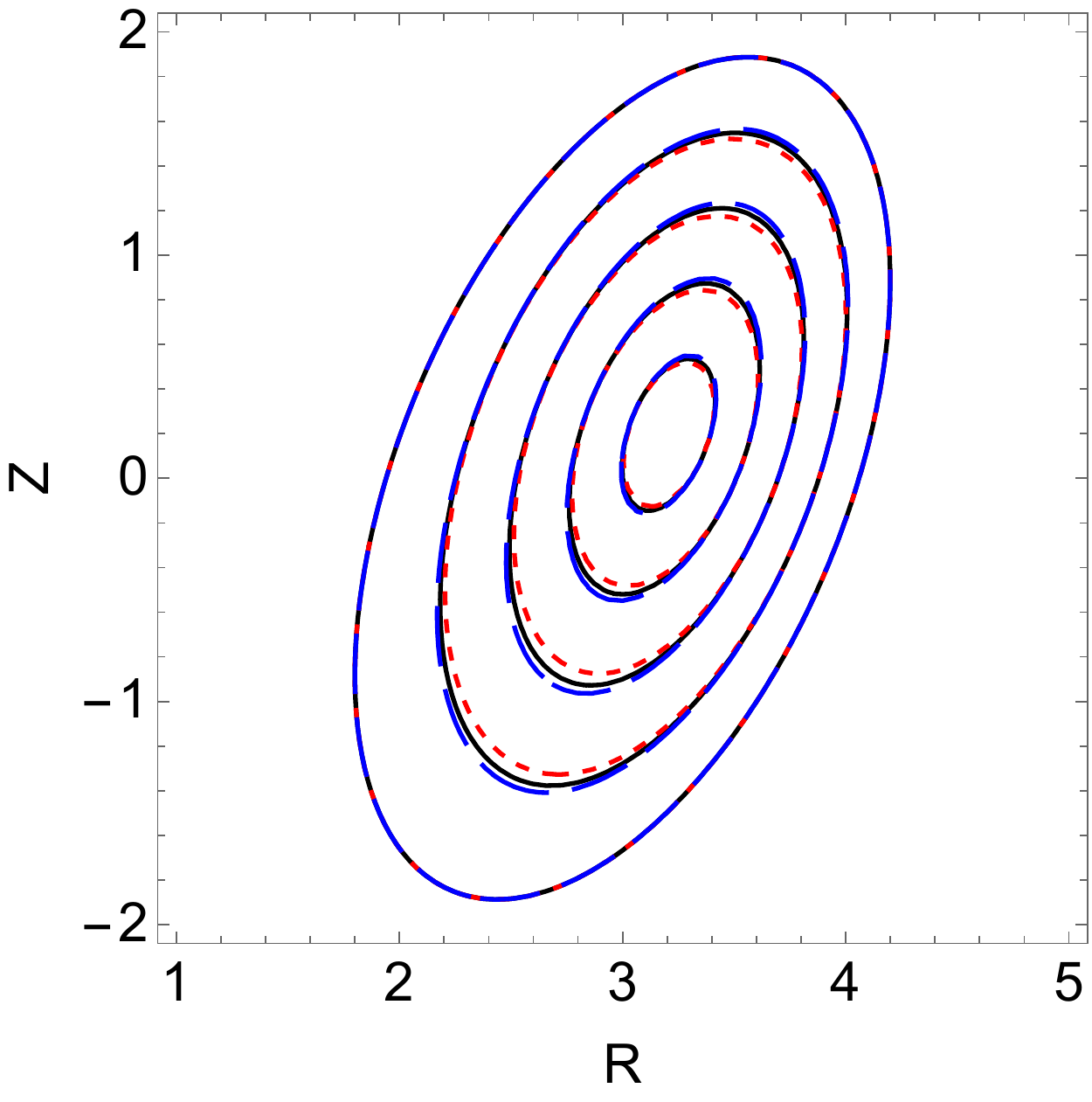}
 \caption{Calculated flux surfaces for $f_{N} \psi_{0 b} = f_{N p} \psi_{0 b} = 0$ (black, solid), $f_{N} \psi_{0 b} = f_{N p} \psi_{0 b} = 0.4$ (red, dotted), and $f_{N} \psi_{0 b} = f_{N p} \psi_{0 b} = - 0.4$ (blue, dashed).}
 \label{fig:gradShafCalcSol}
\end{figure}

In figure \ref{fig:gradShafCalcSol} we plot the calculated flux surfaces resulting from three different current profiles, setting $f_{N p} = f_{N}$. We use inputs of $R_{0} = 3$, $a = 1$ (where we have normalized all lengths to the minor radius), $\kappa_{b} = 2$, and
\begin{align}
   \frac{j_{N p}}{j_{N}} \approx \frac{\pi a^{2} \kappa_{b} R_{0}}{I_{p}} \frac{p_{\text{axis}}}{\psi_{0 b}} \approx 0.7 \label{eq:jNconstantsRatio}
\end{align}
using projections for ITER \cite{AymarITERSummary2001}. Additionally, we choose to plot the case of $\theta_{\kappa b} = \pi / 8$ because nonlinear gyrokinetic simulations have shown this value to be optimal for generating rotation \cite{BallMomUpDownAsym2014}. Note that the $\psi_{0 b}$ appearing in equation \refEq{eq:jNconstantsRatio} is part of $p_{\text{axis}} / \psi_{0 b}$, so it is fixed for all three profiles and can be calculated for a constant current profile from equation \refEq{eq:boundaryFluxConst}. In figure \ref{fig:gradShafCalcSol} we see that the current profile has an effect on the penetration of elongation from the boundary to the magnetic axis. This indicates that hollower current profiles better support elongation throughout the plasma, which is consistent with previous theoretical work \cite{BallMomUpDownAsym2014, BallMastersThesis2013, RodriguesMHDupDownAsym2014, BizarroUpDownAsymGradShafEq2014, BallShapingPenetration2015} as well as EFIT equilibrium reconstruction on simulated experimental data (see figure 5(b) of reference \cite{LaoShapeAndCurrent1985}). However, given these parameters, the Shafranov shift is not visibly altered, even with the extreme changes in the current profile.

\begin{figure}
 \begin{center}
   (a) \hspace{0.36\textwidth}
  
  \includegraphics[height=0.45\textwidth]{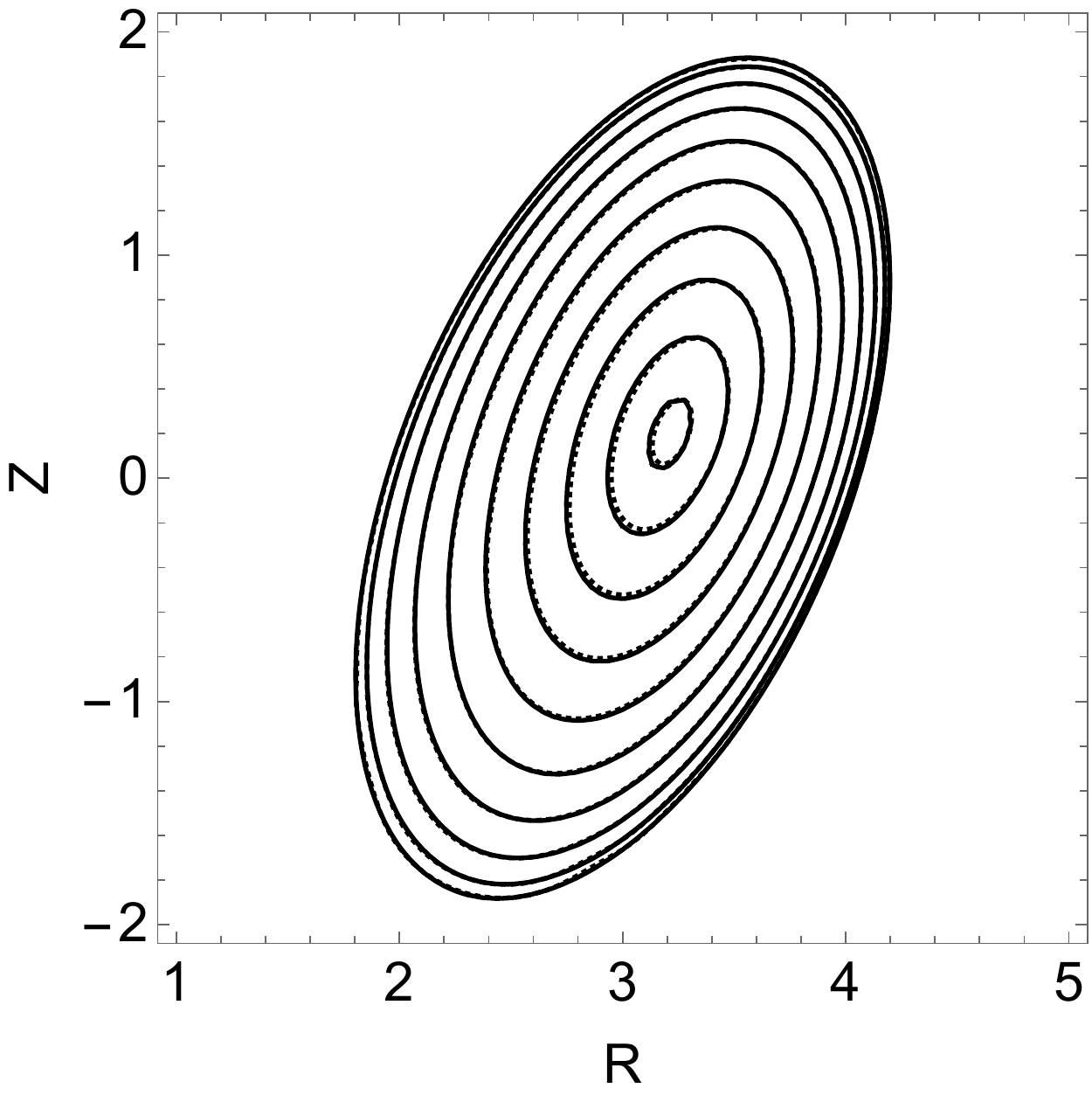}
  
  (b) \hspace{0.37\textwidth} (c) \hspace{0.36\textwidth}
  
  \includegraphics[height=0.45\textwidth]{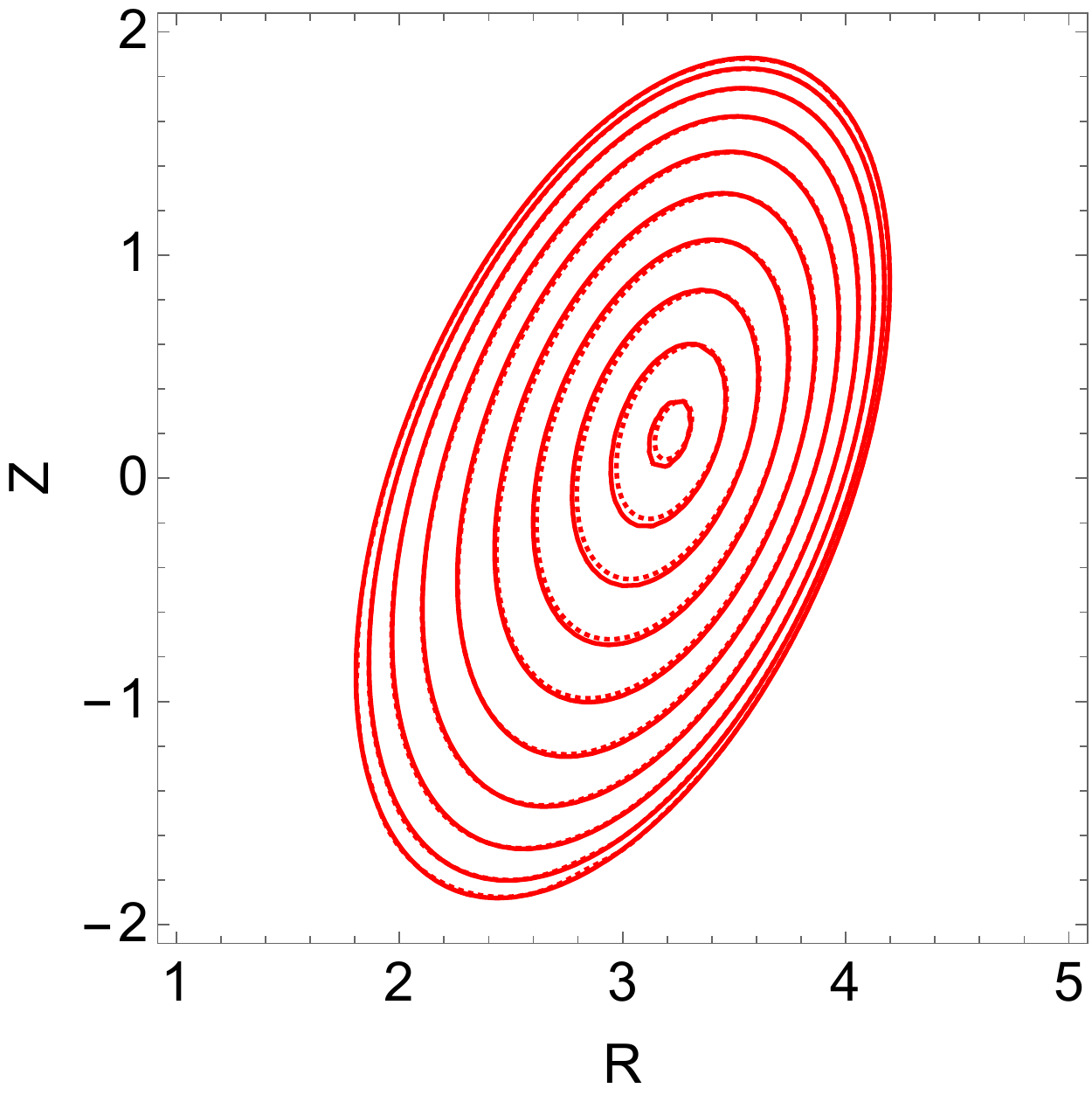}
  \includegraphics[height=0.45\textwidth]{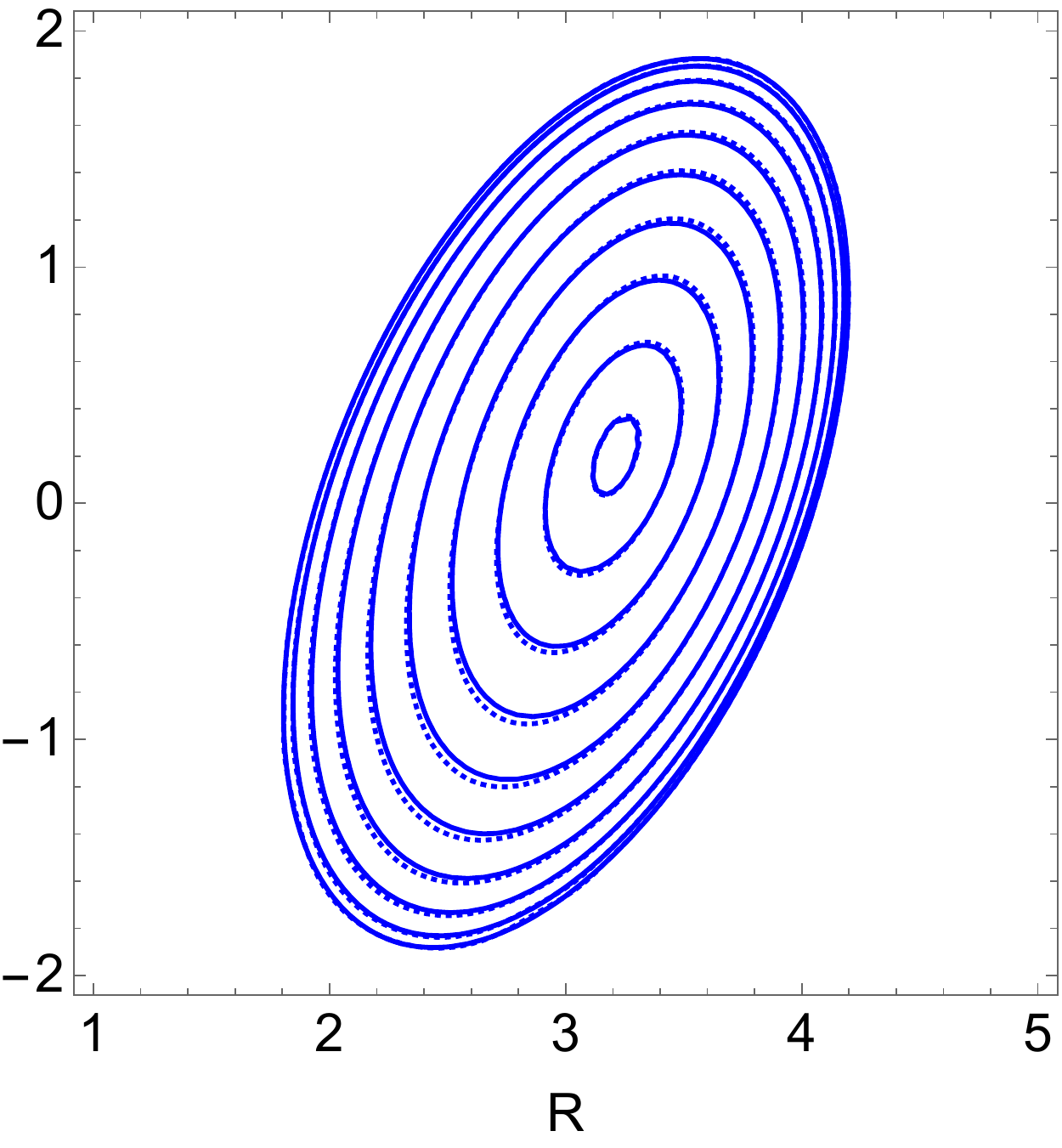}
 \end{center}
 \caption{Flux surfaces calculated by both ECOM (dotted) and analytically (solid) for (a) $f_{N} \psi_{0 b} = f_{N p} \psi_{0 b} = 0$ (black), (b) $f_{N} \psi_{0 b} = f_{N p} \psi_{0 b} = 0.4$ (red), and (c) $f_{N} \psi_{0 b} = f_{N p} \psi_{0 b} = - 0.4$ (blue).}
 \label{fig:gradShafSolComp}
\end{figure}

In order to verify our calculation, we compared our results with the ECOM code \cite{LeeECOM2015}, a fixed boundary equilibrium solver capable of modeling up-down asymmetric configurations. In figure \ref{fig:gradShafSolComp} we see a direct graphical comparison between ECOM and the results of our calculation that were shown in figure \ref{fig:gradShafCalcSol}. The two sets of results agree well, especially for the constant and hollow current profile cases. The most significant source of error is expected to be finite aspect ratio effects in our analytic calculation, which arise from the assumption that $\epsilon = 1 / 3 \ll 1$. Hence, since we carried out the analytic calculation to lowest and next order in the aspect ratio expansion, we expect to have an $\epsilon^{2} \sim 10\%$ error. We also note that we do not expect the $\Order{\epsilon^{2} B_{0}}$ solution (i.e. the largest order that we omitted) to modify the Shafranov shift in a configuration with an elliptical boundary. This is because reference \cite{HakkarainenEquilibrium1990} demonstrates that toroidicity only introduces $m = 0$ and $m = 2$ modes at order $\Order{\epsilon^{2} B_{0}}$. Furthermore, equation \refEq{eq:psiNextOrderSolConst} demonstrates that a lowest order shaping effect $m$ introduces only $m - 1$ and $m + 1$ modes to order $\Order{\epsilon B_{0}}$. This suggests that only $m - 2$, $m$, and $m + 2$ modes will appear to $\Order{\epsilon^{2} B_{0}}$. Therefore, we expect that the $m = 1$ mode will not appear at $\Order{\epsilon^{2} B_{0}}$, so the Shafranov shift will not be changed.

\begin{figure}
 \centering
 \includegraphics[width=0.45\textwidth]{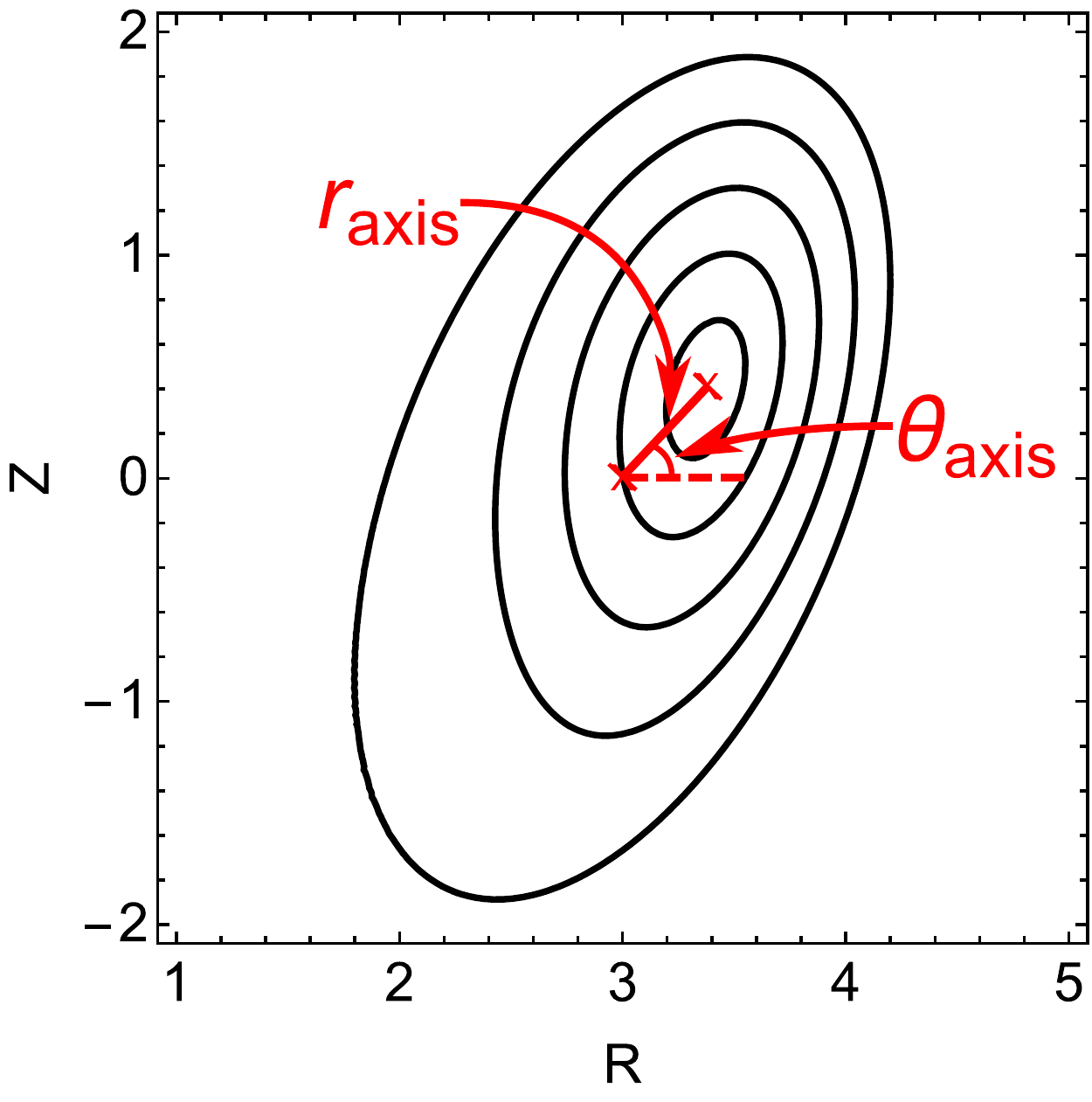}
 \caption{Example flux surfaces showing the geometric meaning of the parameters $r_{\text{axis}}$ and $\theta_{\text{axis}}$, the minor radial and poloidal locations of the magnetic axis respectively.}
 \label{fig:geoMagAxisLoc}
\end{figure}

\subsection{Location of the magnetic axis}
\label{subsec:ECOMcomparison}

We can obtain the Shafranov shift from our calculation by numerically solving the equation
\begin{align}
   \left. \Nabla \left( \psi_{0} \left( r, \theta \right) + \psi_{1} \left( r, \theta \right) \right) \right|_{r = r_{\text{axis}}, \theta = \theta_{\text{axis}}} = 0 \label{eq:magAxisCondition}
\end{align}
using equations \refEq{eq:gradShafLowestOrderSolsTilt}, \refEq{eq:gradShafNextOrderSols}, \refEq{eq:inhomoTermsC0} through \refEq{eq:inhomoTermsTm}, \refEq{eq:psiFourierNextOrderCoeffs}, and \refEq{eq:tiltAngleSol} as well as our numerical solutions for $N_{0, m}$, $C_{1, m}$, and $S_{1, m}$. Here $r_{\text{axis}}$ and $\theta_{\text{axis}}$ are the minor radial and poloidal location of the magnetic axis respectively, as indicated in figure \ref{fig:geoMagAxisLoc}. For the special case of a tilted elliptical boundary with a constant toroidal current profile (i.e. $f_{N} = 0$) we can exactly solve equation \refEq{eq:magAxisCondition} as shown in \ref{app:exactMagAxisLoc}. Equations \refEq{eq:magAxisPoloidalLocExact} and \refEq{eq:magAxisRadialLocExactCondition} give the exact location of the magnetic axis when considering the poloidal flux to lowest order and next order in $\epsilon \ll 1$.

\begin{figure}
 \hspace{0.04\textwidth} (a) \hspace{0.4\textwidth} (b) \hspace{0.25\textwidth}
 \begin{center}
  \includegraphics[width=0.45\textwidth]{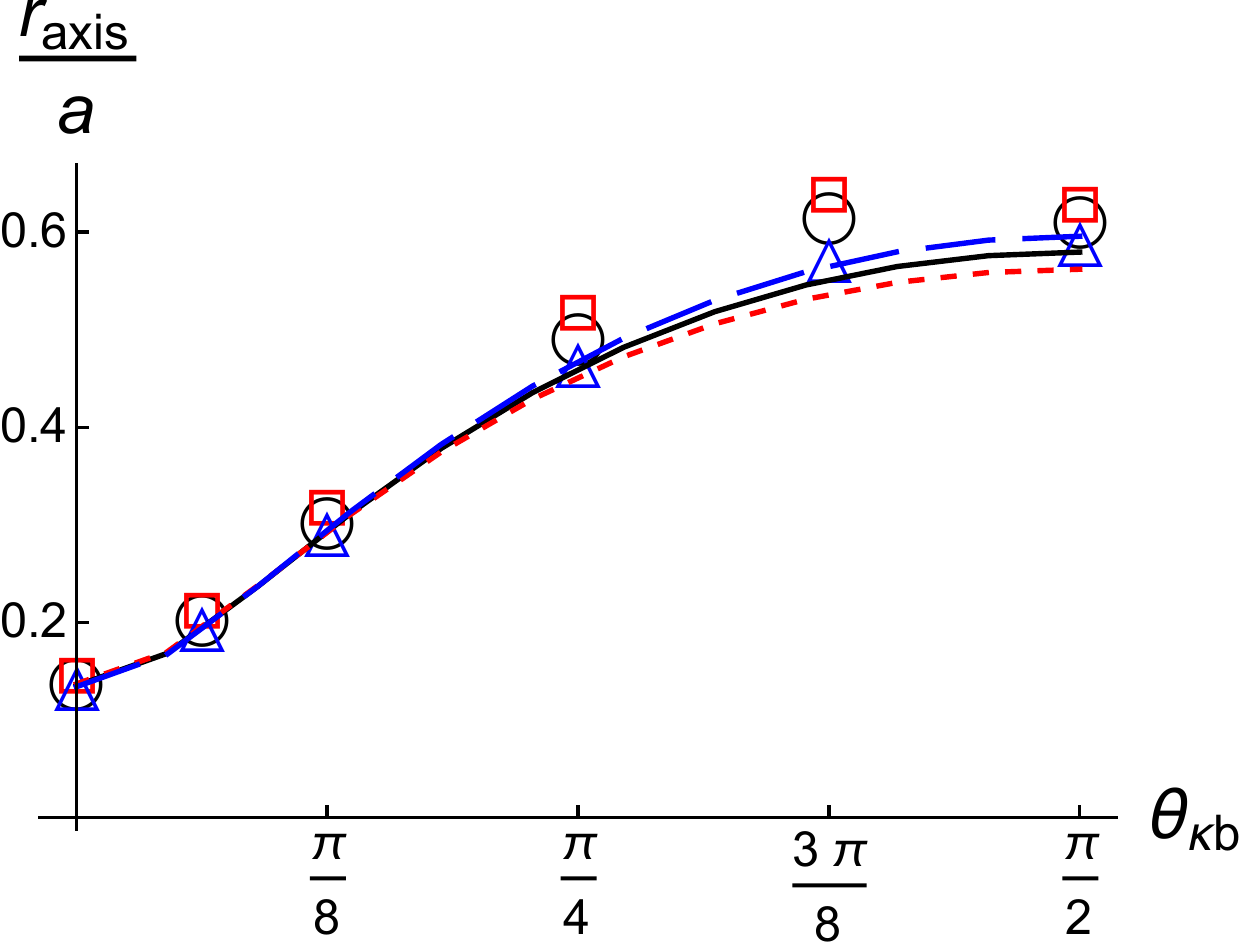}
  \includegraphics[width=0.45\textwidth]{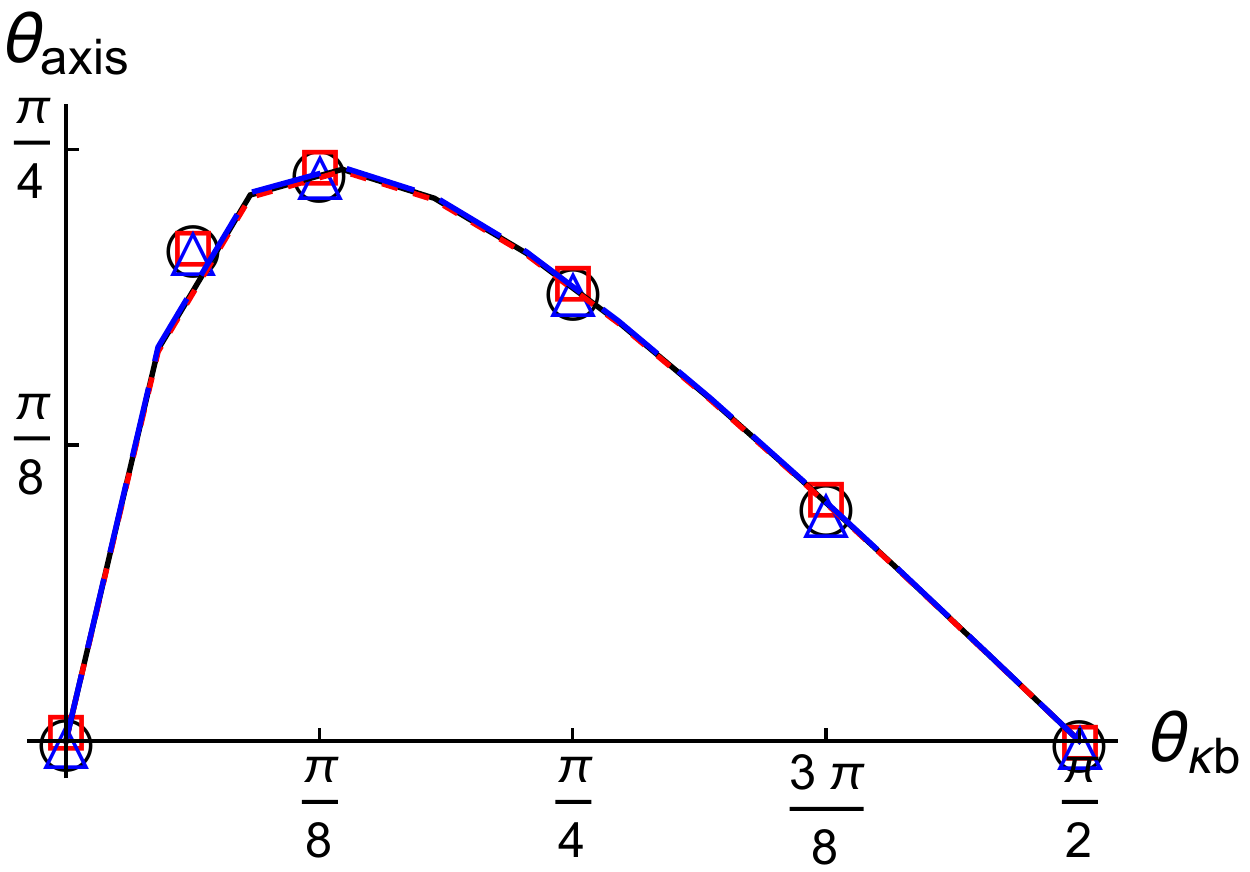}
 \end{center}
 \caption{The (a) minor radial and (b) poloidal location of the magnetic axis for constant ($f_{N} \psi_{0 b} = f_{N p} \psi_{0 b} = 0$) (black, solid, circles), linear peaked ($f_{N} \psi_{0 b} = f_{N p} \psi_{0 b} = 0.4$) (red, dotted, squares), and linear hollow ($f_{N} \psi_{0 b} = f_{N p} \psi_{0 b} = - 0.4$) (blue, dashed, triangles) current/pressure gradient profiles, calculated analytically (lines) and by ECOM (points).}
 \label{fig:ShafShiftLoc}
\end{figure}

In figure \ref{fig:ShafShiftLoc} we show the location of the magnetic axis as we vary the shape of the current/pressure profile (by changing $f_{N}$ and keeping $f_{N p} = f_{N}$), while holding the geometry, $I_{p}$, and $p_{\text{axis}} / \psi_{0 b}$ fixed. For the most part, we see reasonable quantitative agreement between our theoretical results and ECOM. However, the two calculations disagree on the trend of $r_{\text{axis}}$ with $f_{N} \psi_{0 b}$ at large tilt angles. We do not think this is significant as it appears to be a breakdown in our inverse aspect ratio expansion. The two calculations become consistent if the aspect ratio is directly increased or if smaller tilt angles are used (where the effective aspect ratio is larger).

An important property of figure \ref{fig:ShafShiftLoc}, which is supported by both the analytic and ECOM calculations, is the insensitivity of the Shafranov shift to extreme changes in the shape of the current profile. Both the magnitude and the direction of the Shafranov shift change very little between the different current profiles. This is especially true in the domain of $\theta_{\kappa b} \in \left[ 0, \pi / 4 \right]$, which is the range of tilt angles that seem most promising for implementing in an experiment \cite{CamenenPRLExp2010, BallMomUpDownAsym2014}. This result allows us to simplify our treatment of the Shafranov shift. The gyrokinetic simulations we will present in section \ref{sec:gyrokineticSims} are formally inconsistent because they do not assume constant current and pressure gradient profiles, but they use the Shafranov shift of equilibria with constant current and pressure gradient profiles. However, this inconsistency is not important because the Shafranov shift only depends weakly on the shape of the current and pressure gradient profiles. As we will see in figure \ref{fig:momHeatFluxRatioWithShift}, the turbulent momentum flux driven by the Shafranov shift is approximately linear in the size of the shift, so small errors in the shift will only lead to small errors in the momentum flux.

Also from figure \ref{fig:ShafShiftLoc}, we learn that the tilt angle has a large effect, not just on the direction of the Shafranov shift, but also its magnitude. This is intuitive because we know that, for an ellipse with $\kappa = 2$, the midplane chord length is twice as long in the $\theta_{\kappa b} = \pi / 2$ geometry as it is in the $\theta_{\kappa b} = 0$ geometry. Lastly, we see that the direction of the Shafranov shift varies considerably, but it is purely outwards for the $0$ and $\pi / 2$ tilt angles as expected. Importantly, it does not align with the tilt angle of the ellipse, so it breaks the mirror symmetry of the configuration.

\begin{figure}
 \hspace{0.04\textwidth} (a) \hspace{0.4\textwidth} (b) \hspace{0.25\textwidth}
 \begin{center}
  \includegraphics[width=0.45\textwidth]{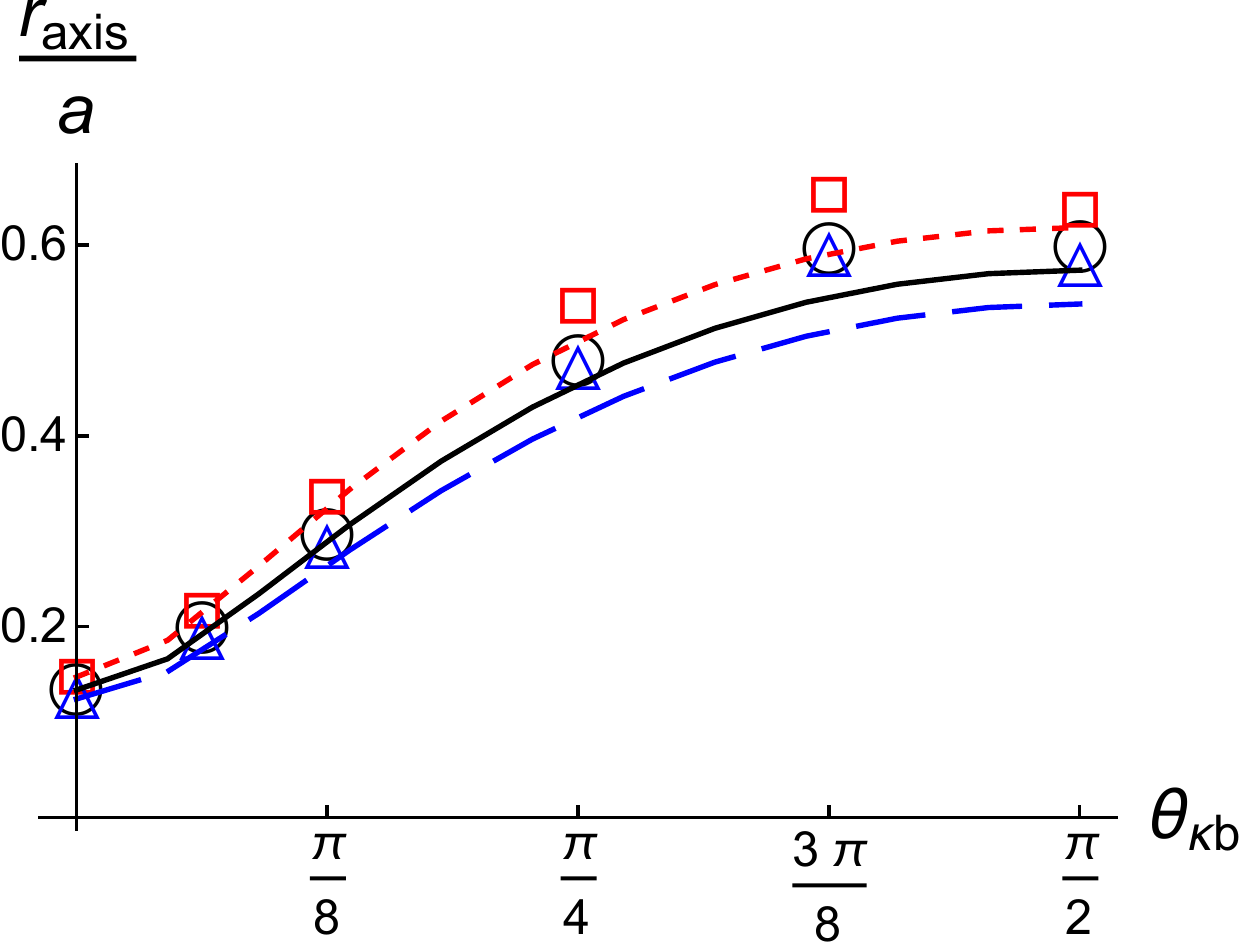}
  \includegraphics[width=0.45\textwidth]{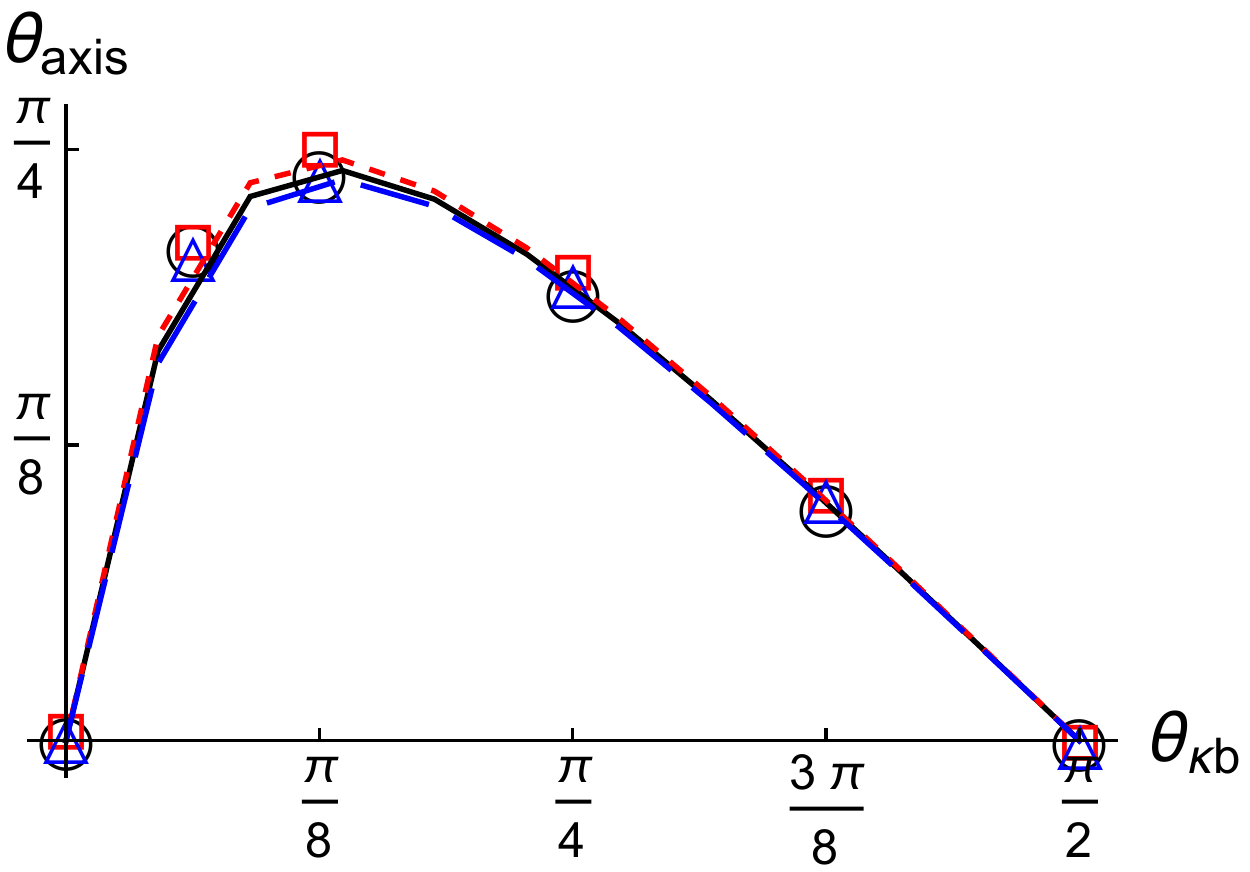}
 \end{center}
 \caption{The (a) minor radial and (b) poloidal location of the magnetic axis for constant ($f_{N p} \psi_{0 b} = 0$) (black, solid, circles), linear peaked ($f_{N p} \psi_{0 b} = 0.4$) (red, dotted, squares), and linear hollow ($f_{N p} \psi_{0 b} = - 0.4$) (blue, dashed, triangles) pressure gradient profiles, calculated analytically (lines) and by ECOM (points) for a constant current profile.}
 \label{fig:ShafShiftLocPressureProf}
\end{figure}

In figure \ref{fig:ShafShiftLocPressureProf} we show the location of the magnetic axis as we vary the shape of the pressure profile (by changing $f_{N p}$) with a constant current profile (i.e. $f_{N} = 0$), while holding the geometry, $I_{p}$, and $p_{\text{axis}} / \psi_{0 b}$ fixed. We see good quantitative agreement between the calculation given in \ref{app:exactMagAxisLoc} and ECOM. Figure \ref{fig:ShafShiftLocPressureProf} indicates that varying the shape of the pressure profile has little effect on the Shafranov shift. We note that, in general, varying the pressure profile has a large effect on the magnitude of the Shafranov shift, but not when $I_{p}$ and $p_{\text{axis}} / \psi_{0 b}$ are held constant. This is important as it justifies using our MHD results for the Shafranov shift with a constant $d p / d \psi$ profile as input for gyrokinetic simulations that are based on ITER, which has a constant $d p / d r_{\psi}$ profile \cite{AymarITERSummary2001}. Even though this is formally inconsistent, our analysis suggests the Shafranov shift in a configuration with constant $d p / d \psi$ will be a reasonable estimate of the Shafranov shift in a configuration with constant $d p / d r_{\psi}$ (as long as the geometry, $I_{p}$, and $p_{\text{axis}} / \psi_{0 b}$ are the same). As we will show in figure \ref{fig:momHeatFluxRatioWithShift}, the momentum flux is approximately linear in the size of the Shafranov shift. Hence, the small error introduced by using the Shafranov shift calculated with a constant pressure gradient profile (in $\psi$) will not lead to large differences in the momentum flux.

\section{Nonlinear gyrokinetic simulations}
\label{sec:gyrokineticSims}

In this section we will use the results from section \ref{sec:MHDequil} in order to perform nonlinear gyrokinetic simulations that include the effect of a realistic Shafranov shift on plasma turbulence. Since the size of the Shafranov shift is closely connected to the plasma pressure, we also included the effect of $\beta'$ on the magnetic equilibrium. We will use a modified version of GS2 \cite{KotschenreutherGS21995} to self-consistently calculate the time-averaged radial flux of toroidal angular momentum $\left\langle \Pi_{\zeta i} \right\rangle_{t}$ and the time-averaged radial flux of energy $\left\langle Q_{i} \right\rangle_{t}$ for ions. These calculations use a local equilibrium specified by an up-down asymmetric generalization of the Miller geometry model \cite{MillerGeometry1998}.

\subsection{Input parameters}
\label{subsec:inputParameters}

In this work, we will use a flux surface of interest with Cyclone base case parameters (unless otherwise specified) \cite{DimitsCycloneBaseCase2000}: a minor radius of $\rho_{0} = 0.54$, a major radius of $R_{c 0} / a = 3$ (i.e. the major radial location of the center of the flux surface of interest), a safety factor of $q = 1.4$, a magnetic shear of $\hat{s} \equiv \left( \rho_{0} / q \right) d q / d \rho = 0.8$, a temperature gradient of $d \Ln{T_{s}} / d \rho = - 2.3$, and a density gradient of $d \Ln{n_{s}} / d \rho = - 0.733$ (where the subscript $s$ indicates either the ion or electron species). Here $\rho \equiv r_{\psi} / a$ is the normalized minor radial flux surface label, $r_{\psi}$ is a real-space flux surface label that indicates the minimum distance of each flux surface from its center, $\rho_{0} \equiv r_{\psi 0} / a$ is the value of $\rho$ on the flux surface of interest, and $r_{\psi 0}$ is the value of $r_{\psi}$ on the flux surface of interest. We note that taking $\hat{s} \neq 0$ can be formally inconsistent with a constant toroidal current profile (as it is in the large aspect ratio limit for circular flux surfaces). However, from figure \ref{fig:ShafShiftLoc} we know that we can vary the current profile without affecting the Shafranov shift much, as long as we keep $I_{p}$ and $p_{\text{axis}} / \psi_{0 b}$ fixed. Because of this freedom, we can use the Shafranov shift calculated assuming constant current and pressure gradient profiles for the Cyclone base case. Many of our simulations will model elliptical flux surfaces, all of which have an elongation of $\kappa = 2$. Furthermore, all turbulent fluxes calculated by GS2 will be normalized to gyroBohm values of
\begin{align}
   \Pi_{gB} &\equiv \rho_{\ast}^{2} n_{i} a m_{i} v_{th, i}^{2} \\
   Q_{gB} &\equiv \rho_{\ast}^{2} n_{i} T_{i} v_{th, i} ,
\end{align}
where $\rho_{\ast} \equiv \rho_{i} / a$ is the ratio of the ion gyroradius to the tokamak minor radius, $n_{i}$ is the ion density, $m_{i}$ is the ion mass, $T_{i}$ is the local ion temperature, and $v_{th, i} \equiv \sqrt{2 T_{i} / m_{i}}$ is the local ion thermal speed. All simulations used at least 48 grid points in the poloidal angle, 127 grid points in the wavenumber of the radial direction, 22 grid points in the wavenumber of the direction within the flux surface (but still perpendicular to the magnetic field), 12 grid points in the energy, and 10 grid points in the untrapped pitch angle. The large number of poloidal grid points was needed to properly resolve the strong flux surface shaping.

\begin{figure}
 \centering
 \includegraphics[width=0.45\textwidth]{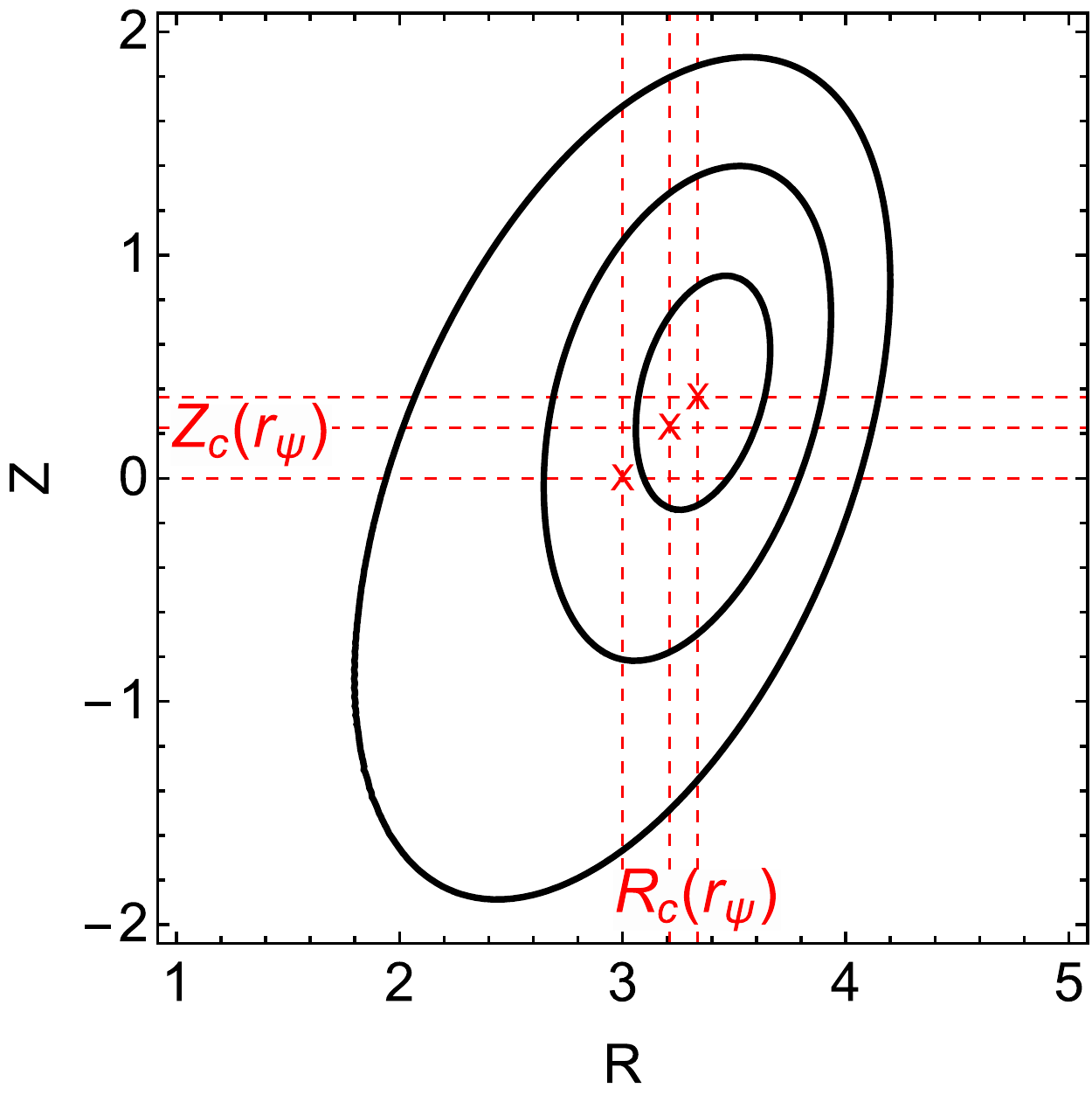}
 \caption{Three example flux surfaces (black, solid) at different values of $r_{\psi}$ with their geometric center (red, crosses). This illustrates the meaning of the parameters $R_{c} \left( r_{\psi} \right)$ (red, dashed, vertical) and $Z_{c} \left( r_{\psi} \right)$ (red, dashed, horizontal), the major radial and axial locations of the center of each flux surface respectively.}
 \label{fig:geoFluxSurfCenter}
\end{figure}

\begin{figure}
 \centering
 \includegraphics[width=0.6\textwidth]{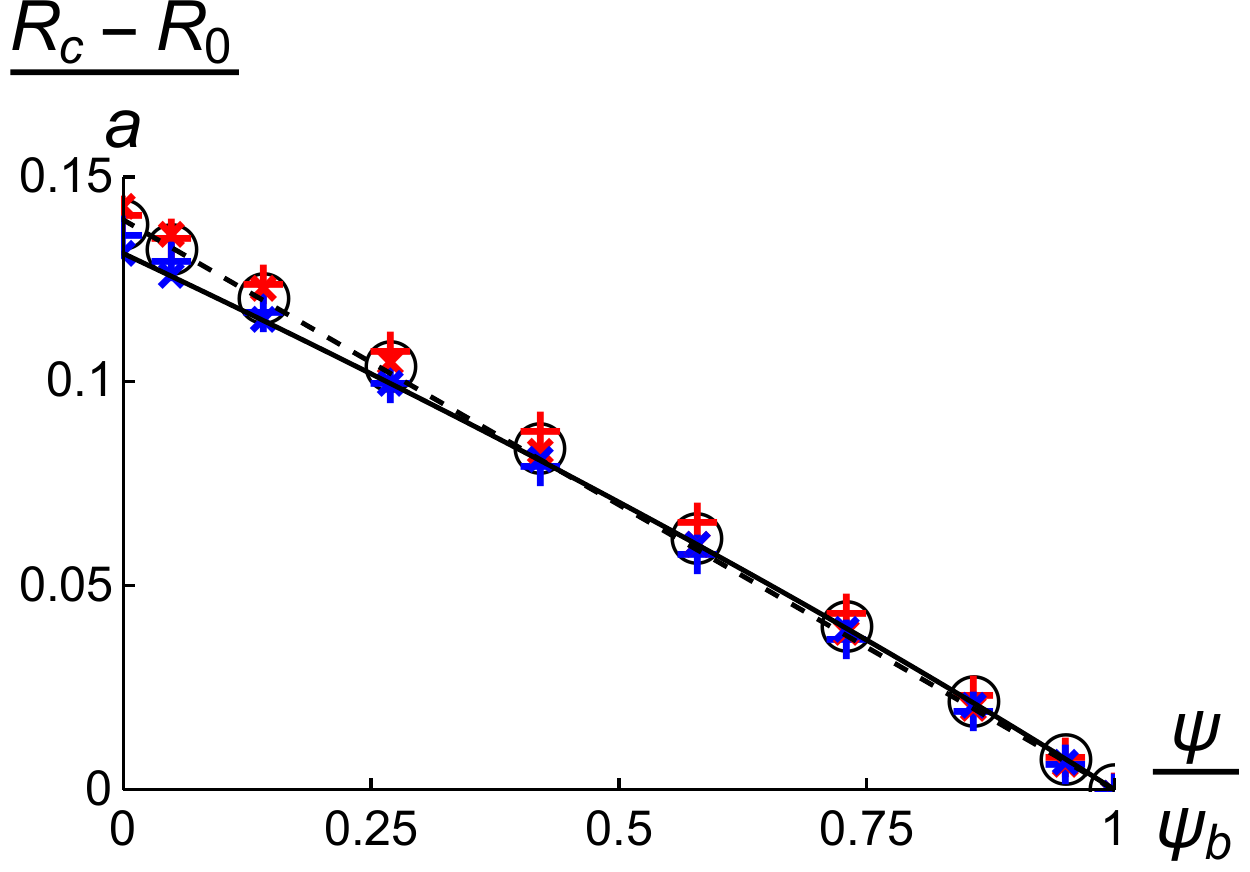}
 \caption{The shift in the center of flux surfaces (relative to the center of the boundary flux surface $R_{0}$) as a function of normalized poloidal flux for geometries with $\theta_{\kappa b} = 0$. The points are calculated by ECOM for a constant current profile ($f_{N} \psi_{0 b} = f_{N p} \psi_{0 b} = 0$) (black, circles), a linear peaked current profile ($f_{N} \psi_{0 b} = f_{N p} \psi_{0 b} = 0.4$) (red, pluses), a linear hollow current profile ($f_{N} \psi_{0 b} = f_{N p} \psi_{0 b} = - 0.4$) (blue, pluses), a linear peaked pressure profile ($f_{N} \psi_{0 b} = 0$ and $f_{N p} \psi_{0 b} = 0.4$) (red, crosses), and a linear hollow pressure profile ($f_{N} \psi_{0 b} = 0$ and $f_{N p} \psi_{0 b} = - 0.4$) (blue, crosses). Also shown is our analytic solution (solid line) and a linear best fit (dashed line).}
 \label{fig:shiftWithRadius}
\end{figure}

The Miller geometry specification in GS2 captures the Shafranov shift through local values of $d R_{c} / d r_{\psi}$ and $d Z_{c} / d r_{\psi}$. Here $R_{c} \left( r_{\psi} \right)$ and $Z_{c} \left( r_{\psi} \right)$ indicate the location of the center of each flux surface as shown in figure \ref{fig:geoFluxSurfCenter}. In order to model a realistic geometry, we will calculate local values of $d R_{c} / d r_{\psi}$ and $d Z_{c} / d r_{\psi}$ for arbitrary tilt angle from our global MHD results. Specifically, we will use the dependence of the global Shafranov shift on tilt angle calculated for constant current and $d p / d \psi$ profiles (i.e. the solid black line shown in figure \ref{fig:ShafShiftLoc}).

First we will assume that $d R_{c} / d \psi$ and $d Z_{c} / d \psi$ are constant from the boundary flux surface to the magnetic axis. In figure \ref{fig:shiftWithRadius}, we plot our analytic solution (using the coefficients calculated in \ref{app:exactMagAxisLoc}) and ECOM results to show that this assumption holds, regardless of the shape of the pressure and current profiles. Additionally, using equations \refEq{eq:gradShafLowestOrderSolsTiltConst} and \refEq{eq:fourierShapingConst} we see that
\begin{align}
   \psi \propto r_{\psi}^{2}
\end{align}
for a constant current profile and an exactly elliptical boundary. Therefore, using that $\psi = \psi_{b}$ at $r_{\psi} = a$, one can calculate the constant of proportionality and show
\begin{align}
   \frac{d \psi}{d r_{\psi}} = 2 \frac{\psi_{b}}{a} \rho .
\end{align}
Hence, the local Shafranov shift can be written as
\begin{align}
   \left. \frac{d R_{c}}{d r_{\psi}} \right|_{r_{\psi 0}} =& \left. \frac{d \psi}{d r_{\psi}} \right|_{r_{\psi 0}} \frac{d R_{c}}{d \psi} = \left( 2 \frac{\psi_{b}}{a} \rho_{0} \right) \frac{R_{0} - R_{c} \left( 0 \right)}{\psi_{b} - 0} = - 2 \rho_{0} \frac{r_{\text{axis}}}{a} \Cos{\theta_{\text{axis}}} \label{eq:radialShafShiftLocal} \\
   \left. \frac{d Z_{c}}{d r_{\psi}} \right|_{r_{\psi 0}} =& \left. \frac{d \psi}{d r_{\psi}} \right|_{r_{\psi 0}} \frac{d Z_{c}}{d \psi} = \left( 2 \frac{\psi_{b}}{a} \rho_{0} \right) \frac{0 - Z_{c} \left( 0 \right)}{\psi_{b} - 0} = - 2 \rho_{0} \frac{r_{\text{axis}}}{a} \Sin{\theta_{\text{axis}}} ,  \label{eq:axialShafShiftLocal}
\end{align}
where the coordinate system is defined such that the boundary flux surface is centered at $\left( R = R_{0}, Z = 0 \right)$. Therefore, we are able to calculate $\left. d R_{c} / d r_{\psi} \right|_{r_{\psi 0}}$ and $\left. d Z_{c} / d r_{\psi} \right|_{r_{\psi 0}}$ for an ITER-like pressure profile using equations \refEq{eq:radialShafShiftLocal} and \refEq{eq:axialShafShiftLocal} as well as the constant current results shown in figure \ref{fig:ShafShiftLoc}.

GS2 also requires a local value of
\begin{align}
   \beta' \equiv \frac{2 \mu_{0} a}{B_{0}^{2}} \frac{d p}{d r_{\psi}} \label{eq:betaPrimeDef}
\end{align}
because it constructs the poloidal magnetic field to be consistent with the Grad-Shafranov equation. We will find that the momentum transport is quite sensitive to $\beta'$, so it is an important parameter. In keeping with rough projections for ITER \cite{AymarITERSummary2001}, we use a pressure profile that is linear in $r_{\psi}$. This allows us to estimate that
\begin{align}
   \beta' \approx - \frac{2 \mu_{0} p_{\text{axis}}}{B_{0}^{2}} \approx - 0.06 , \label{eq:betaPrimeEstimate}
\end{align}
using an ITER-like value for $p_{\text{axis}}$. Since we are running electrostatic simulations the value of $\beta$ itself has no effect.

We note that assuming a constant $\beta'$ (i.e. $d p / d r_{\psi}$) profile is formally inconsistent with the constant $d p / d \psi$ profile used in the MHD calculation of the Shafranov shift. Hence, using the results shown in figure \ref{fig:ShafShiftLoc} together with equation \refEq{eq:betaPrimeEstimate} is not formally valid. However, figure \ref{fig:ShafShiftLocPressureProf} shows that the magnitude and direction of the Shafranov shift is insensitive to large changes in the shape of the pressure profile at constant $R_{0}$, $a$, $\kappa_{b}$, $I_{p}$, and $p_{\text{axis}} / \psi_{0 b}$. This suggests that, since we have kept the proper parameters fixed, the mismatch between the pressure profile of the simulation and the pressure profile used to calculate the Shafranov shift will not have much effect.

\subsection{Parameter scan results}
\label{subsec:results}

A total of four scans in $\theta_{\kappa}$, the tilt angle of the flux surface of interest, were performed at
\begin{enumerate}
   \item[(1)] $\beta' = 0$ with no Shafranov shift, \\
   \item[(2)] $\beta' = 0$ with a modest Shafranov shift (approximately half the ITER-like Shafranov shift), \\
   \item[(3)] $\beta' = 0$ with an ITER-like Shafranov shift, and \\
   \item[(4)] an ITER-like $\beta' = -0.06$ with an ITER-like Shafranov shift.
\end{enumerate}
These scans were chosen to directly determine the independent influences of the Shafranov shift and $\beta'$, while minimizing the total number of simulations. The magnitude and direction of the local ITER-like Shafranov shift was kept consistent with equations \refEq{eq:radialShafShiftLocal} and \refEq{eq:axialShafShiftLocal}. Additionally, a single simulation was performed with $\beta' = -0.06$ and no Shafranov shift in order to isolate the effect of $\beta'$.

Four scans in $\rho_{0}$, the minor radial coordinate of the flux surface of interest, were performed at
\begin{enumerate}
   \item[(1)] $\beta' = 0$ with no Shafranov shift, \\
   \item[(2)] $\beta' = 0$ with an ITER-like Shafranov shift, \\
   \item[(3)] an ITER-like $\beta' = -0.06$ with no Shafranov shift, and \\
   \item[(4)] an ITER-like $\beta' = -0.06$ with an ITER-like Shafranov shift.
\end{enumerate}
All simulations had elliptical flux surfaces with $\theta_{\kappa} = \pi / 8$. These scans were done in order to investigate the balance between the Shafranov shift, which we expect to enhance the momentum transport, and $\beta'$, which our GS2 simulations will reveal to reduce the momentum transport.  For these scans we kept $\beta'$ constant to be consistent with ITER (according to equation \refEq{eq:betaPrimeEstimate}) and again calculated the local Shafranov shift at each minor radius according to equations \refEq{eq:radialShafShiftLocal} and \refEq{eq:axialShafShiftLocal}.

Lastly a small scan was performed with circular flux surfaces in which $\theta_{\text{axis}}$, the direction of the Shafranov shift, was varied. This is unphysical, but it was done to explicitly isolate the effect of a pure flux surface Shafranov shift.

\subsubsection{Elliptical boundary tilt scans.}
\label{subsubsec:tiltedEllipticalScans}

\begin{figure}
 \centering
 \includegraphics[width=0.7\textwidth]{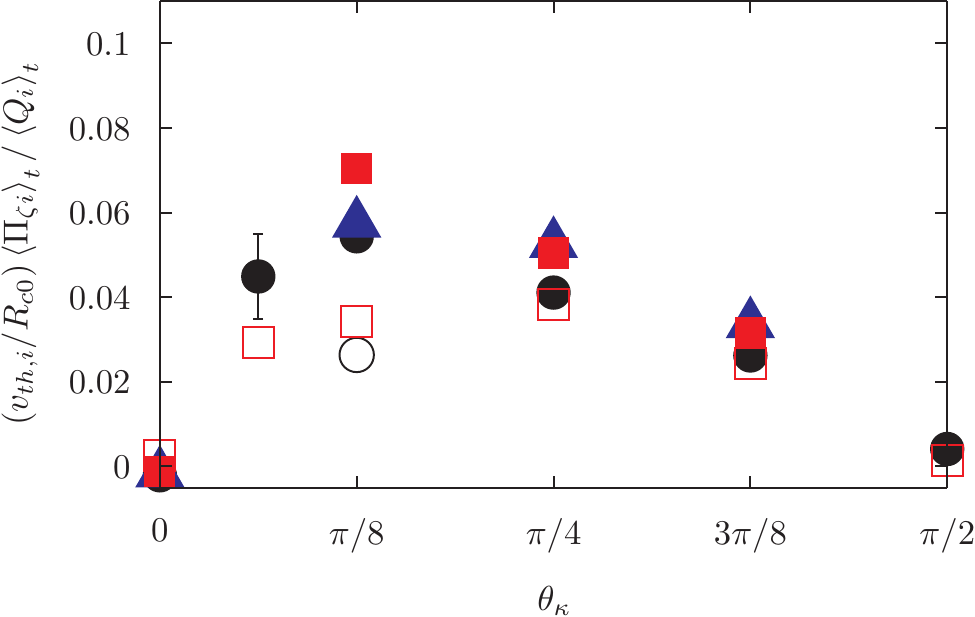}
 \caption{The ion momentum transport for flux surfaces with no shift (black, circles), a modest shift (blue, triangles), and an ITER-like shift (red, squares) for $\beta' = 0$ (filled) and an ITER-like $\beta'$ (empty).}
 \label{fig:momHeatFluxRatio}
\end{figure}

Figure \ref{fig:momHeatFluxRatio} shows the ratio of the time-averaged ion momentum flux to the time-averaged ion energy flux, calculated by GS2 for the tilted elliptical scans. As we will show in section \ref{subsec:betaProfile}, this quantity indicates the strength of momentum transport and is roughly proportional to the level of rotation (see equation \refEq{eq:rotationGradEst}). Figure \ref{fig:momHeatFluxRatio} also provides an estimate of the statistical error in the data. This error arises from performing a finite time-average over noisy turbulent quantities. It was estimated by repeating several simulations and computing the average difference between the corresponding results.

Figure \ref{fig:momHeatFluxRatio} demonstrates that the presence of an ITER-like Shafranov shift increases the momentum transport, here by approximately $30 \%$. As discussed in section \ref{sec:introduction}, this is expected because the Shafranov shift provides an additional source of up-down asymmetry and breaks both the mirror and tilting symmetry of the flux surfaces. However, we see that a non-zero $\beta'$ significantly reduces the momentum transport. We will investigate this result in section \ref{subsec:betaPrime} by studying at the magnitude of the up-down symmetry-breaking in the gyrokinetic equation. These two effects counteract one another and for ITER-like values at $\theta_{\kappa} = \pi / 8$ and $\rho_{0} = 0.54$ the shift is overshadowed by $\beta'$, leading to a net reduction in the momentum transport of about $30 \%$. In performing this scan, we added two simulations at $\pi / 16$ in order to better resolve the steep gradient that appears at small tilt angles. Additionally, we removed two simulations at $\pi / 2$ to save computational time because we had already confirmed that up-down symmetric shapes drive no rotation, even with a Shafranov shift.

\begin{figure}
 \centering
 \includegraphics[width=0.7\textwidth]{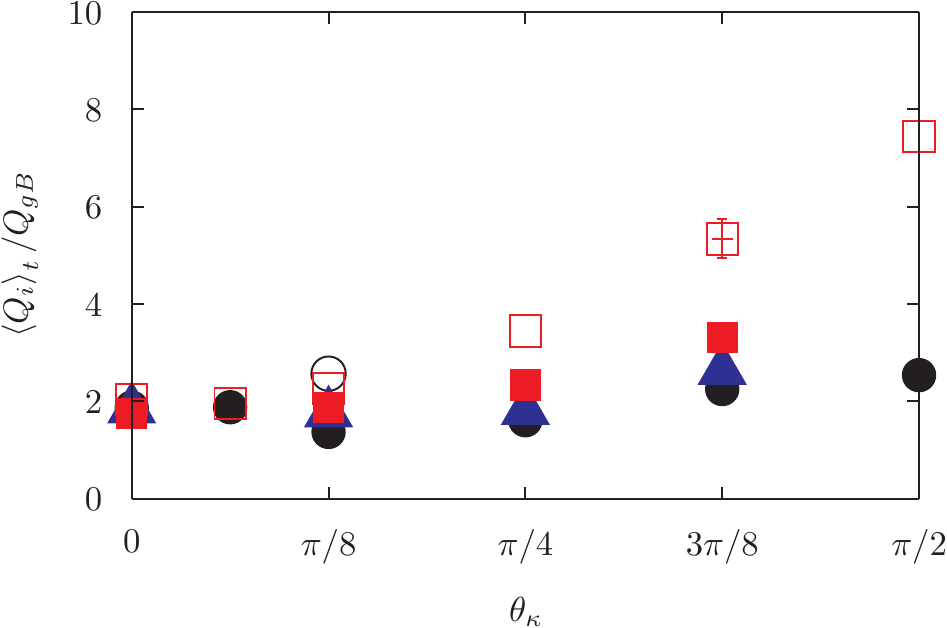}
 \caption{The ion energy flux for flux surfaces with no shift (black, circles), a modest shift (blue, triangles), and an ITER-like shift (red, squares) for $\beta' = 0$ (filled) and an ITER-like $\beta'$ (empty). In this and subsequent figures, whenever a single set of error bars is shown, it gives a representative estimate of the error for each data point.}
 \label{fig:heatFlux}
\end{figure}

Figure \ref{fig:heatFlux} shows the ion energy flux. We see that it is fairly insensitive to the effects of both the Shafranov shift and $\beta'$ in the domain of $\theta_{\kappa} \in \left[ 0, \pi / 8 \right]$. At more extreme tilt angles we see that $\beta'$ dramatically increases the energy flux, as does the shift (albeit to a lesser extent).

\subsubsection{Minor radial scans.}
\label{subsubsec:minorRadialScans}

These scans keep $\beta'$, $d \Ln{T_{s}} / d \rho$, $d \Ln{n_{s}} / d \rho$, $q$, and $\hat{s}$ constant with minor radius. We chose to keep $\beta'$ constant to be consistent with ITER (according to equation \refEq{eq:betaPrimeEstimate}). The others were kept fixed in order to make comparisons with previous results more straightforward. However, constant values for $d \Ln{T_{s}} / d \rho$ and $d \Ln{n_{s}} / d \rho$ is not an unreasonable approximation to many experiments, especially in the core of tokamaks \cite{BarnesTrinity2010}. The local shift is calculated at each minor radius to be consistent with equations \refEq{eq:radialShafShiftLocal} and \refEq{eq:axialShafShiftLocal}, which result from the global MHD calculation.

\begin{figure}
 \centering
 \includegraphics[width=0.7\textwidth]{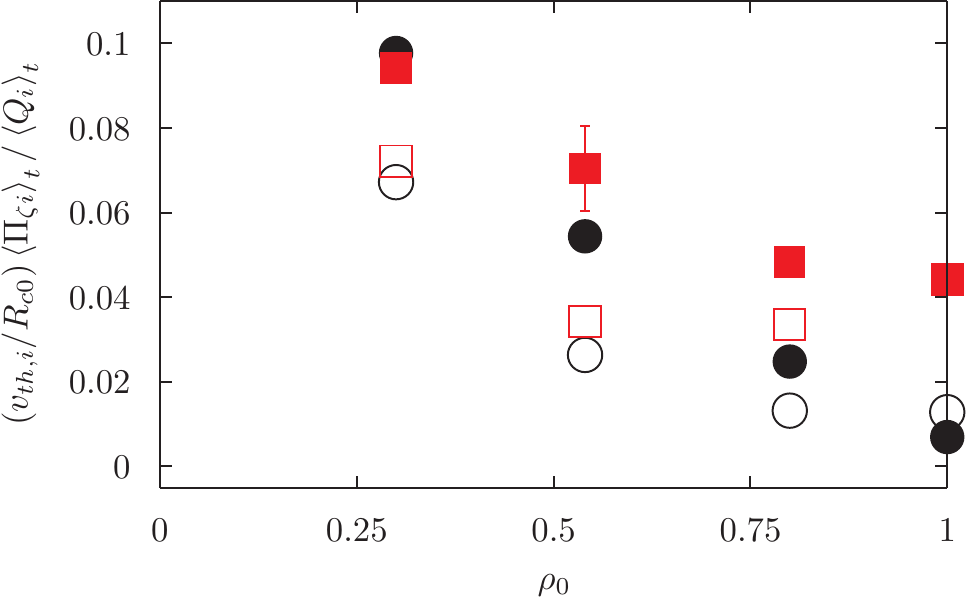}
 \caption{The radial dependence of the momentum transport for flux surfaces with no shift (black, circles) and a strong shift (red, squares) varied according to equations \refEq{eq:radialShafShiftLocal} and \refEq{eq:axialShafShiftLocal}, for $\beta' = 0$ (filled) and an ITER-like $\beta'$ (empty).}
 \label{fig:momHeatFluxRatioWithMinorRadius}
\end{figure}

The minor radial dependence of the momentum flux is shown in figure \ref{fig:momHeatFluxRatioWithMinorRadius}. Note that at $\rho_{0} = 1$ the momentum transport in the shifted configurations with and without $\beta'$ are indistinguishable. Comparing the two scans with $\beta' = 0$, we see that the difference in the momentum transport from the two scans increases with minor radius. The only difference between the scans is the presence of the local Shafranov shift, which also increases with minor radius. Hence, this reinforces a result of figure \ref{fig:momHeatFluxRatio}: the Shafranov shift increases the momentum transport. Similarly, comparing the two scans with no shift reinforces the fact that $\beta'$ reduces the momentum transport (which we also observed in figure \ref{fig:momHeatFluxRatio}). Additionally, comparing the no shift, $\beta' = 0$ case to the ITER-like shift, ITER-like $\beta'$ case demonstrates the counteracting effects of the shift and $\beta'$ on the momentum transport. Because the shift is weak at small values of $\rho_{0}$, the net effect of the shift and $\beta'$ is to lower the momentum transport. However, at large values of $\rho_{0}$ the shift is stronger, but $\beta'$ remains the same. Here the net effect of the shift and $\beta'$ is to enhance the momentum transport.

Lastly, a dominant trend appearing in figure \ref{fig:momHeatFluxRatioWithMinorRadius} is the roughly linear decrease of the momentum transport with minor radius. It is most clearly seen in the data series with no shift and $\beta' = 0$, because the only difference between the four simulations is the value of the minor radius. This trend (i.e. an increase in the momentum transport with increasing aspect ratio) is not currently understood as nearly all simulations of intrinsic rotation from up-down asymmetry were performed using $r_{\psi 0} / R_{c 0} \approx 1 / 6$. However, it was also observed in several simulations performed at $r_{\psi 0} / R_{c 0} \approx 1 / 12$ and $r_{\psi 0} / R_{c 0} \approx 1 / 3$ in reference \cite{BallMomUpDownAsym2014}.

\begin{figure}
 \centering
 \includegraphics[width=0.7\textwidth]{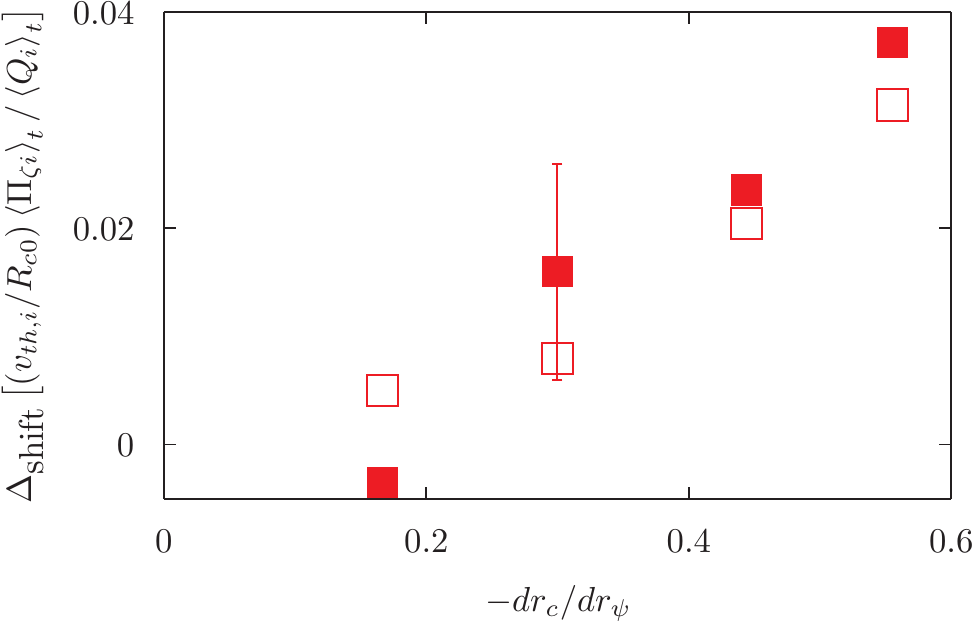}
 \caption{The change in the momentum transport caused by introducing a local Shafranov shift with a magnitude of $-d r_{c} / d r_{\psi}$ for $\beta' = 0$ (filled) and an ITER-like $\beta'$ (empty).}
 \label{fig:momHeatFluxRatioWithShift}
\end{figure}

In figure \ref{fig:momHeatFluxRatioWithShift} we show $\Delta_{\text{shift}} \left[ \left( v_{th, i} / R_{c 0} \right) \left\langle \Pi_{\zeta i} \right\rangle_{t} / \left\langle Q_{i} \right\rangle_{t} \right]$, the change in the momentum transport due to the Shafranov shift, where $r_{c} \equiv \sqrt{ \left( R_{c} - R_{0} \right)^{2} + Z_{c}^{2}}$ and $\Delta_{\text{shift}} \left[ x \right]$ is defined to be the value of $x$ when the Shafranov shift is included minus the value of $x$ when the Shafranov shift is omitted. This figure uses the same data as figure \ref{fig:momHeatFluxRatioWithMinorRadius}, but more clearly demonstrates that the momentum transport is not sensitive to small changes in the Shafranov shift. Rather it increases smoothly and fairly linearly with the strength of the Shafranov shift, irrespective of the value of $\beta'$.

\subsubsection{Circular flux surface scan.}
\label{subsubsec:circularScan}

\begin{figure}
 (a) \hspace{0.27\textwidth} (b) \hspace{0.28\textwidth} (c)
 \begin{center}
  \includegraphics[height=0.24\textwidth]{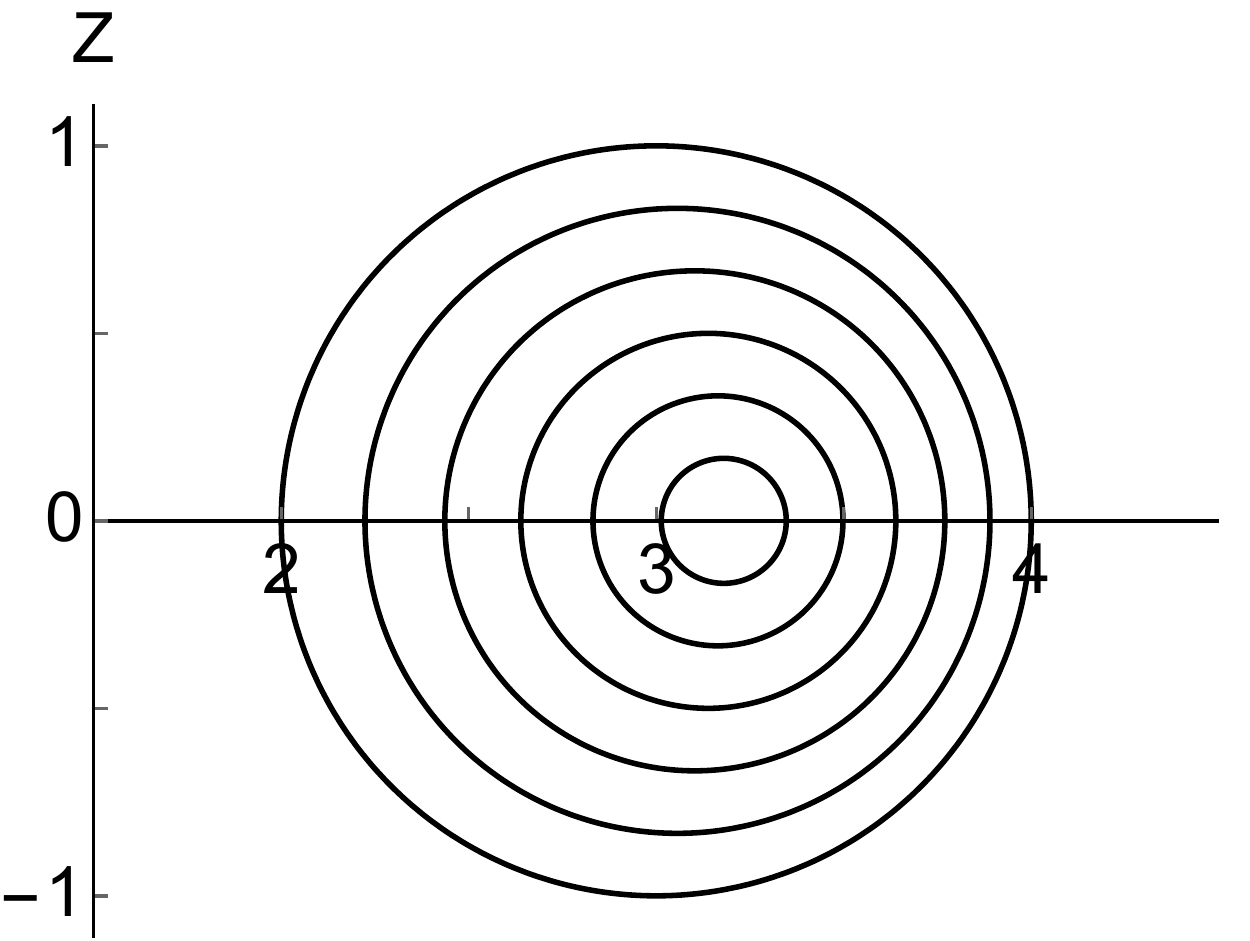}
  \includegraphics[height=0.24\textwidth]{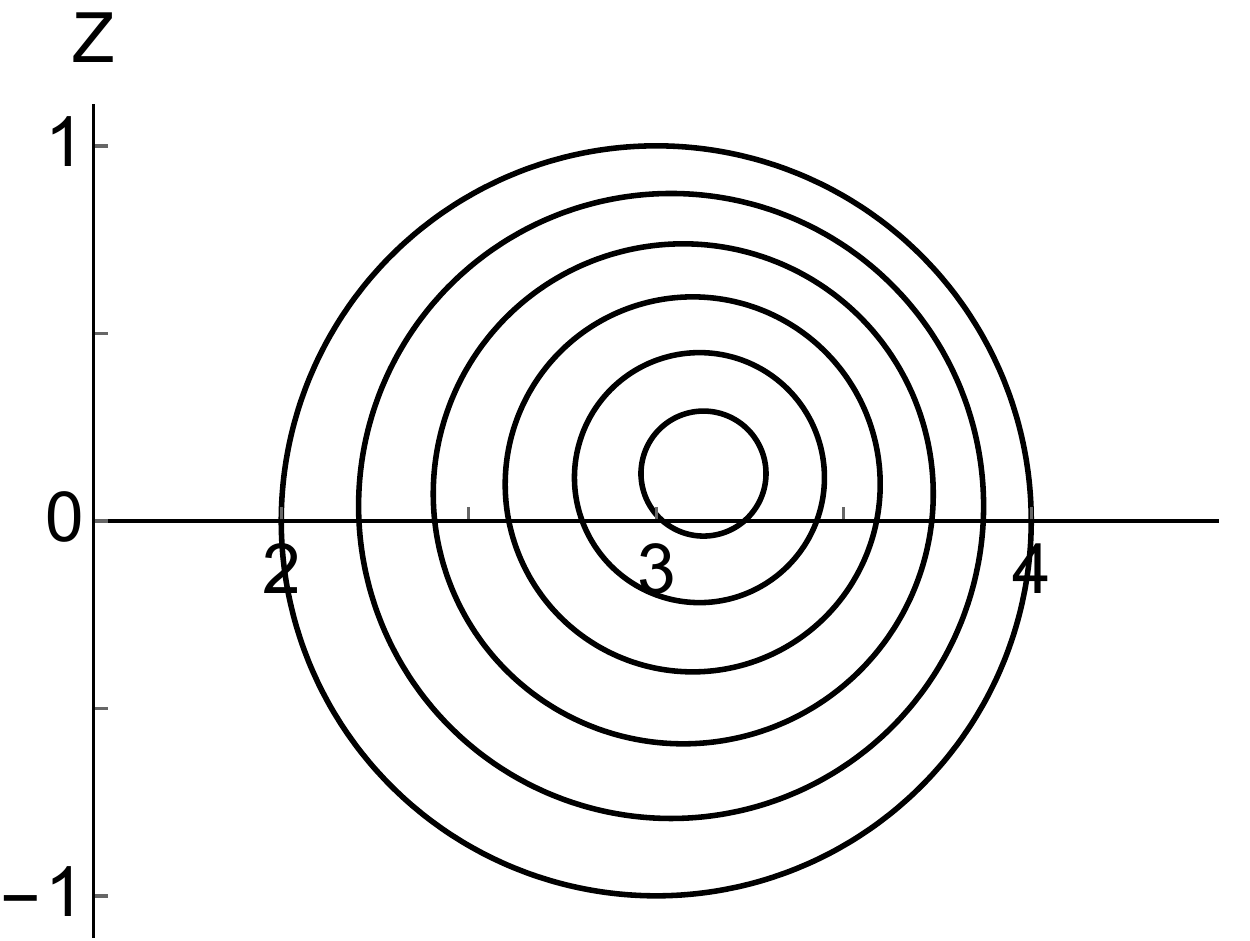}
  \includegraphics[height=0.24\textwidth]{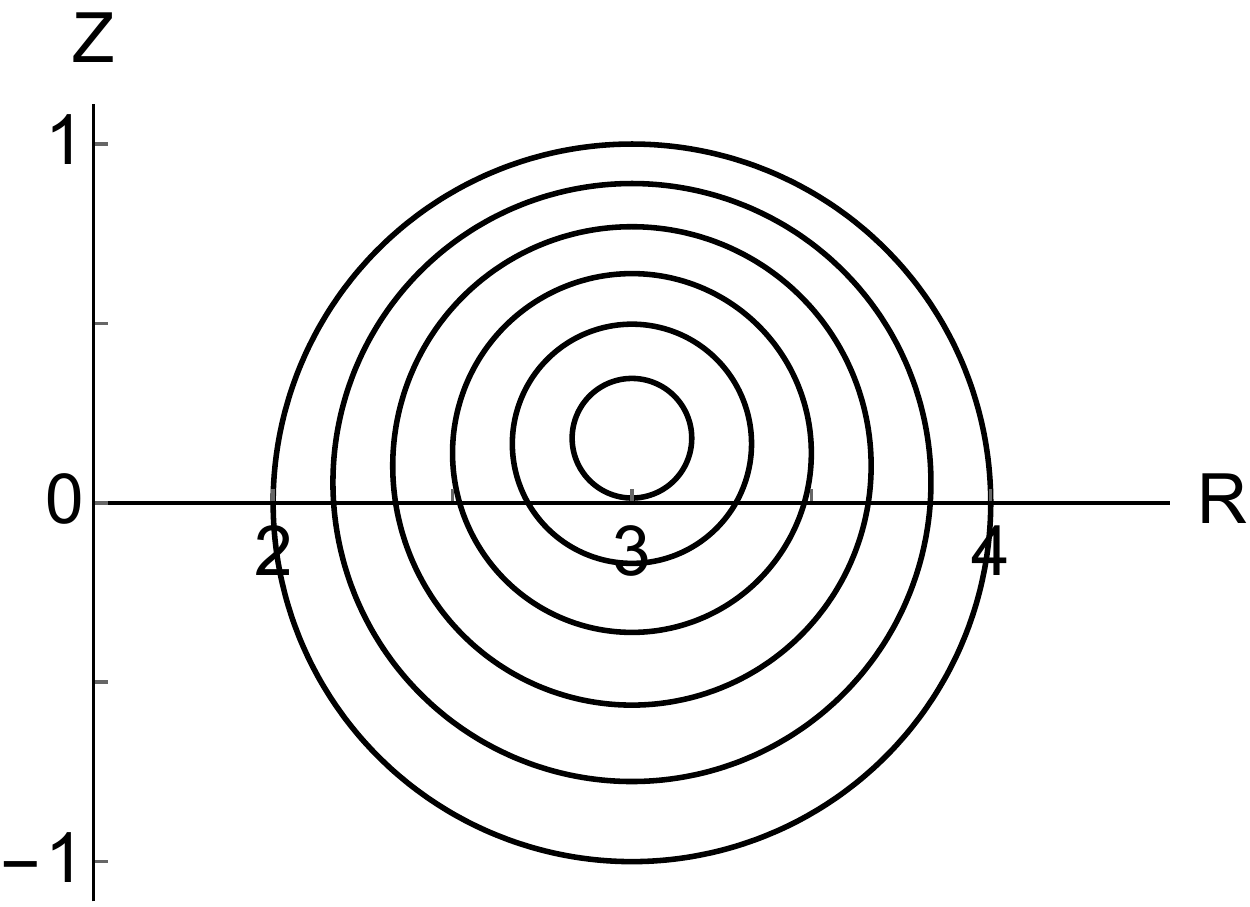}
 \end{center}
 \caption{The magnetic geometry for circular flux surfaces with an ITER-like (a) horizontal shift, (b) diagonal shift, or (c) vertical shift.}
 \label{fig:circScanGeo}
\end{figure}

To completely isolate the effect of the Shafranov shift on momentum transport we also ran simulations with shifted circular flux surfaces as shown in figure \ref{fig:circScanGeo}. To create up-down asymmetry and drive momentum transport we varied the direction of the tilt by changing the parameter $\theta_{\text{axis}}$ with the magnitude of the shift fixed at $\sim 30 \%$ larger than an untilted ITER-like machine. Scanning $\theta_{\text{axis}}$ is unphysical because circular flux surfaces can only ever have a shift in the outboard radial direction, which corresponds to $\theta_{\text{axis}} = 0$. Though unphysical, this scan will help clarify the influence of the Shafranov shift.

\begin{figure}
 \centering
 \includegraphics[width=0.7\textwidth]{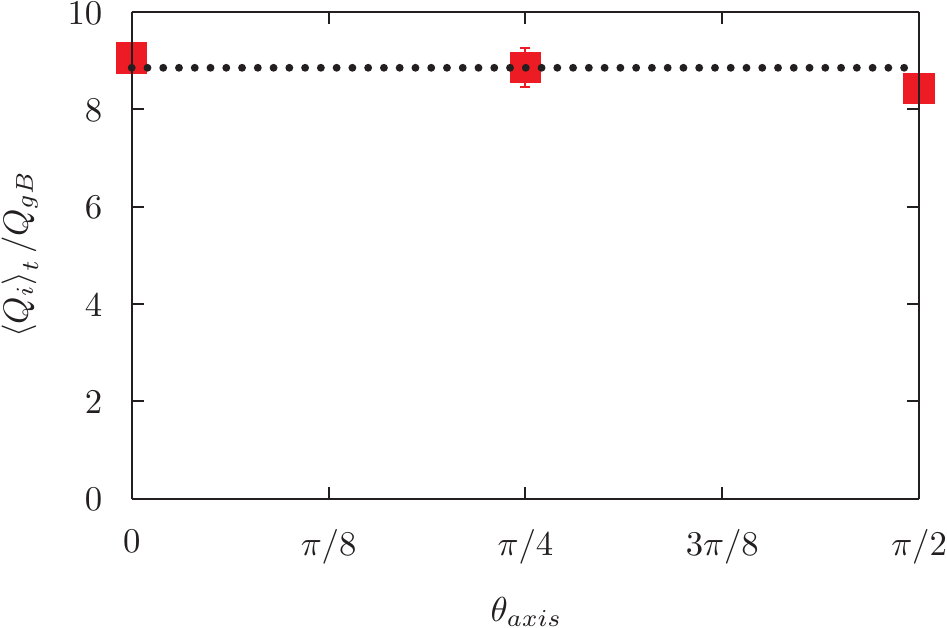}
 \caption{The energy flux for circular flux surfaces with no shift (black, dotted line) and an ITER-like shift (red, square points) as a function of the direction of the Shafranov shift. All simulations have $\beta' = 0$.}
 \label{fig:heatFluxCirc}
\end{figure}

Figure \ref{fig:heatFluxCirc} shows that the presence and direction of the Shafranov shift has little effect on the ion energy flux from circular flux surfaces. This behavior is similar to the tilted elliptical results (see figure \ref{fig:heatFlux}) in the range of $\theta_{\kappa} \in \left[ 0, \pi / 8 \right]$, but different from the tilted elliptical results in the range of $\theta_{\kappa} \in \left[ \pi / 8, \pi / 2 \right]$. This is consistent because the magnitude of the shift in the circular equilibria is similar to that of the elliptical equilibria in the range of $\theta_{\kappa} \in \left[ 0, \pi / 8 \right]$, but considerably less than the magnitude of the shift present in the elliptical equilibria with larger tilt angles. Therefore, both figures indicate that the shift present in the circular and minimally-tilted elliptical flux surfaces is not strong enough to modify the energy flux significantly.

\begin{figure}
 \centering
 \includegraphics[width=0.7\textwidth]{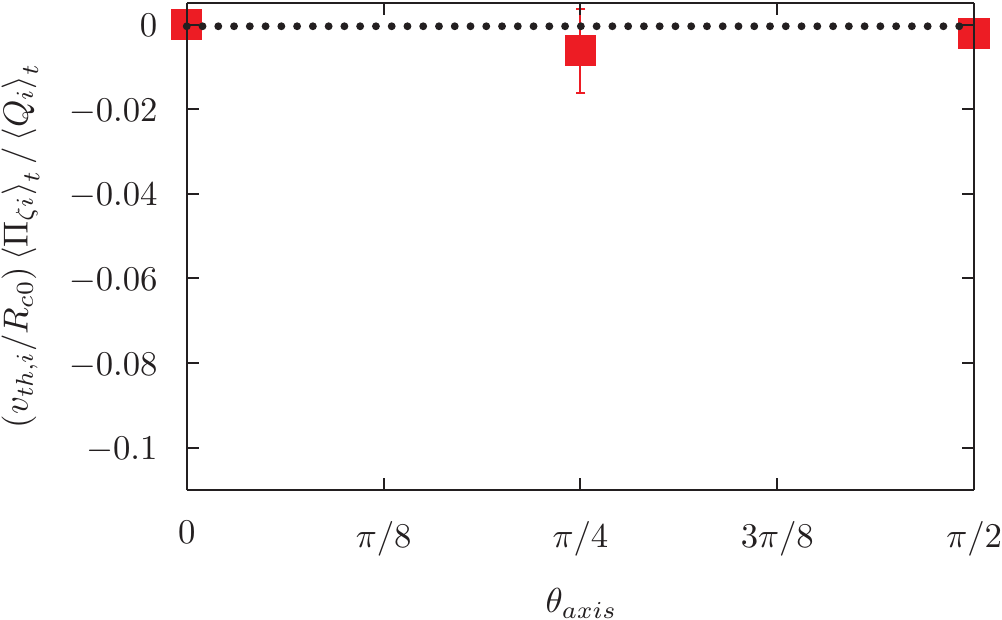}
 \caption{The momentum flux for circular flux surfaces with no shift (black, dotted line) and an ITER-like shift (red, square points) as a function of the direction of the Shafranov shift. All simulations have $\beta' = 0$. Note that we have kept the range of the vertical axis the same as in figures \ref{fig:momHeatFluxRatio} and \ref{fig:momHeatFluxRatioWithMinorRadius} for ease of comparison.}
 \label{fig:momHeatFluxRatioCirc}
\end{figure}

\begin{figure}
 \centering
 \includegraphics[width=0.7\textwidth]{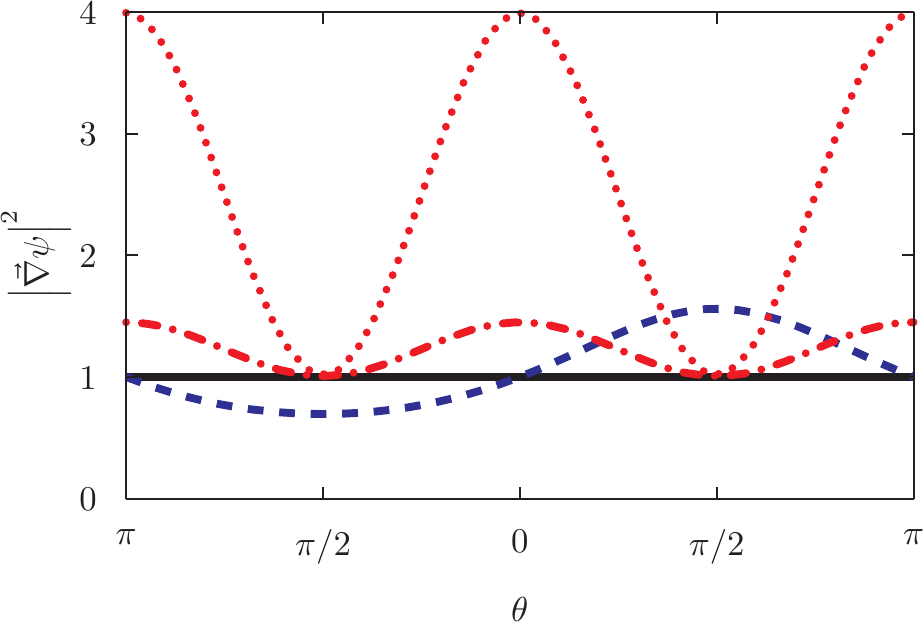}
 \caption{The geometric coefficient $\left| \Nabla \psi \right|^{2}$ for unshifted circular flux surfaces (black, solid), circular flux surfaces with a strong vertical shift (blue, dashed), and unshifted flux surfaces with a vertical elongation of $\kappa = 2$ (red, dotted) or $\kappa = 1.2$ (red, dash-dotted) normalized to the unshifted circular value.}
 \label{fig:gds22}
\end{figure}

Figure \ref{fig:momHeatFluxRatioCirc} shows the effect of a strong Shafranov shift on momentum transport. We see that a pure shift in circular flux surfaces (even when it is diagonal or vertical) drives minimal rotation compared to that generated by elliptical flux surfaces (as shown in figure \ref{fig:momHeatFluxRatio}). This is somewhat surprising since the shift is an $m = 1$ shaping effect and we expect the momentum flux to scale as $\Exp{-m}$ in mirror symmetric configurations \cite{BallMirrorSymArg2016}. However, there are two important caveats. Firstly, the exponential scaling is only true in the limit of $m \gg 1$, which is clearly not satisfied for $m = 1$. Secondly, the Shafranov shift has a relatively minor effect on the magnetic equilibrium compared with elongating the flux surfaces to $\kappa = 2$ (even when the shift is $30 \%$ stronger than that expected in ITER). This can be quantified by looking at the geometric coefficients that appear in the gyrokinetic equations (see \ref{app:geoCoeff} and reference \cite{BallMirrorSymArg2016} for more details on these coefficients). The geometric coefficients are the only way the magnetic geometry enters the local gyrokinetic model, so we know they must control the momentum transport. Plotting the geometric coefficient $\left| \Nabla \psi \right|^{2}$ as an example produces figure \ref{fig:gds22}, which shows that elongating an unshifted circular configuration to $\kappa = 2$ causes a $300 \%$ change, while introducing the Shafranov shift only causes a $50 \%$ change. To fairly compare the ability of the Shafranov shift and elongation to drive rotation we should control for the effect on the magnetic equilibrium. From figure \ref{fig:gds22} we see that an elliptical configuration with $\kappa = 1.2$ has a similar effect on $\left| \Nabla \psi \right|^{2}$ as the pure Shafranov shift. Performing a nonlinear gyrokinetic simulation of a tilted elliptical configuration with $\kappa = 1.2$ and $\theta_{\kappa} = \pi / 8$ demonstrates that, like a pure Shafranov shift, it generates little momentum transport. This suggests that the Shafranov shift and elongation drive similar levels of rotation when they alter the geometric coefficients to a similar degree. Elongation is capable of driving much more rotation than a pure Shafranov shift, because it can have a much larger effect on the geometric coefficients. The effect of the Shafranov shift on the geometric coefficients is constrained through a practical limit on the maximum value of $\beta$. This proves to be more restrictive than the vertical stability limit, which constrains the externally-applied elongation.

\subsection{Effect of the value of $\beta'$}
\label{subsec:betaPrime}

\begin{figure}
 \begin{center}
   (a) \hspace{0.4\textwidth} (b) \hspace{0.35\textwidth}
  
  \includegraphics[height=0.3\textwidth]{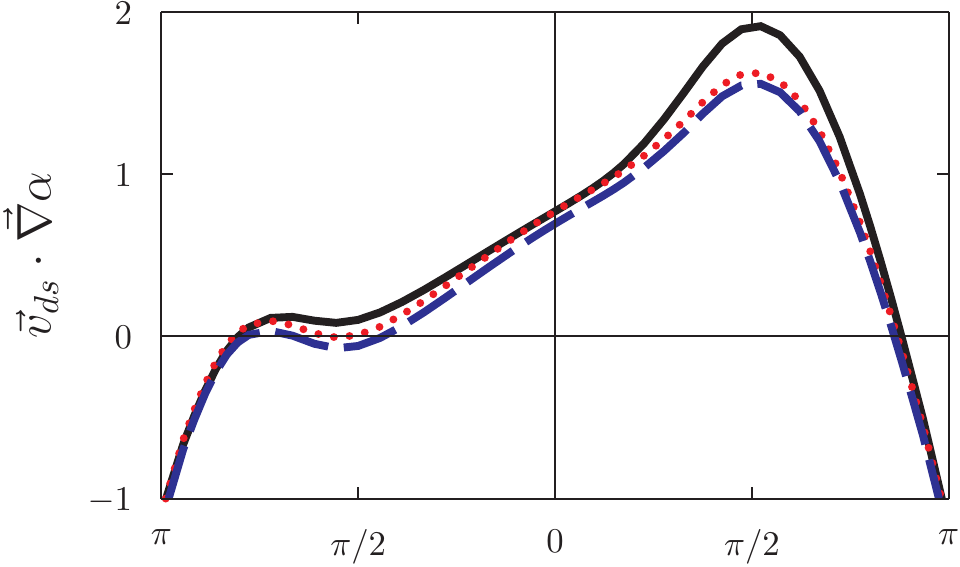}
  \includegraphics[height=0.3\textwidth]{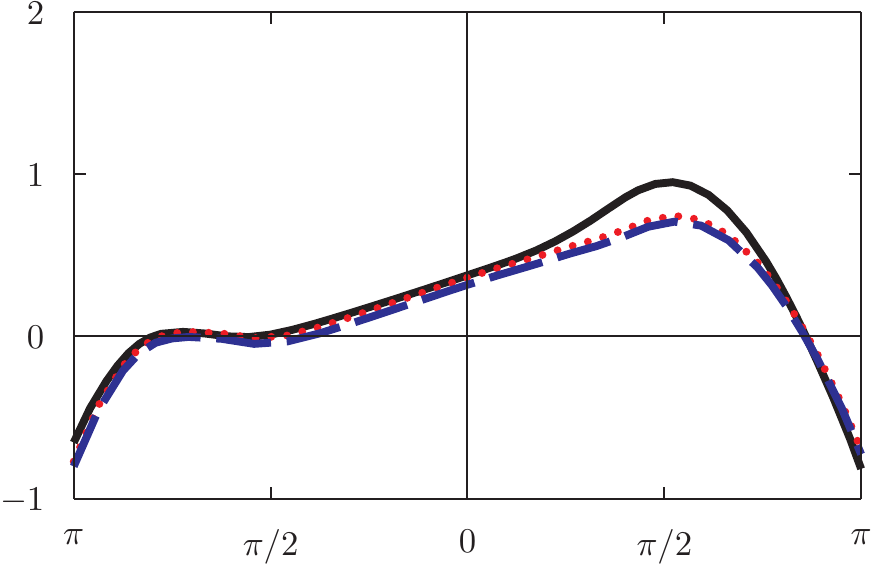}
  
   (c) \hspace{0.4\textwidth} (d) \hspace{0.35\textwidth}
  
  \includegraphics[height=0.35\textwidth]{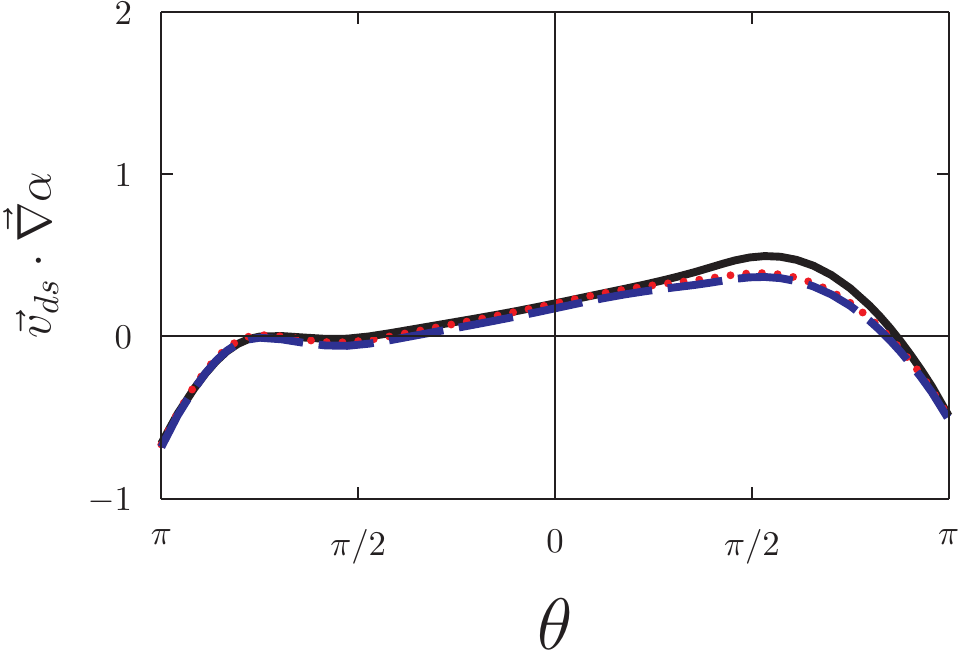}
  \includegraphics[height=0.35\textwidth]{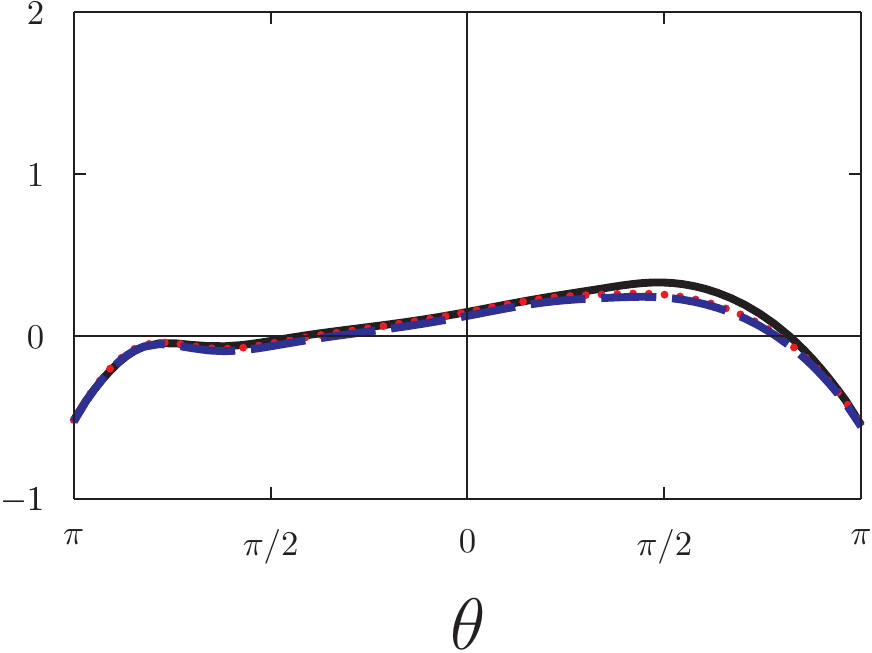}
 \end{center}
 \caption{The geometric coefficient $\vec{v}_{d s} \cdot \Nabla \alpha$ in units of $v_{th, s}^{2} / \left( a^{2} \Omega_{s} \right)$ without Shafranov shift at (a) $\rho_{0} = 0.3$, (b) $\rho_{0} = 0.54$, (c) $\rho_{0} = 0.8$, and (d) $\rho_{0} = 1$ for no $\beta'$, $w_{||}^{2} = v_{th, s}^{2}$, $w_{\perp}^{2} = 0$ (black, solid); an ITER-like $\beta'$, $w_{||}^{2} = v_{th, s}^{2}$, $w_{\perp}^{2} = 0$ (red, dotted); and an ITER-like $\beta'$, $w_{||}^{2} = 0$, $w_{\perp}^{2} = 2 v_{th, s}^{2}$ (blue, dashed).}
 \label{fig:driftCoeff}
\end{figure}

In section \ref{subsec:results} we included the effect of the Shafranov shift in nonlinear, local gyrokinetic simulations and found that it enhanced momentum transport as expected. Since the magnitude of the shift depends on the plasma pressure, we also included a non-zero $\beta'$. While the Shafranov shift alters the spacing between flux surfaces, $\beta'$ enters through the right-hand side of the Grad-Shafranov equation and alters the local magnetic shear (i.e. the radial derivative of the magnetic field line pitch angle). We found that $\beta'$ strongly reduced the momentum flux, often entirely canceling the enhancement due to the Shafranov shift. Consequently, it is important to understand how $\beta'$ alters the geometric coefficients of gyrokinetics.

In \ref{app:geoCoeff} we discuss how $\beta'$ enters into the analytic expressions for the geometric coefficients. We show that $\beta'$ vanishes in the large aspect ratio limit (for the orderings of equation \refEq{eq:gradShafOrderings}), like the Shafranov shift. This means that for large aspect ratio tokamaks $\beta'$ can be ignored and the results of reference \cite{BallMomUpDownAsym2014} (which ignores $\beta'$) apply. However, the Shafranov shift also vanishes in this limit, so it cannot be used to enhance the momentum transport.

Figure \ref{fig:driftCoeff} uses the geometries from figure \ref{fig:momHeatFluxRatioWithMinorRadius} to show the quantitative effect of $\beta'$ on the geometric coefficient $\vec{v}_{d s} \cdot \Nabla \alpha$ (defined by equation \refEq{eq:driftVelAlpha}) with different values of $w_{||}$ and $w_{\perp}$. Here $\vec{w}$ is the velocity coordinate in the frame rotating with the background plasma flow, the $||$ subscript indicates parallel to the magnetic field, the $\perp$ subscript indicates perpendicular to the magnetic field, $\vec{v}_{d s}$ is the guiding center particle magnetic drifts,
\begin{align}
   \alpha \equiv& ~ \zeta - I \left( \psi \right) \left. \int_{\theta_{\alpha} \left( \psi \right)}^{\theta} \right|_{\psi} d \theta' \left( R^{2} \vec{B} \cdot \Nabla \theta' \right)^{-1} \label{eq:alphaDef}
\end{align}
is the coordinate within the flux surface and perpendicular to the magnetic field, and $\theta_{\alpha} \left( \psi \right)$ is a free function. Previous work seems to indicate that $\vec{v}_{d s} \cdot \Nabla \alpha$ may be the most important geometric coefficient for understanding intrinsic rotation transport due to up-down asymmetry \cite{BallMomFluxScaling2016}. We see that including a non-zero $\beta'$ tends to reduce the up-down asymmetry of $\vec{v}_{d s} \cdot \Nabla \alpha$, which is consistent with the observed reduction in momentum transport.

\subsection{Effect of the $\beta$ profile}
\label{subsec:betaProfile}

In order to estimate a realistic value for $\beta'$, we used the on-axis value of $\beta$ predicted for ITER and assumed $\beta$ was linear with minor radius $r_{\psi}$. This gave a reasonable order of magnitude estimate. However, since the momentum transport is strongly and adversely affected by $\beta'$ it is worthwhile to discuss the implications of different radial profiles of $\beta'$. For example, we expect that in H-mode operation $\beta'$ would be larger at the plasma edge and smaller in the core compared to L-mode. Unfortunately, since intrinsic rotation is ultimately driven by the gradients in density and temperature, $\beta'$ is necessary, even though including the effect of $\beta'$ in the geometric coefficients reduces the momentum flux. To see the relationship between $\beta'$ and the rotation gradient we will follow the analysis of reference \cite{BallMomUpDownAsym2014}.

First, we neglect the momentum pinch (which can only ever enhance the level of rotation) and assume that diffusion is the only mechanism balancing the intrinsic source to get
\begin{align}
   \left\langle \Pi_{\zeta i} \right\rangle_{t} \approx D_{\Pi i} n_{i} m_{i} R_{c}^{2} \frac{d \Omega_{\zeta i}}{d r_{\psi}} , \label{eq:momFluxEstimate}
\end{align}
where $\left\langle \Pi_{\zeta i} \right\rangle_{t}$ is the time-averaged intrinsic ion momentum flux source term arising from up-down asymmetry (i.e. the momentum flux calculated by GS2 for $\Omega_{\zeta i} = d \Omega_{\zeta i} / d r_{\psi} = 0$), $D_{\Pi i}$ is the momentum diffusivity (i.e. the kinematic viscosity), $R_{c}$ is the major radial location of the center of a given flux surface, $\Omega_{\zeta i} \equiv u_{\zeta i} / R$ is the ion rotation frequency, and $u_{\zeta i}$ is the ion bulk toroidal velocity. We take the energy flux to be the diffusion of a temperature gradient \cite{FreidbergFusionEnergy2007pg452} according to
\begin{align}
   \left\langle Q_{i} \right\rangle_{t} \approx - D_{Q i} n_{i} \frac{d T_{i}}{d r_{\psi}} , \label{eq:heatFluxEstimate}
\end{align}
where $\left\langle Q_{i} \right\rangle_{t}$ is the time-averaged energy flux calculated by GS2. Combining these two equations through the turbulent ion Prandtl number $Pr_{i} \equiv D_{\Pi i} / D_{Q i} \approx 0.7$ \cite{BallMomUpDownAsym2014} gives
\begin{align}
   \frac{1}{v_{th, i}} \frac{d u_{\zeta i}}{d r_{\psi}} \approx \frac{-1}{2 Pr_{i}} \left( \frac{v_{th, i}}{R_{c}} \frac{\left\langle \Pi_{\zeta i} \right\rangle_{t}}{\left\langle Q_{i} \right\rangle_{t}} \right) \frac{d}{d r_{\psi}} \Ln{T_{i}} . \label{eq:rotationGradEstSimple}
\end{align}
where we used that $T_{i} = m_{i} v_{th, i}^{2} / 2$. Doing this is useful because the Prandtl number is expected to be unaffected by changes in tokamak parameters. We will introduce the Alfv\'{e}n Mach number, $M_{A} \equiv \left| u_{\zeta i} \right| \sqrt{\mu_{0} n_{i} m_{i}} / B_{0}$, because it is the relevant quantity for stabilizing MHD modes, such as resistive wall modes. Neglecting the density gradient (because $d \Ln{T_{i}} / d r_{\psi}$ is three times larger than $d \Ln{n_{i}} / d r_{\psi}$) as well as assuming $n_{e} = n_{i}$ and $T_{e} = T_{i}$ allows equation \refEq{eq:rotationGradEstSimple} to be rewritten as
\begin{align}
   M_{A} \left( \rho \right) \approx \left| \int_{1}^{\rho} d \rho' \frac{1}{2 \sqrt{2} Pr_{i} \left( \rho' \right)} \left( \frac{v_{th, i} \left( \rho' \right)}{R_{c} \left( \rho' \right)} \frac{\left\langle \Pi_{\zeta i} \left( \rho' \right) \right\rangle_{t}}{\left\langle Q_{i} \left( \rho' \right) \right\rangle_{t}} \right) \frac{\beta' \left( \rho' \right)}{\sqrt{\beta \left( \rho' \right)}} \right| . \label{eq:rotationGradEst}
\end{align}
We wrote this expression in terms of $\left( v_{th, i} / R_{c} \right) \left\langle \Pi_{\zeta i} \right\rangle_{t} / \left\langle Q_{i} \right\rangle_{t}$ because it is the normalized parameter that indicates how strongly a given geometry drives rotation.

\begin{figure}
 \begin{center}
   (a) \hspace{0.4\textwidth} (b) \hspace{0.35\textwidth}
  
  \includegraphics[width=0.43\textwidth]{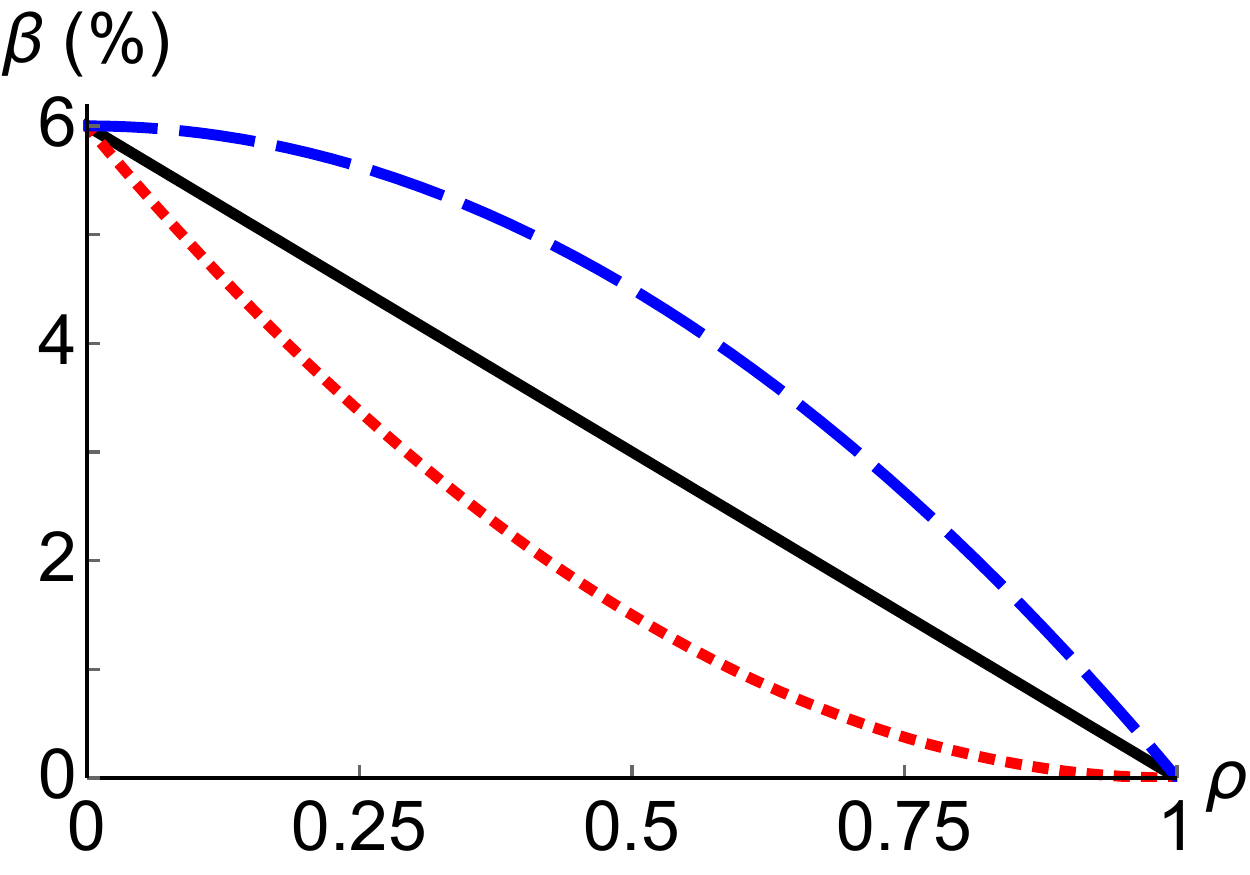}
  \includegraphics[width=0.47\textwidth]{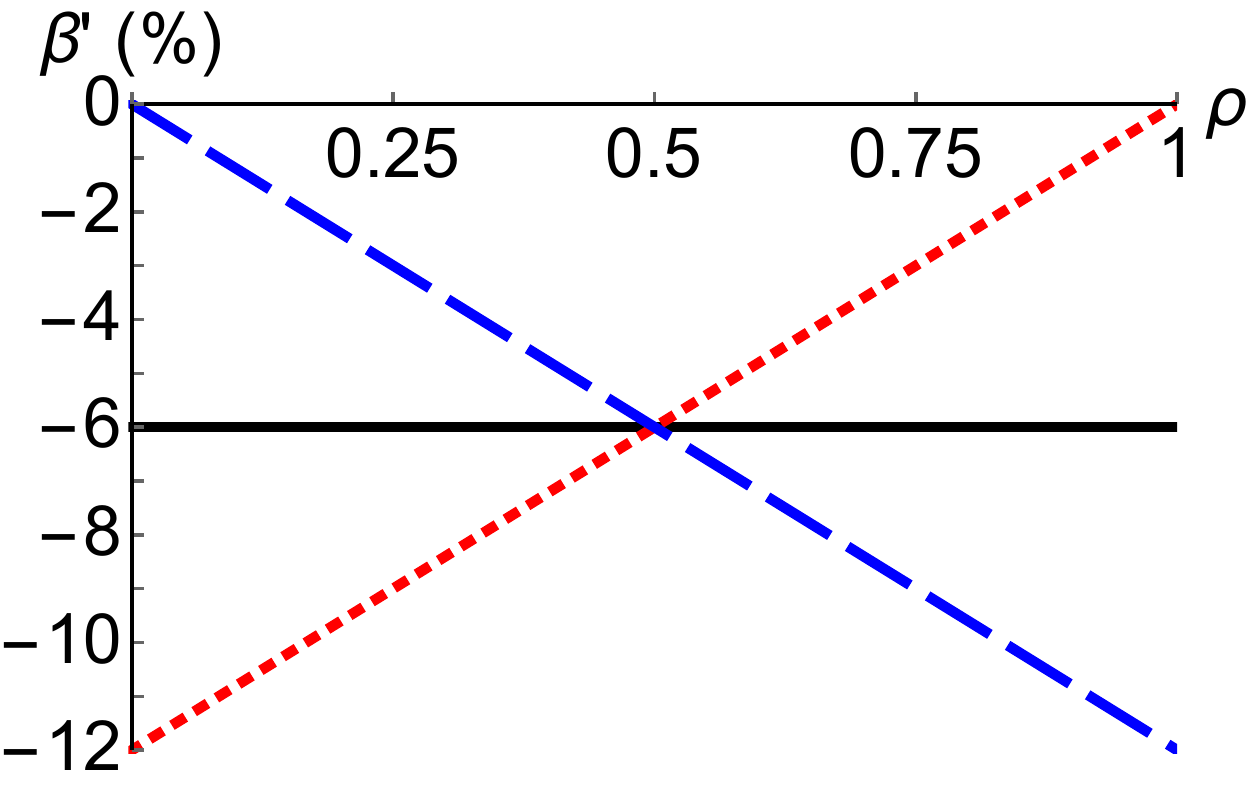}
  
   (c) \hspace{0.4\textwidth} (d) \hspace{0.35\textwidth}
  
  \includegraphics[width=0.45\textwidth]{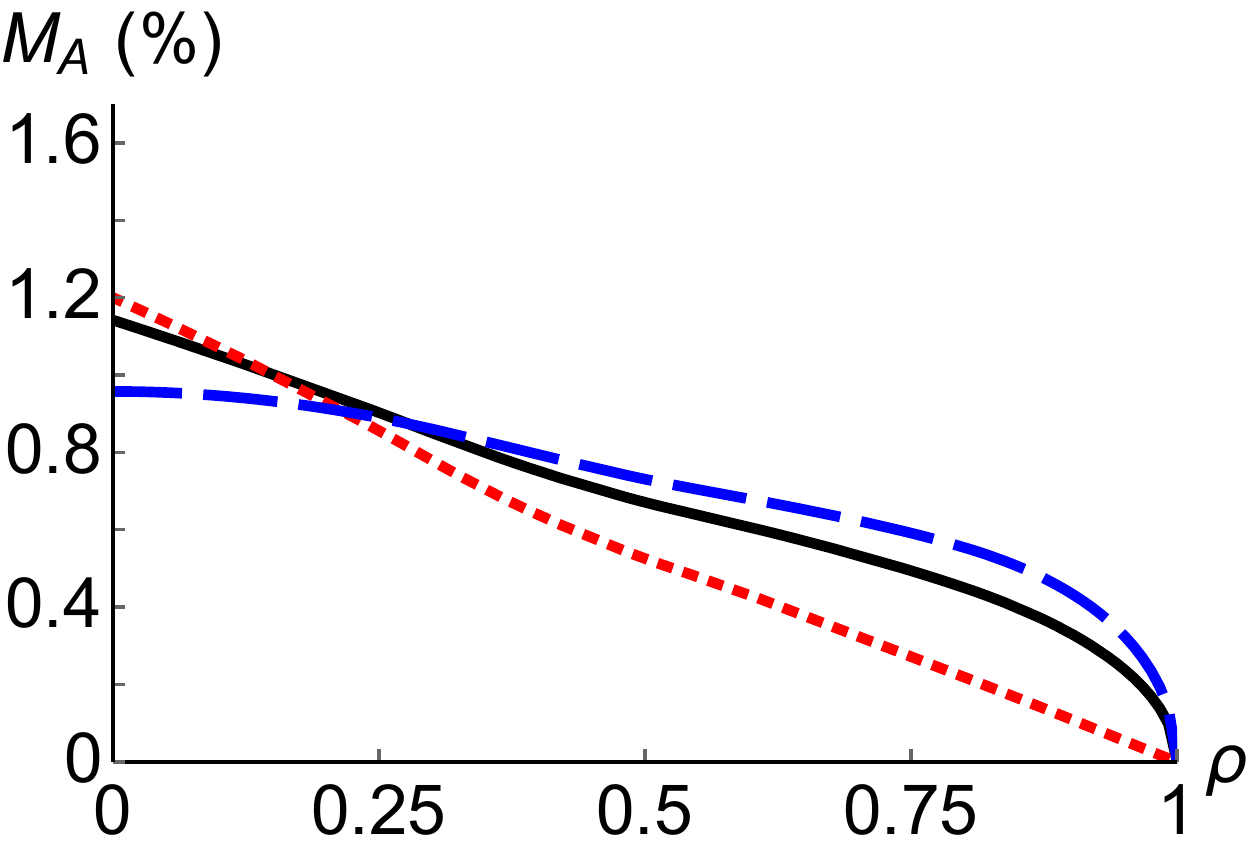}
    \includegraphics[width=0.45\textwidth]{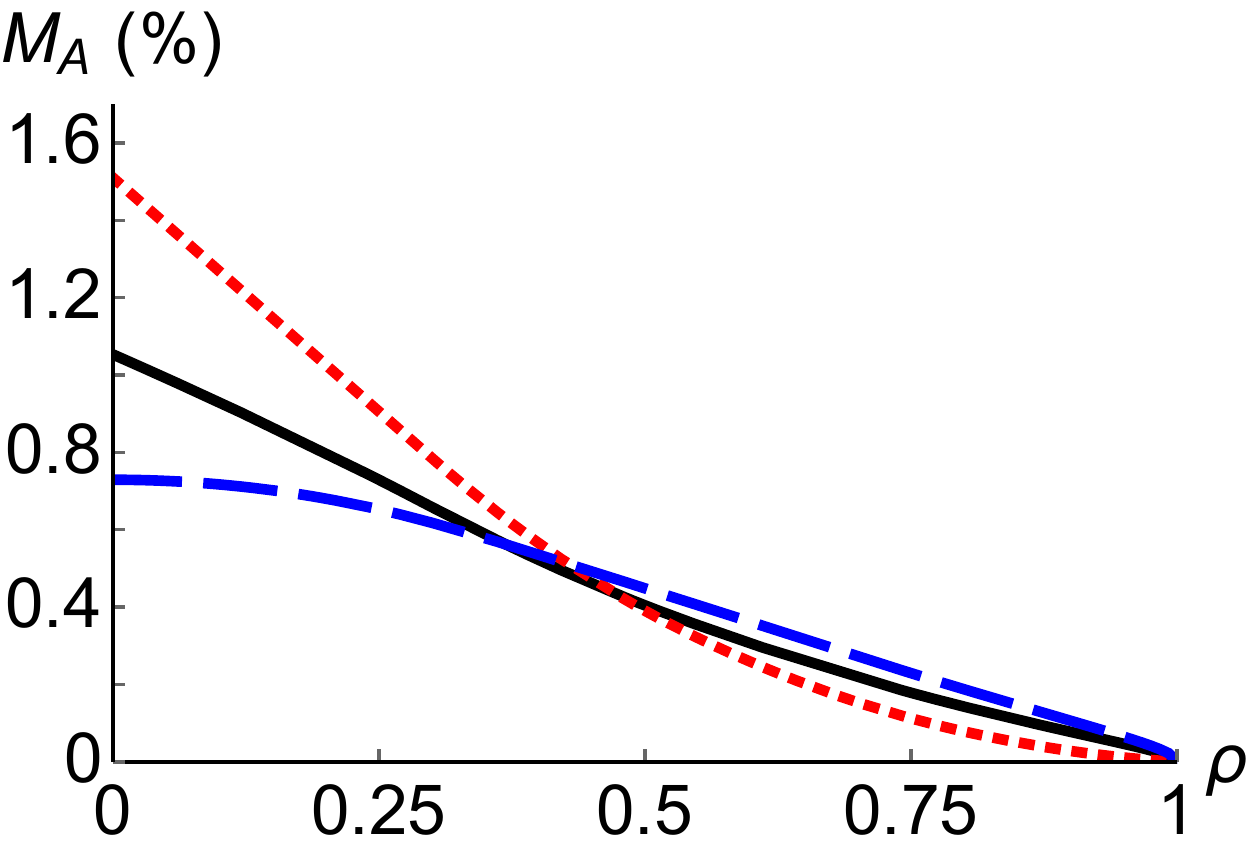}
 \end{center}
 \caption{Example (a) $\beta$ profiles with their corresponding (b) $\beta'$ and (c,d) Alfv\'{e}n Mach number profiles, estimated using the data from figure \ref{fig:momHeatFluxRatioWithMinorRadius} both (c) with and (d) without the effects of the pressure profile on the magnetic equilibrium, for constant $\beta'$ (black, solid), linear peaked $\beta'$ (red, dotted), and linear hollow $\beta'$ (blue, dashed) profiles.}
 \label{fig:MachNumberProfile}
\end{figure}

Equation \refEq{eq:rotationGradEst} shows several competing dependencies on $\beta$ and $\beta'$, both explicitly and through $\left( v_{th, i} / R_{c} \right) \left\langle \Pi_{\zeta i} \right\rangle_{t} / \left\langle Q_{i} \right\rangle_{t}$. Hence, it is difficult to analytically determine the $\beta$ profile that maximizes rotation. However, we can perform a bilinear interpolation of the data in figure \ref{fig:momHeatFluxRatioWithMinorRadius} to approximate the functional form of $G \left( \rho, \beta' \right) \equiv \left( v_{th, i} / R_{c} \right) \left\langle \Pi_{\zeta i} \right\rangle_{t} / \left\langle Q_{i} \right\rangle_{t}$. We note that the dependence on $\rho$ also includes a change in the strength of the Shafranov shift according to equations \refEq{eq:radialShafShiftLocal} and \refEq{eq:axialShafShiftLocal}. To estimate the function $G \left( \rho, \beta' \right)$ between data points at $\left( \rho_{1}, \beta'_{1} \right)$, $\left( \rho_{1}, \beta'_{2} \right)$, $\left( \rho_{2}, \beta'_{1} \right)$, and $\left( \rho_{2}, \beta'_{2} \right)$ we use
\begin{align}
   G \left( \rho, \beta' \right) &\approx \frac{\beta'_{2} - \beta'}{\beta'_{2} - \beta'_{1}} \left( \frac{\rho_{2} - \rho}{\rho_{2} - \rho_{1}} G \left( \rho_{1}, \beta'_{1} \right) + \frac{\rho - \rho_{1}}{\rho_{2} - \rho_{1}} G \left( \rho_{2}, \beta'_{1} \right) \right) \\
   &+ \frac{\beta' - \beta'_{1}}{\beta'_{2} - \beta'_{1}} \left( \frac{\rho_{2} - \rho}{\rho_{2} - \rho_{1}} G \left( \rho_{1}, \beta'_{2} \right) + \frac{\rho - \rho_{1}}{\rho_{2} - \rho_{1}} G \left( \rho_{2}, \beta'_{2} \right) \right) . \nonumber
\end{align}
In the region $\rho < 0.3$ we do not have data, so we assume that $\left( v_{th, i} / R_{c} \right) \left\langle \Pi_{\zeta i} \right\rangle_{t} / \left\langle Q_{i} \right\rangle_{t}$ is constant in $\rho$. This assumption is conservative compared to a linear extrapolation using the data at $\rho = 0.3$ and $\rho = 0.54$. Furthermore, it leads to zero slope on-axis, which is consistent with constant asymptotic behavior in the large aspect ratio limit. To calculate the rotation profile for geometries with a strong Shafranov shift (shown in figure \ref{fig:MachNumberProfile}(c)) we used only the red square points in figure \ref{fig:momHeatFluxRatioWithMinorRadius}. Alternatively, if we assume that $\beta$ is very low (i.e. $\beta' = 0$) we can ignore the effect of the pressure profile on the magnetic geometry (i.e. neglect the Shafranov shift and assume pressure gradient term in the Grad-Shafranov equation is much smaller than the toroidal field flux function term). In this case the rotation profile (shown in figure \ref{fig:MachNumberProfile}(d)) can be calculated by a 1-D interpolation of the filled black circles because the turbulent transport becomes independent of $\beta'$.

Figure \ref{fig:MachNumberProfile} shows that both the Shafranov shift and the shape of the $\beta$ profile have a significant effect on the rotation profile. A broader $\beta$ profile consistently produces a broader rotation profile, but with a lower on-axis Mach number. This means that the $\beta$ profile that maximizes the {\it on-axis} Mach number is not necessarily optimal because broad rotation profiles are expected to be significantly more effective at stabilizing resistive wall modes \cite{LiuITERrwmStabilization2004}. Additionally, figures \ref{fig:MachNumberProfile}(c) and (d) indicate that stronger plasma pressure effects (i.e. Shafranov shift and $\beta'$) will cause up-down asymmetry to drive broader intrinsic rotation profiles. The reason for this can be seen in figure \ref{fig:momHeatFluxRatioWithMinorRadius}. Adding both the Shafranov shift and $\beta'$ (to go from the filled black circles to the empty red squares) reduces the core momentum transport, while enhancing the edge momentum transport. Lastly, we see that the largest rotation gradient occurs at the edge of the peaked pressure profile because the integral over the momentum flux in equation \refEq{eq:rotationGradEst} is weighted towards regions with small $\beta$ and large $\beta'$. This indicates that, even though the up-down asymmetry of a single-null divertor is usually limited to the edge, it may still drive significant rotation (especially in H-mode operation).

\section{Conclusions}
\label{sec:conclusions}

This paper focuses on two competing effects influencing the momentum transport: the Shafranov shift and $\beta'$. Together the two effects reduce momentum transport in the core, enhance it near the edge, and roughly cancel when averaged over the entire device. Using the nonlinear gyrokinetic simulations shown in figure \ref{fig:momHeatFluxRatioWithMinorRadius}, we estimate the rotation profile when these two effects are included (i.e. figure \ref{fig:MachNumberProfile}(c)) and when they are omitted (i.e. figure \ref{fig:MachNumberProfile}(d)). Comparing these profiles demonstrates that the on-axis value of the rotation is roughly unchanged, but the rotation profile is broadened (which is expected to be advantageous for stabilizing resistive wall modes). The magnitude of the on-axis rotation was found to be $\sim 1 \%$ (without including any enhancement due to the momentum pinch effect), which is in the range of what is needed to stabilize resistive wall modes in a large device like ITER (i.e. $0.5\% - 5\%$) \cite{LiuITERrwmStabilization2004}.

As anticipated a strong Shafranov shift was found to enhance the momentum transport in up-down asymmetric configurations because the shift itself becomes up-down asymmetric. The magnitude and direction of the shift was found to be insensitive to the shape of both the toroidal current (for a pressure profile that is a uniform fraction of the current profile) and pressure (for a uniform current profile) profiles at fixed geometry, plasma current, and average $d p / d \psi$.

On the other hand, it was found that the effect of $\beta'$ on the magnetic equilibrium significantly reduces the momentum transport, often entirely canceling the effect of the Shafranov shift. Consequently, the shape and magnitude of the rotation profile is sensitive to the radial profile of $\beta$. By studying the geometric coefficients, we found that, like the Shafranov shift, $\beta'$ appears to $\Order{\epsilon}$. However, unlike the Shafranov shift it tends to reduce the up-down asymmetry of the geometric coefficients.

\ack

J.B. and F.I.P. were funded in part by the RCUK Energy Programme (grant number EP/I501045). Computing time for this work was provided by the Helios supercomputer at IFERC-CSC under the projects SPIN, TRIN, MULTEIM, and GKMSC. The authors also acknowledge the use of ARCHER through the Plasma HEC Consortium EPSRC grant number EP/L000237/1 under the projects e281-gs2 and e281-rotation. J.P.L. was supported by the U.S. Department of Energy, Office of Science, Fusion Energy Sciences under Award No. DE-FG02-91ER-54109. A.J.C. was supported by the U.S. Department of Energy, Office of Science, Fusion Energy Sciences under Award Nos. DE-FG02-86ER53223 and DE-SC0012398.

\appendix

\section{Location of the magnetic axis for a constant current profile}
\label{app:exactMagAxisLoc}

In order to find the location of the magnetic axis for a constant toroidal current profile we will start with equation \refEq{eq:gradShafLowestOrderSolsTiltConst}. By requiring that $\psi_{0} \left( r_{b} \left( \theta \right), \theta \right) = \psi_{0 b}$ be constant on a tilted elliptical boundary parameterized by equation \refEq{eq:boundarySurf}, we find that 
\begin{align}
   \psi_{0 b} &= \frac{j_{N}}{2} \frac{a^{2} \kappa_{b}^{2}}{\kappa_{b}^{2} + 1} \label{eq:boundaryFluxConst} \\
   N_{0, 2} &= \frac{j_{N}}{4} \frac{\kappa_{b}^{2} - 1}{\kappa_{b}^{2} + 1} \label{eq:fourierShapingConst}
\end{align}
and $\theta_{t 0, 2} = \theta_{\kappa b}$ (according to equation \refEq{eq:tiltAngleSol}). All other lowest order Fourier coefficients are zero.

Calculating the next order Fourier coefficients from the boundary condition (i.e. requiring that $\psi_{1} \left( r_{b} \left( \theta \right), \theta \right) = \psi_{1 b}$ is constant) is algebraically intensive. We start with equation \refEq{eq:psiNextOrderSolConst}, the next order solution of the poloidal flux for a constant current profile. Note that while the current profile is assumed to be constant, we are allowing for a pressure gradient that is linear in $\psi$. First, we will postulate that the fifth, third, and first Fourier harmonics are the only ones required to match the boundary condition. All other next order Fourier coefficients are set to zero. Then we change to the shifted poloidal angle $\theta_{s} \equiv \theta + \theta_{\kappa b}$ in order to align the coordinate system with the minor and major axes of the elliptical boundary flux surface. Next we change from polar coordinates to Cartesian coordinates in the poloidal plane (i.e. $r = \sqrt{X^{2} + Y^{2}}$ and $\theta_{s} = \ArcTan{Y/X}$). This converts $\psi_{1} \left( r, \theta_{s} \right)$ into $\psi_{1} \left( X, Y \right)$, a fifth-order polynomial that contains products of $X$ and $Y$. Instead of equation \refEq{eq:boundarySurf}, we use
\begin{align}
   \left( \frac{X}{a} \right)^{2} + \left( \frac{Y}{\kappa_{b} a} \right)^{2} = 1 ,
\end{align}
the traditional Cartesian formula for an ellipse, as the boundary condition. Solving for $Y \left( X \right)$ and substituting it into $\psi_{1} \left( X, Y \right)$ allows us to eliminate all appearances of $X^{2}$, $X^{4}$, $Y^{2}$, and $Y^{4}$. We are left with a fifth-order polynomial that only has six terms, one proportional to each of $X^{5}$, $Y^{5} \left( X \right)$, $X^{3}$, $Y^{3} \left( X \right)$, $X$, and $Y \left( X \right)$. Since we have already made use of the boundary condition, we know that the whole polynomial must be constant. Requiring that the coefficients of the six terms be zero gives
\begin{align}
   C_{1, m} &= A_{C m} \Cos{m \theta_{\kappa b}} - A_{S m} \Sin{m \theta_{\kappa b}} \\
   S_{1, m} &= - A_{S m} \Cos{m \theta_{\kappa b}} - A_{C m} \Sin{m \theta_{\kappa b}} ,
\end{align}
where
\begin{align}
   A_{C 5} &\equiv \left( \kappa_{b}^{2} - 1 \right) \frac{f_{N p} j_{N p}}{48 R_{0}} \frac{\left( \kappa_{b}^{2} - 1 \right) j_{N} - \left( 7 \kappa_{b}^{2} + 5 \right) N_{0, 2}}{5 \kappa_{b}^{4} + 10 \kappa_{b}^{2} + 1} \Cos{\theta_{\kappa b}} \\
   A_{S 5} &\equiv - \left( - \kappa_{b}^{2} + 1 \right) \frac{f_{N p} j_{N p}}{48 R_{0}} \frac{\left( - \kappa_{b}^{2} + 1 \right) j_{N} + \left( 5 \kappa_{b}^{2} + 7 \right) N_{0, 2}}{\kappa_{b}^{4} + 10 \kappa_{b}^{2} + 5} \Sin{\theta_{\kappa b}} \\
   A_{C 3} &\equiv \frac{1}{4 R_{0}} \frac{1}{3 \kappa_{b}^{2} + 1} \left( \left( \kappa_{b}^{2} - 1 \right) \left( \frac{j_{N}}{4} + j_{N p} + N_{0, 2} \right) \right. \\
   &+ \left. \frac{1}{3} \frac{\left( - 5 \kappa_{b}^{4} + 2 \kappa_{b}^{2} + 3 \right) j_{N} + 4 \left( 5 \kappa_{b}^{4} + 4 \kappa_{b}^{2} + 3 \right) N_{0, 2}}{5 \kappa_{b}^{4} + 10 \kappa_{b}^{2} + 1} \kappa_{b}^{2} a^{2} f_{N p} j_{N p} \right) \Cos{\theta_{\kappa b}} \nonumber \\
   A_{S 3} &\equiv \frac{1}{4 R_{0}} \frac{1}{\kappa_{b}^{2} + 3} \left( \left( - \kappa_{b}^{2} + 1 \right) \left( \frac{j_{N}}{4} + j_{N p} - N_{0, 2} \right) \right. \\
   &+ \left. \frac{1}{3} \frac{\left( 3 \kappa_{b}^{4} + 2 \kappa_{b}^{2} - 5 \right) j_{N} - 4 \left( 3 \kappa_{b}^{4} + 4 \kappa_{b}^{2} + 5 \right) N_{0, 2}}{\kappa_{b}^{4} + 10 \kappa_{b}^{2} + 5} \kappa_{b}^{2} a^{2} f_{N p} j_{N p} \right) \Sin{\theta_{\kappa b}} \nonumber \\
   A_{C 1} &\equiv - \frac{1}{4 R_{0}} \frac{\kappa_{b}^{2} a^{2}}{3 \kappa_{b}^{2} + 1} \left( \left( j_{N} + 4 j_{N p} + 4 N_{0, 2} \right) \right. \\
   &- \left. \frac{4}{3} \frac{2 \left( \kappa_{b}^{2} + 1 \right) j_{N} + \left( \kappa_{b}^{2} + 7 \right) N_{0, 2}}{5 \kappa_{b}^{4} + 10 \kappa_{b}^{2} + 1} \kappa_{b}^{2} a^{2} f_{N p} j_{N p} \right) \Cos{\theta_{\kappa b}} \nonumber \\
   A_{S 1} &\equiv \frac{1}{4 R_{0}} \frac{\kappa_{b}^{2} a^{2}}{\kappa_{b}^{2} + 3} \left( \left( j_{N} + 4 j_{N p} - 4 N_{0, 2} \right) \right. \\
   &- \left. \frac{4}{3} \frac{2 \left( \kappa_{b}^{2} + 1 \right) j_{N} - \left( 7 \kappa_{b}^{2} + 1 \right) N_{0, 2}}{\kappa_{b}^{4} + 10 \kappa_{b}^{2} + 5} \kappa_{b}^{2} a^{2} f_{N p} j_{N p} \right) \Sin{\theta_{\kappa b}} \nonumber
\end{align}
and $A_{C m} = A_{S m} = 0$ for all other $m$. These coefficients reduce to those found in section 2.1.2 of reference \cite{BallMastersThesis2013} when $f_{N p} = 0$ as expected.

The above equations give the full solution to the Grad-Shafranov equation to lowest and next order in aspect ratio for a constant toroidal current profile, linear (in $\psi$) pressure gradient, and tilted elliptical boundary. We want to substitute these solutions into equation \refEq{eq:magAxisCondition} and solve for $r_{\text{axis}}$ and $\theta_{\text{axis}}$, the minor radial and poloidal locations of the magnetic axis. The simplest approach is to first expand equation \refEq{eq:magAxisCondition} to lowest order in $\epsilon \ll 1$ and change to Cartesian coordinates to find
\begin{align}
   \left. \Nabla \psi_{0} \left( R, Z \right) \right|_{R = R_{\text{axis} 0}, Z = Z_{\text{axis} 0}} + \left. \Nabla \psi_{1} \left( R, Z \right) \right|_{R = R_{0}, Z = 0} = 0 , \label{eq:magAxisConditionExpansion}
\end{align}
where $R_{\text{axis} 0}$ and $Z_{\text{axis} 0}$ are the lowest order solutions for the major radial and axial locations of the magnetic axis respectively. The solution to this,
\begin{align}
   R_{\text{axis} 0} - R_{0} =& ~ \frac{1}{2} \frac{S_{0, 2} S_{1, 1} - \left( \frac{j_{N}}{4} - C_{0, 2} \right) C_{1, 1}}{\left( \frac{j_{N}}{4} \right)^{2} - C_{0, 2}^{2} - S_{0, 2}^{2}} \label{eq:lowestOrderMagAxisRadialLoc} \\
   Z_{\text{axis} 0} =& ~ \frac{1}{2} \frac{S_{0, 2} C_{1, 1} - \left( \frac{j_{N}}{4} + C_{0, 2} \right) S_{1, 1}}{\left( \frac{j_{N}}{4} \right)^{2} - C_{0, 2}^{2} - S_{0, 2}^{2}} . \label{eq:lowestOrderMagAxisAxialLoc}
\end{align}
is easy to find and gives the location of the magnetic axis to first order in $\epsilon \ll 1$. However, this turns out to be a fairly poor approximation and does not produce close agreement with the numerical results from ECOM. However, if we solve equation \refEq{eq:magAxisCondition} exactly we get a much better approximation that matches ECOM. The crucial step to solving equation \refEq{eq:magAxisCondition} exactly is to guess that the lowest order solution for the location of the magnetic axis in equations \refEq{eq:lowestOrderMagAxisRadialLoc} and \refEq{eq:lowestOrderMagAxisAxialLoc} has the exactly correct tilt angle, i.e.
\begin{align}
   \theta_{\text{axis}} =& ~ \theta_{\text{axis} 0} = \ArcTan{\frac{Z_{\text{axis} 0}}{R_{\text{axis} 0} - R_{0}}} .  \label{eq:magAxisPoloidalLocExact}
\end{align}
We can see that this is indeed true by substituting equation \refEq{eq:magAxisPoloidalLocExact} into equation \refEq{eq:magAxisCondition}, which produces a quartic equation of the form
\begin{align}
   d_{4} r_{\text{axis}}^{4} + d_{2} r_{\text{axis}}^{2} + d_{1} r_{\text{axis}} + d_{0} = 0 \label{eq:magAxisRadialLocExactCondition}
\end{align}
with coefficients
\begin{align}
   d_{4} &\equiv - \frac{5 f_{N p} j_{N}}{8 R_{0}} \left[ \frac{j_{N} \Cos{\theta_{\text{axis} 0}}}{6} + \frac{C_{0, 2} \Cos{\theta_{\text{axis} 0}} + S_{0, 2}  \Sin{\theta_{\text{axis} 0}}}{3} \right. \\
   &+ \left. \frac{C_{0, 2} \Cos{3 \theta_{\text{axis} 0}} + S_{0, 2}  \Sin{\theta_{3 \text{axis} 0}}}{2} \right] + 5 \left( C_{1, 5} \Cos{5 \theta_{\text{axis} 0}} + S_{1, 5}  \Sin{5 \theta_{\text{axis} 0}} \right) \nonumber \\
   d_{2} &\equiv \frac{3}{4 R_{0}} \left[ \left( \frac{j_{N} + 4 j_{N p}}{4} + C_{0, 2} \right) \Cos{\theta_{\text{axis} 0}} + S_{0, 2}  \Sin{\theta_{\text{axis} 0}} \right] \\
   &+ 3 \left( C_{1, 3} \Cos{3 \theta_{\text{axis} 0}} + S_{1, 3}  \Sin{3 \theta_{\text{axis} 0}} \right) \nonumber \\
   d_{1} &\equiv 2 \left( \frac{j_{N}}{4} + C_{0, 2} \Cos{2 \theta_{\text{axis} 0}} + S_{0, 2}  \Sin{2 \theta_{\text{axis} 0}} \right) \\
   d_{0} &\equiv C_{1, 1} \Cos{\theta_{\text{axis} 0}} + S_{1, 1}  \Sin{\theta_{\text{axis} 0}} .
\end{align}
The exact location of the magnetic axis is given by solution of this quartic and equation \refEq{eq:magAxisPoloidalLocExact}. Quartics have a very complicated analytic solution, so in practice it is simpler to solve computationally. However, for the special case of $f_{N p} = 0$ we see that $d_{4} = 0$ and the quartic reduces to a quadratic solved by
\begin{align}
   r_{\text{axis}} =& ~ \frac{- d_{1} + \sqrt{d_{1}^{2} - 4 d_{2} d_{0}}}{2 d_{2}} . \label{eq:magAxisRadialLocExact}
\end{align}

\section{Dependence of the gyrokinetic geometric coefficients on $\beta'$}
\label{app:geoCoeff}

In this appendix, we will study the sensitivity of the momentum flux to $\beta'$ by investigating how the gyrokinetic equation changes with $\beta'$. The magnetic geometry only enters the electrostatic local gyrokinetic model (in the absence of rotation) through eight geometric coefficients \cite{BallMirrorSymArg2016}: $\hat{b} \cdot \Nabla \theta$, $B$,
\begin{align}
   v_{d s \psi} \equiv& ~ \vec{v}_{d s} \cdot \Nabla \psi = - \frac{I \left( w_{||}^{2} + \mu B / m_{s} \right)}{\Omega_{s} B} \hat{b} \cdot \Nabla \theta \frac{\partial B}{\partial \theta} \label{eq:driftVelPsi} , \\
   v_{d s \alpha} \equiv& ~ \vec{v}_{d s} \cdot \Nabla \alpha = - \frac{w_{||}^{2} + \mu B / m_{s}}{\Omega_{s}} \left[ \frac{\partial B}{\partial \psi} - \frac{\partial B}{\partial \theta} \frac{ \hat{b} \cdot \left( \Nabla \theta \times \Nabla \alpha \right) }{B} \right] - \frac{\mu_{0} w_{||}^{2}}{\Omega_{s} B} \frac{d p}{d \psi} , \label{eq:driftVelAlpha} \\
   a_{s ||} \equiv& - \frac{\mu}{m_{s}} \hat{b} \cdot \Nabla \theta \left. \frac{\partial B}{\partial \theta} \right|_{\psi} , \label{eq:parAccelDef}
\end{align}
$\left| \Nabla \psi \right|^{2}$, $\Nabla \psi \cdot \Nabla \alpha$, and $\left| \Nabla \alpha \right|^{2}$. Here $\hat{b} \equiv \vec{B} / B$ is the magnetic field unit vector, $\vec{v}_{d s}$ is the guiding center particle drift velocity, $\alpha$ (defined by equation \refEq{eq:alphaDef}) is the coordinate within the flux surface and perpendicular to the magnetic field, $\mu \equiv m_{s} w_{\perp}^{2} / 2 B$ is the magnetic moment, $m_{s}$ is the particle mass of species $s$, and $\Omega_{s}$ is the gyrofrequency.

The calculation of the geometric coefficients in GS2 is done in the context of the Miller local equilibrium \cite{MillerGeometry1998}. This must be done carefully as the Miller model takes the flux surface shape and its radial derivative as input, but all second order radial derivatives are calculated through the Grad-Shafranov equation. It is through these second order radial derivatives (as well as the explicit dependence appearing in $v_{d s \alpha}$) that $\beta'$ enters the geometric coefficients. Additionally, we note that we keep the safety factor, the magnetic shear, the background gradients, and the geometry fixed as we change $\beta'$. Therefore, while the Shafranov shift directly enters the flux surface geometry and affects all of the geometric coefficients, the effect of $\beta'$ is limited to a few coefficients. The parameter $\beta'$, which is a normalized form of $d p / d r_{\psi}$ (see equation \refEq{eq:betaPrimeDef}), only enters into three coefficients: $v_{d s \alpha}$, $\Nabla \psi \cdot \Nabla \alpha$, and $\left| \Nabla \alpha \right|^{2}$. We will start with equations derived in reference \cite{BallMomFluxScaling2016} to show precisely how $\beta'$ enters and that its effect is small in the inverse aspect ratio $\epsilon \ll 1$, when using the ohmically heated tokamak ordering (see equation \refEq{eq:gradShafOrderings}).

First we combine equations (B.16) and (6) from reference \cite{BallMomFluxScaling2016} to get
\begin{align}
   \underbrace{I \frac{d I}{d \psi}}_{B_{0}} =& ~ \bigg( \underbrace{\frac{2 \pi q}{I^{3}}}_{R_{0}^{-3} B_{0}^{-3}} + \underbrace{\left. \oint_{0}^{2 \pi} \right|_{\psi} d \theta' \frac{1}{R^{4} B_{p}^{2} \vec{B} \cdot \Nabla \theta'}}_{\epsilon^{-2} R_{0}^{-3} B_{0}^{-3}} \bigg)^{-1} \Bigg( \underbrace{\frac{2 \pi}{I} \frac{d q}{d \psi}}_{\epsilon^{-2} R_{0}^{-3} B_{0}^{-2}} \label{eq:IdIdpsi} \\
   &- \left. \oint_{0}^{2 \pi} \right|_{\psi} d \theta' \underbrace{\frac{1}{R^{2} \vec{B} \cdot \Nabla \theta'}}_{R_{0}^{-1} B_{0}^{-1}} \bigg[ \underbrace{\frac{\mu_{0}}{B_{p}^{2}} \frac{d p}{d \psi}}_{\epsilon^{-2} R_{0}^{-2} B_{0}^{-1}} - \underbrace{\frac{2}{R^{2} B_{p}} \left( \dlpdthetaPrime \right)^{-1} \frac{\partial Z}{\partial \theta'}}_{\epsilon^{-1} R_{0}^{-2} B_{0}^{-1}} + \underbrace{\frac{2 \kappa_{p}}{R B_{p}}}_{\epsilon^{-2} R_{0}^{-2} B_{0}^{-1}} \bigg] \Bigg) , \nonumber
\end{align}
where the curly braces below the different terms give their ordering in $\epsilon \ll 1$, $l_{p}$ is the poloidal arc length such that $\partial l_{p} / \partial \theta = \sqrt{ \left( \partial R / \partial \theta \right)^{2} + \left( \partial Z / \partial \theta \right)^{2}}$, $\kappa_{p} \equiv - \left( \hat{b}_{p} \cdot \Nabla \hat{b}_{p} \right) \cdot \Nabla \psi / \left| \Nabla \psi \right|$ is the curvature of the poloidal magnetic field, and $\hat{b}_{p} \equiv \vec{B}_{p} / B_{p}$ is the poloidal field unit vector. We see that introducing $\beta'$ creates a lowest order modification to $I \left( d I / d \psi \right)$. Next, using equation \refEq{eq:toroidalCur} of this paper, we can find that the right-hand side of the Grad-Shafranov equation can be written as
\begin{align}
   \underbrace{\mu_{0} j_{\zeta} R}_{B_{0}} &= - \bigg( \underbrace{\frac{2 \pi q}{I^{3}}}_{R_{0}^{-3} B_{0}^{-3}} + \underbrace{\left. \oint_{0}^{2 \pi} \right|_{\psi} d \theta' \frac{1}{R^{4} B_{p}^{2} \vec{B} \cdot \Nabla \theta'}}_{\epsilon^{-2} R_{0}^{-3} B_{0}^{-3}} \bigg)^{-1} \label{eq:betaPrimeInhomoGradShaf} \\
   &\times \Bigg[ \underbrace{\mu_{0} R^{2} \frac{d p}{d \psi}}_{B_{0}} \Bigg( \underbrace{\frac{2 \pi q}{I^{3}}}_{R_{0}^{-3} B_{0}^{-3}} + \underbrace{\left. \oint_{0}^{2 \pi} \right|_{\psi} d \theta' \frac{1}{R^{4} B_{p}^{2} \vec{B} \cdot \Nabla \theta'}}_{\epsilon^{-2} R_{0}^{-3} B_{0}^{-3}} - \underbrace{\frac{R_{0}^{2}}{R^{2}} \left. \oint_{0}^{2 \pi} \right|_{\psi} d \theta' \frac{1}{R_{0}^{2} R^{2} B_{p}^{2} \vec{B} \cdot \Nabla \theta'}}_{\epsilon^{-2} R_{0}^{-3} B_{0}^{-3}} \Bigg) \nonumber \\
   &+ \underbrace{\frac{2 \pi}{I} \frac{d q}{d \psi}}_{\epsilon^{-2} R_{0}^{-3} B_{0}^{-2}} - \left. \oint_{0}^{2 \pi} \right|_{\psi} d \theta' \underbrace{\frac{1}{R^{2} \vec{B} \cdot \Nabla \theta'}}_{R_{0}^{-1} B_{0}^{-1}} \bigg( \underbrace{\frac{2 \kappa_{p}}{R B_{p}}}_{\epsilon^{-2} R_{0}^{-2} B_{0}^{-1}} - \underbrace{\frac{2}{R^{2} B_{p}} \left( \dlpdthetaPrime \right)^{-1} \frac{\partial Z}{\partial \theta'}}_{\epsilon^{-1} R_{0}^{-2} B_{0}^{-1}} \bigg) \Bigg] , \nonumber
\end{align}
which explicitly includes a term proportional to the pressure gradient (i.e. $\beta'$). However, to lowest order in aspect ratio the coefficient of this term is zero as it is composed of a safety factor term that is small and two integral terms that cancel with each other (because $R = R_{0} + O \left( \epsilon R_{0} \right)$). All other quantities in equation \refEq{eq:betaPrimeInhomoGradShaf} do not contain the pressure gradient and can be calculated directly from the flux surface geometry provided to the Miller model. Therefore, $\beta'$ only introduces an $O \left( \epsilon B_{0} \right)$ modification to $\mu_{0} j_{\zeta} R$.

We will see that the toroidal current density (i.e. $\mu_{0} j_{\zeta} R$) will appear in several places in the geometric coefficients. Equation (B.6) from reference \cite{BallMomFluxScaling2016} gives the radial derivative of the poloidal field as
\begin{align}
   \underbrace{\frac{\partial B_{p}}{\partial \psi}}_{a^{-1} R_{0}^{-1}} =& ~ \underbrace{\frac{\mu_{0} j_{\zeta} R}{R^{2} B_{p}}}_{a^{-1} R_{0}^{-1}} - \underbrace{B_{p} \left( \dlpdtheta \right)^{-1} \frac{\partial}{\partial \psi} \left( \dlpdtheta \right)}_{a^{-1} R_{0}^{-1}} \label{eq:dBpdpsi} \\
   &+ \underbrace{\left( \dlpdtheta \right)^{-1} \frac{\partial}{\partial \theta} \left( B_{p} \left( \dlpdtheta \right)^{-1} \frac{\partial \vec{r}}{\partial \psi} \cdot \frac{\partial \vec{r}}{\partial \theta} \right)}_{a^{-1} R_{0}^{-1}} . \nonumber
\end{align}
Although the toroidal current term appears as $O \left( a^{-1} R_{0}^{-1} \right)$, the effect of $\beta'$ on $\partial B_{p} / \partial \psi$ is small by an order (i.e. $O \left( \epsilon a^{-1} R_{0}^{-1} \right)$) because $\beta'$ does not enter $\mu_{0} j_{\zeta} R$ to lowest order. We can directly differentiate $B_{\zeta} = I / R$ to get
\begin{align}
   \underbrace{\frac{\partial B_{\zeta}}{\partial \psi}}_{a^{-1} R_{0}^{-1}} =& -\underbrace{ \frac{I}{R^{2}} \frac{\partial R}{\partial \psi}}_{a^{-1} R_{0}^{-1}} + \underbrace{\frac{1}{R} \frac{d I}{d \psi}}_{\epsilon a^{-1} R_{0}^{-1}} . \label{eq:dBtordpsi}
\end{align}
Ordering these two terms we see that the effect of $d I / d \psi$ is small, so the effect of $\beta'$ on $\partial B_{\zeta} / \partial \psi$ through equation \refEq{eq:IdIdpsi} is small by one order, entering at $O \left( \epsilon a^{-1} R_{0}^{-1} \right)$.

Using equations (B.14) and (6) from reference \cite{BallMomFluxScaling2016} gives
\begin{align}
   \underbrace{\Nabla \alpha}_{a^{-1}} =& \Bigg( - \left. \int_{\theta_{\alpha}}^{\theta} \right|_{\psi} d \theta' \underbrace{\frac{I}{R^{2} \vec{B} \cdot \Nabla \theta'}}_{1} \bigg[ \underbrace{\frac{1}{I} \frac{d I}{d \psi}}_{\epsilon^{2} a^{-2} B_{0}^{-1}} - \underbrace{\frac{\mu_{0} j_{\zeta} R}{R^{2} B_{p}^{2}}}_{a^{-2} B_{0}^{-1}} - \underbrace{\frac{2}{R^{2} B_{p}} \left( \dlpdthetaPrime \right)^{-1} \frac{\partial Z}{\partial \theta'}}_{\epsilon a^{-2} B_{0}^{-1}} + \underbrace{\frac{2 \kappa_{p}}{R B_{p}}}_{a^{-2} B_{0}^{-1}} \bigg] \nonumber \\
   &+ \underbrace{\left[ \frac{I \Nabla \psi \cdot \Nabla \theta'}{R^{4} B_{p}^{2} \vec{B} \cdot \Nabla \theta'} \right]_{\theta' = \theta_{\alpha}}^{\theta' = \theta}}_{a^{-2} B_{0}^{-1}} + \underbrace{\left( \frac{I}{R^{2} \vec{B} \cdot \Nabla \theta'} \right)_{\theta' = \theta_{\alpha}} \frac{d \theta_{\alpha}}{d \psi}}_{a^{-2} B_{0}^{-1}} \Bigg) \underbrace{\Nabla \psi}_{a B_{0}} \label{eq:gradAlphaFinal} \\
   &- \underbrace{\frac{I}{R^{2} \vec{B} \cdot \Nabla \theta} \Nabla \theta}_{a^{-1}} + \underbrace{\Nabla \zeta}_{\epsilon a^{-1}} . \nonumber
\end{align}
By ordering the various terms we find that the $d I / d \psi$ term is small by two orders in $\epsilon \ll 1$. However, the $\mu_{0} j_{\zeta} R$ term enters to lowest order, therefore the effect of $\beta'$ on $\Nabla \alpha$ is only small by one order (i.e. $\Order{\epsilon a^{-1}}$). The dependence of the coefficients $\Nabla \psi \cdot \Nabla \alpha$ and $\left| \Nabla \alpha \right|^{2}$ on $\Nabla \alpha$ is apparent. Hence $\beta'$ does not enter $\Nabla \psi \cdot \Nabla \alpha$ and $\left| \Nabla \alpha \right|^{2}$ to lowest order in $\epsilon \ll 1$. Instead it enters to next order due to the quantity $\mu_{0} j_{\zeta} R$, which is given by equation \refEq{eq:betaPrimeInhomoGradShaf}. The geometric coefficient $v_{d s \alpha}$ is more complicated. Substituting equation \refEq{eq:dBtordpsi} into equation \refEq{eq:driftVelAlpha} gives
\begin{align}
   \underbrace{v_{d s \alpha}}_{a^{-1} R_{0}^{-1} v_{th, s}^{2} \Omega_{s}^{-1}} =& - \underbrace{\frac{w_{||}^{2}}{\Omega_{s}}}_{v_{th, s}^{2} \Omega_{s}^{-1}} \Bigg( - \underbrace{\frac{\mu_{0} j_{\zeta} R}{R^{2} B}}_{\epsilon a^{-1} R_{0}^{-1}} - \underbrace{\frac{I^{2}}{R^{3} B} \frac{\partial R}{\partial \psi}}_{a^{-1} R_{0}^{-1}} + \underbrace{\frac{B_{p}}{B} \frac{\partial B_{p}}{\partial \psi}}_{\epsilon a^{-1} R_{0}^{-1}} - \underbrace{\frac{\partial B}{\partial \theta} \frac{ \hat{b} \cdot \left( \Nabla \theta \times \Nabla \alpha \right)}{B}}_{a^{-1} R_{0}^{-1}} \Bigg) \\
   &- \underbrace{\frac{\mu B}{m_{s} \Omega_{s}}}_{v_{th, s}^{2} \Omega_{s}^{-1}} \Bigg( \underbrace{\frac{I}{R^{2} B} \frac{d I}{d \psi}}_{\epsilon a^{-1} R_{0}^{-1}} - \underbrace{\frac{I^{2}}{R^{3} B} \frac{\partial R}{\partial \psi}}_{a^{-1} R_{0}^{-1}} + \underbrace{\frac{B_{p}}{B} \frac{\partial B_{p}}{\partial \psi}}_{\epsilon a^{-1} R_{0}^{-1}} - \underbrace{\frac{\partial B}{\partial \theta} \frac{ \hat{b} \cdot \left( \Nabla \theta \times \Nabla \alpha \right)}{B}}_{a^{-1} R_{0}^{-1}} \Bigg) \nonumber .
\end{align}
We see that $\beta'$ will enter into the $\mu_{0} j_{\zeta} R$ term as well as both $\partial B_{p} / \partial \psi$ terms, but ordering these three terms reveals that the effect of $\beta'$ is $\Order{\epsilon^{2} a^{-1} R_{0}^{-1} v_{th, s}^{2} \Omega_{s}^{-1}}$. The parameter $\beta'$ has a much larger $O \left( \epsilon a^{-1} R_{0}^{-1} v_{th, s}^{2} \Omega_{s}^{-1} \right)$ effect through the two $\Nabla \alpha$ terms as well as the $d I / d \psi$ term. Figure \ref{fig:driftCoeff} illustrates the relative magnitudes of these two effects for a few typical geometries. The difference between the dotted red line and the dashed blue line indicates the effect of the $d I / d \psi$ term, while the difference between the solid black line and the dotted red line indicates the effect of $\mu_{0} j_{\zeta} R$ acting through $\Nabla \alpha$. We see that the effect of $\mu_{0} j_{\zeta} R$ seems to dominate.

In conclusion, $\beta'$ only enters into three of the geometric coefficients: $v_{d s \alpha}$, $\Nabla \psi \cdot \Nabla \alpha$, and $\left| \Nabla \alpha \right|^{2}$. The dominant effect of $\beta'$ on $\Nabla \psi \cdot \Nabla \alpha$ and $\left| \Nabla \alpha \right|^{2}$ is contained in the quantity $\mu_{0} j_{\zeta} R$ and is small in $\epsilon \ll 1$. The drift coefficient $v_{d s \alpha}$ also depends on $\beta'$ to next order because of $\mu_{0} j_{\zeta} R$. However, it has another separate dependence through the quantity $d I / d \psi$ that is formally the same size in $\epsilon \ll 1$, but in practice this appears to be a weak effect. These dependences are the only way that the gyrokinetic model knows about $\beta'$. Hence they must be responsible for the significant reduction in the momentum transport.

\section*{References}
\bibliographystyle{unsrt}
\bibliography{references.bib}

\end{document}